\documentclass[reprint, amsmath, amssymb, aps, prx, twocolumn, superscriptaddress]{revtex4-2}

\usepackage{graphicx} 
\usepackage{dcolumn}  
\usepackage{bm}       
\usepackage{xcolor}   
\usepackage{tikz}
\usepackage{makecell}

\usepackage{algorithm}
\usepackage{algpseudocode}
\usepackage{hyperref} 

\usetikzlibrary{positioning, arrows.meta, calc}

\usepackage[margin=1.5cm]{geometry}
\usepackage{subcaption}
\graphicspath{{output/}}

\usetikzlibrary{positioning, fit, arrows.meta, calc, backgrounds}

\usepackage{multirow}

\usepackage{soul}

\definecolor{bgblue}{RGB}{68, 75, 105}      
\definecolor{initcol}{RGB}{124, 98, 234}    
\definecolor{growthcol}{RGB}{155, 89, 182}  
\definecolor{vqecol}{RGB}{93, 173, 226}     
\definecolor{compcol}{RGB}{72, 201, 176}    
\definecolor{prunecol}{RGB}{88, 214, 141}   
\definecolor{ept2col}{RGB}{46, 204, 113}    
\definecolor{condcol}{RGB}{52, 152, 219}    
\definecolor{projcol}{RGB}{241, 148, 138}

\newcommand{\methodename}{VULCAN }
\newcommand{\methodenamens}{VULCAN}

\begin{document}

\title{Grow, Compress, and Perturb: A Hybrid Quantum-Classical Pathway to Scalable Quantum Chemistry on Quantum Computers}

\author{Hugo D. Nogueira}
\affiliation{Sorbonne Universit\'e, Laboratoire de Chimie Th\'eorique, UMR 7616 CNRS, 75005 Paris, France}

\author{Siwar Badreddine}
\affiliation{Qubit Pharmaceuticals, Advanced Research Department, 75014 Paris, France}

\author{Olivier Adjoua}
\affiliation{Sorbonne Universit\'e, Laboratoire de Chimie Th\'eorique, UMR 7616 CNRS, 75005 Paris, France}

\author{Diata Traore}
\affiliation{Qubit Pharmaceuticals, Advanced Research Department, 75014 Paris, France}

\author{Guillaume Michel}
\affiliation{Qubit Pharmaceuticals, Advanced Research Department, 75014 Paris, France}

\author{Jean-Philip Piquemal}
\email{jean-philip.piquemal@sorbonne-universite.fr}
\affiliation{Sorbonne Universit\'e, Laboratoire de Chimie Th\'eorique, UMR 7616 CNRS, 75005 Paris, France}
\affiliation{Qubit Pharmaceuticals, Advanced Research Department, 75014 Paris, France}

\date{\today}

\begin{abstract}

While Variational Quantum Eigensolver (VQE) variants are among the leading candidates for electronic structure calculations on quantum computers, gradient-based operator selection limits their scalability. To solve this, we introduce the Hamiltonian-gradient-free \methodename (Variational Upbuilding, Layered Compression, Adaptive N-body perturbation) layered VQE framework. Our method dynamically generates ansatz operators using Variational Double Bracket Flow (VDBF) equation. Combined with a  minimal complete pool, whose size
scales linearly with the number of qubits, it creates a pool 4–25 times smaller than QEB (Qubit-Excitation-Based) pools. By appending entire layers simultaneously, Hamiltonian gradient measurements are completely eliminated. Circuit depth is managed via a grow-then-compress scheme using efficient state-overlap measurements, and qubit requirements are reduced by projecting the Hamiltonian onto a dominant-determinant subspace. Using high-performance emulation up to 32 qubits, we achieve chemical
accuracy at just 5--20\% of the measurement cost of batched ADAPT-VQE for
hydrogen chains, BeH$_2$ and H$_2$O. For the 32-qubit C$_3$H$_6$, where batched
ADAPT-VQE does not reach chemical accuracy, a single VULCAN layer combined with
a GPU-accelerated stochastic Epstein--Nesbet (ENPT2) correction does. This stochastic post-processing step enables early termination at chemical accuracy. Discussions of hardware implementations are reported alongside an initial proof-of-principle experiment on a Quantinuum System Model H2 trapped-ion quantum processor.

\end{abstract}

\maketitle


\section{Introduction}

Accurate calculation of ground state energies for strongly correlated many-body systems represents a major computational bottleneck for classical computing, as the exponential growth of the Hilbert space renders exact classical diagonalization intractable  for large-scale electronic structure problems. Quantum computing offers an alternative by mapping exponentially large electronic states onto a linear number of qubits. Fully quantum algorithms such as Quantum Phase Estimation (QPE) \cite{lee2021even} achieve this in principle, but their reliance on deep circuits and large numbers of logical qubits currently places them beyond the capabilities of the Noisy Intermediate-Scale Quantum (NISQ) devices \cite{preskill2018quantum}. The Variational Quantum Eigensolver (VQE) \cite{peruzzo2014variational} has consequently emerged as the leading hybrid quantum-classical algorithm for near-term quantum computation. 

Despite its promise, the practical deployment of VQE is hindered by fundamental bottlenecks \cite{haidar2023open,harville2026recent}, particularly in ansatz design. Standard hardware-efficient ansätze  construct highly expressive but unstructured variational manifolds, which severely limit trainability. This lack of physical structure  leads to the barren plateau phenomenon, where gradients vanish exponentially with system size, and to  a critical sensitivity to the initial reference state.

The ADAPT-VQE \cite{Grimsley2019} family of algorithms \cite{tang2021qubit,Yordanov2021,feniou2023overlap} addresses this by dynamically constructing problem-tailored circuits. Iteratively adding only the most relevant operators from a predefined pool restricts the variational manifold to physically relevant subspaces, yielding highly compact ansätze that remain resilient to barren plateaus. This adaptive construction introduces a severe trade-off: selecting each operator requires evaluating a polynomially scaling number of energy gradients. Recent measurement protocols have elegantly reduced this  cost \cite{anastasiou2023really}, but the overall measurement overhead remains high, and under shot noise and hardware error these gradient evaluations become noisy enough to produce sub-optimal operator choices. With this in mind, some of us introduced a gradient-approximated variant of ADAPT-VQE denoted GGA-VQE (Greedy Gradient-free Adaptive VQE)\cite{feniou2025greedy} to mitigate statistical noise and bypass the measurement-heavy classical optimization of standard ADAPT-VQE. GGA-VQE uses an analytical 1D energy landscape extrapolation (requiring only a few measurements) to select a pool operator and its optimal angle simultaneously, freezing previous parameters rather than re-optimizing the entire circuit. However, this greedy, local-only optimization strategy makes the algorithm prone to trapping in shallow local minima for strongly correlated systems, where it requires additional post-processing sweeps, such as sequential re-optimization, and it still suffers from an $O(M)$ measurement overhead per step across candidate operators in the pool unless special Hamiltonian symmetries can be exploited.
In that context, we introduce \methodename (Variational Upbuilding, Layered Compression, Adaptive N-body perturbation) -- a hybrid quantum-classical
framework based on a ``grow, compress, and perturb" architecture that directly
mitigates these bottlenecks. Its hardware efficiency comes from removing the
gradient-screening step altogether: no energy gradients are measured to select
operators, which reduces the number of Pauli-term evaluations needed on the
quantum computer. To do so, we leverage some ingredients that have recently been proposed.

First, Minimal Complete Pools (MCPs) \cite{Viswanathan2026,Shkolnikov2023,tang2021qubit,feniou2025greedy} provide a robust mathematical framework for constructing generator sets. While scaling only linearly with the number of qubits $N$, the use of MCPs has long been limited by the intractable search for the required generators. This has changed recently with the availability of a general mathematical framework \cite{Viswanathan2026} scaling polynomially with the number of qubits to enable the fast verification of the completeness of a user-defined pool of Pauli string operators with respect to a target Lie algebra.  However, our previous work integrating MCPs into both adaptive \cite{Viswanathan2026} and fixed Lie-algebraic \cite{Viswanathan2026, Haidar2025} frameworks to reduce quantum resources showed that their restricted size shifts the
computational burden onto the choice of \textit{starter} operators. Identifying the optimal starters remains an open challenge in the pursuit of rapid convergence and chemical accuracy for strongly correlated systems. Existing static pre-screening strategies, such as the Unitary Selective Coupled-Cluster (USCC) method \cite{fedorov2022unitary}, can  isolate physically relevant starters from Hamiltonian matrix elements, but require preliminary  VQE runs, whose optimized amplitudes must be computed and stored before screening can begin \cite{Haidar2025,Viswanathan2026}. This reliance on preliminary computational overhead simply replaces one bottleneck with another, rendering static pre-screening highly undesirable for scalable NISQ hardware deployment.
Therefore, in order to eliminate this bottleneck, we propose as an alternative to generate this starter pool dynamically, without any preliminary classical optimization. 
\methodename is inspired by MB-ADAPT-VQE~\cite{Viswanathan2026}, which grows the ansatz by adding, at each iteration, a batch of operators drawn from an MCP built around physically motivated starters.  We keep this MCP-based construction,
but replace both the pre-screening of the starters and the gradient-based
selection, as detailed below, and compress each new layer using state-overlap
measurements to limit the circuit depth.

Furthermore, in the spirit of non-iterative VQE approaches such as NI-DUCC (Non-Iterative Disentangled Unitary Coupled-Cluster) \cite{Haidar2025}, we choose to retain a layered product-of-exponentials ansatz form, while departing from it in two respects: i) the number of layers is grown
until convergence rather than fixed in advance, all parameters are re-optimized after each new layer is appended; and ii) the generators
populating each layer are derived from the Hamiltonian itself rather than
chosen beforehand. Specifically, we employ the Variational Double Bracket Flow (VDBF) equation \cite{wegner1994flow,shrikhande2025rapid}
to expose the variationally useful excitations. Because entire layers are appended at once, the ADAPT selection phase is  bypassed and no Hamiltonian gradient is ever measured. The resulting ansatz reads 
\begin{equation} \label{eq-ans}
    |\psi\rangle =  \prod_{k=1}^{n} \big(  \prod_{j=1}^{M}\text{exp}(i\theta_{j,k}P_j)\big)|\psi_0\rangle\,,
\end{equation}
where $|\psi_0\rangle\ $ is the Hartree--Fock state, the $P_j$ are the $M$ pool operators expressed as Pauli strings, the $\theta_{j,k}$ are the variational parameters, and $n$ is the number of layers. 

Appending complete layers increases the circuit depth quickly, so the approach pairs growth with three mechanisms that trade one hardware resource against another. Depth is controlled by a grow-then-compress scheme inspired by Overlap-ADAPT-VQE \cite{feniou2023overlap}, in which candidate operators are ranked by the gradient of the overlap with a target state rather than by energy gradients. Because our pool consists of single Pauli strings, the magnitude of each such gradient is obtained from a single overlap evaluation, without reference to the Hamiltonian. Qubit count may be reduced further by projecting the Hamiltonian onto the subspace spanned by the dominant determinants of a one-layer ansatz. Finally, a stochastic implementation of the second-order Epstein-Nesbet perturbative correction (ENPT2) \cite{Ryabinkin2021} is introduced to shift the remaining error onto classical GPU-accelerated hardware. Leveraging the optimized VQE energy and wavefunction to preserve quantum resources, it permits an early termination of the quantum optimization. Together these give a tunable trade-off — measurement cost against circuit depth, qubit count, and classical post-processing, along which the user may move according to which resource is scarcest.

We term this framework \methodenamens. In noiseless emulation, on linear hydrogen chains, BeH$_2$, and H$_2$O, \methodename reaches chemical accuracy at 5–20\%  of the Pauli-term evaluation count of batched ADAPT-VQE with QEB pools, and overlap compression brings the circuit
depth below that baseline for four of the seven systems where it was applied. For H$_2$O in the 6-31G basis (26 qubits) and C$_3$H$_6$ (32 qubits), stochastic ENPT2 reaches chemical accuracy after a single layer. Subspace
projection reaches chemical accuracy in exact diagonalization with 23--50\%
fewer qubits. Section~\ref{sec:results} presents these results, and Section~\ref{sec:Methods} details the pool construction, compression loop, projection scheme and perturbative correction.

\section{Results}
\label{sec:results}
\subsection{Algorithmic Workflow and Conceptual Schematic:}

The \methodename workflow is executed in three primary phases, illustrated in Figure \ref{fig:Scheme}, with an optional projection step inserted between the first two.

\begin{itemize}
    \item \textbf{Phase 1:  Operator Pool Initialization (classical).} The algorithm employs the flow equation, which is evaluated as a commutator with the target Hamiltonian, yielding a linear combination of Pauli operators. These are split into individual Pauli strings, operators are added to
selected starters among them until a minimal complete pool of $2N-4$
generators is obtained, and the remaining VDBF operators are then appended.  The pool is produced in seconds on a classical computer and requires no preliminary VQE run.

    \item \textbf{Phase 2: Grow-then-compress loop (hybrid).}
    The algorithm enters an iterative cycle of ansatz construction and compression. Rather than selecting individual operators via Hamiltonian gradients, each iteration appends one complete pool layer to the ansatz and minimizes the energy over all parameters of the ansatz. The resulting state serves as the target for a compression sub-loop, in which
generators are inserted one at a time in order of decreasing overlap gradient
and fitted by infidelity minimization, until the infidelity stops improving
or the compressed layer reaches a set fraction of the grown layer. The
compressed layer is then re-optimized for energy.  The loop terminates when the energy change between iterations falls below $\epsilon$ or a maximum iteration count is reached.
    
    \item \textbf{Optional: Subspace projection.} 
    To maximize viability on NISQ hardware, for systems whose qubit count exceeds the available register, a single ansatz layer is optimized greedily on a classical emulator to identify its dominant excitation patterns. The Hamiltonian is projected
onto the subspace spanned by all combinations of these patterns applied to
the Hartree--Fock state. The original pool is projected onto the subspace: generators whose excitation lies in the span of the selected patterns are kept and rewritten on the reduced register, and the workflow resumes at Phase 2 (growth and compression) on that register. 

    \item \textbf{Phase 3: Classical Post-Processing (classical, GPU).} To achieve a better accuracy without adding quantum gate depth, the algorithm applies a second-order perturbation theory. The optimized energy seeds a second-order Epstein-Nesbet correction, evaluated exactly for small systems and by importance sampling for large ones. Because this step is classical, it adds no quantum resource cost and allows Phase 2 to be terminated early.
    
\end{itemize}

{%

\begin{figure*}[htbp]
\centering

\definecolor{vdbp1frame}{HTML}{3D85C6}
\definecolor{vdbp1fill}{HTML}{CFE2F3}
\definecolor{vdbp2frame}{HTML}{FF9900}
\definecolor{vdbp2fill}{HTML}{FCE5CD}
\definecolor{vdbp3frame}{HTML}{A64D79}
\definecolor{vdbp3fill}{HTML}{EAD1DC}
\definecolor{vdbsubfill}{HTML}{D9D2E9}
\definecolor{vdbedge}{HTML}{595959}
\definecolor{vdbtitle}{HTML}{666666}

\tikzset{
  box/.style={draw=vdbedge, thin, align=center, minimum width=5cm,
               minimum height=1.75cm, text width=4.6cm, font=\sffamily\small},
  side/.style={box, minimum width=2.5cm, text width=2.2cm,
               minimum height=2.85cm},
  frame/.style={draw, line width=1.6pt, rounded corners=0.9cm},
  phase/.style={align=center, font=\sffamily\normalsize},
  flow/.style={draw=vdbedge, thin, -{Latex[length=1.8mm, width=1.4mm]}},
  link/.style={draw=vdbedge, thin},
}

\resizebox{\linewidth}{!}{%
\begin{tikzpicture}

  \node[box, fill=vdbp1fill] (gen)    at (8.0, 0)     {Generation of operators\\ via flow equation};
  \node[box, fill=vdbp1fill] (split)  at (8.0, -2.55) {Splitting into individual\\ Pauli strings};
  \node[box, fill=vdbp1fill] (filter) at (8.0, -5.1)  {Filtering of operators\\ through symmetries};
  \node[box, fill=vdbp1fill] (add)    at (8.0, -7.65) {Add operators until an MCP\\ is obtained};

  \draw[flow] (gen)    -- (split);
  \draw[flow] (split)  -- (filter);
  \draw[flow] (filter) -- (add);

  \node[side, fill=vdbsubfill, dashed] (sub) at (13.225, -3.85)
    {Subspace\\ projection\\ on $k$ qubits\\ (optional)};

  \node[box, fill=vdbp2fill] (grow)  at (18.4, 0)     {Ansatz growth: append\\ complete pool (layer)};
  \node[box, fill=vdbp2fill] (vqe)   at (18.4, -2.55) {Energy VQE: minimize\\ $\langle\psi|H|\psi\rangle$ (gives the overlap target)};
  \node[box, fill=vdbp2fill] (comp)  at (18.4, -5.1)  {Overlap compression:\\ minimize infidelity};
  \node[box, fill=vdbp2fill] (prune) at (18.4, -7.65) {Pruning and second VQE:\\ minimize $\langle\psi|H|\psi\rangle$};
  \node[side, fill=vdbp2fill] (until) at (22.4, -3.85) {Until\\ $|\Delta E|<\varepsilon$ or\\ max\\ iterations\\ reached};

  \draw[flow] (grow) -- (vqe);
  \draw[flow] (vqe)  -- (comp);
  \draw[flow] (comp) -- (prune);
  \draw[link] (prune.east) -| (until.south);
  \draw[flow] (until.north) |- ($(grow.east)+(0,-0.15)$);

  \draw[flow] (add.east) -| (sub.south);
  \draw[flow] (sub.north) |- (grow.west);

  \node[box, fill=vdbp3fill, minimum width=4.6cm, text width=4.3cm] (ept2)
    at (27.8, -4.3) {Additional precision with\\ ENPT2\\ (exact or stochastic)};

  \draw[flow] (prune.south) -- ++(0,-0.3) -| (ept2.south);

  \draw[frame, vdbp1frame] (5.0, 1.3) rectangle (11.15, -9.15);
  \draw[frame, vdbp2frame] (15.3, 1.3) rectangle (24.05, -9.15);
  \draw[frame, vdbp3frame, rounded corners=0.45cm]
    ($(ept2.north west)+(-0.45,0.65)$) rectangle ($(ept2.south east)+(0.45,-0.45)$);

  \node[phase, above=0.1cm] at (8.075, 1.3)  {Phase 1 -- Pool generation\\ Classical};
  \node[phase, above=0.1cm] at (19.675, 1.3) {Phase 2 -- Grow-then-compress\\ Hybrid quantum-classical};
  \node[phase, above=0.1cm] at ($(ept2.north)+(0,0.65)$) {Phase 3 -- Correction\\ Classical, GPU};

\end{tikzpicture}%
}%

\caption{Overview of the algorithmic pipeline. Phase~1 generates the
operator pool classically: the VDBF flow generator is evaluated as a
commutator with the target Hamiltonian and split into individual Pauli
strings; operators are added to selected
starters among them until a minimal complete pool is obtained, and the
remaining VDBF operators are then appended. An optional projection step,
between Phases~1 and~2, selects $k$ excitation patterns from the first-layer
ansatz, takes $H_{\rm proj}$ as the projection of $H$ on the $2^k$
determinants they generate, and projects the pool onto the reduced register.
Phase~2 grows the ansatz one complete layer at a time, re-optimizing  its parameters after each new layer. The optimized state serves as the
target of an overlap-guided compression step, which inserts generators by
decreasing overlap gradient until the infidelity gain becomes negligible or
the compressed layer reaches a set fraction of the grown layer; the
compressed layer is then re-optimized for energy before the next growth
iteration. Phase~3 applies a classical, GPU-accelerated ENPT2 correction,
permitting early termination of Phase~2.}
\label{fig:Scheme}
\end{figure*}
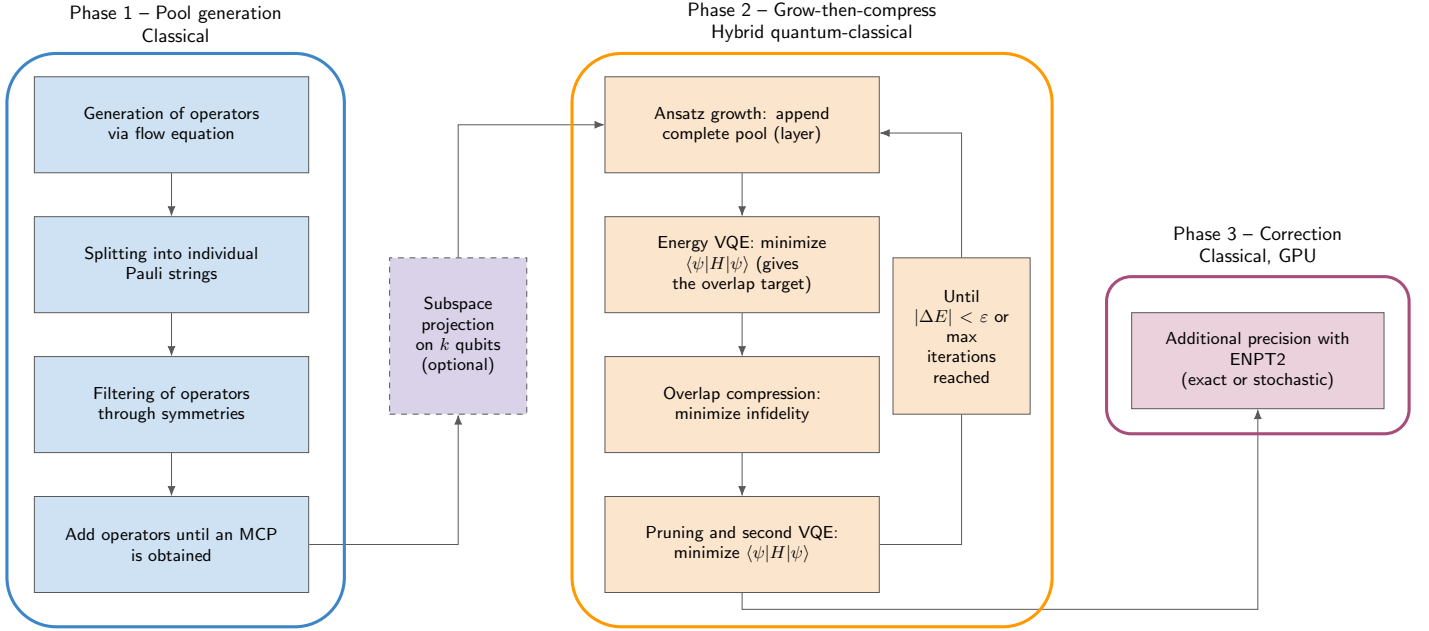
}%

\begin{table*}[htbp]
\centering
\begin{ruledtabular}
\begin{tabular}{lccccc}
\textbf{Molecule} & $\boldsymbol{N}$ & \textbf{MCP} ($2N-4$) &
\textbf{Added excitations} & \textbf{\methodename total} &
\textbf{QEB ($b$)} \\
\hline
BeH$_2$     & 14 & 24 & 73   & 97   & 1092 (10)  \\
H$_6$       & 12 & 20 & 129  & 149  & 561 (10)   \\
H$_8$       & 16 & 28 & 474  & 502  & 1940 (30)  \\
H$_{10}$    & 20 & 36 & 1251 & 1287 & 5035 (50)  \\
H$_2$O      & 26 & 48 & 1278 & 1326 & 15275 (50) \\
C$_3$H$_6$ & 32 & 60 & 1420 & 1480 & 36456 (5) \\
\end{tabular}
\end{ruledtabular}
\caption{Number of qubits $N$ and pool sizes. The \methodename\ pool is the
union of a minimal complete pool (MCP) of $2N-4$ generators, which scales
linearly with $N$ and guarantees that the pool generates the target Lie
algebra, and the excitations added from the VDBF commutator of Eq.~\eqref{eq:comm},
whose number follows the number of Hamiltonian Pauli strings. The number in
parentheses after the QEB pool size is the batch size $b$ used for batched
ADAPT-VQE. Pool sizes are independent of the molecular geometries considered
in this work.}
\label{tab:operator_comparison}
\end{table*}

\subsection{Numerical setup:} \label{sec_setting}

All simulations were performed with Hyperion, a GPU-accelerated quantum emulator optimized for low-level statevector operations \cite{Hyperion}. Hamiltonian integrals were computed with OpenFermion-PySCF, and the Jordan--Wigner transformation maps the fermionic operators to qubit space.  All results are obtained by noiseless state-vector emulation. Unless otherwise stated, the STO-3G basis set is used. 
We  benchmarked \methodename on linear hydrogen chains (H$_6$, H$_8$, H$_{10}$) at
$d = 0.85$ and $3.0$~\AA, BeH$_2$ at $3.0$~\AA, H$_2$O  in the 6-31G basis (26 qubits),
and C$_3$H$_6$  in an active space of 18 electrons in 16 spatial orbitals (32 qubits). The subspace projection results for H$_4$, BeH$_2$ and H$_2$O (STO-3G), and H$_2$O (6-31G, 26 qubits) in Section~\ref{ref::project} demonstrate the reduction in qubit count achieved by projecting the Hamiltonian  onto the $2^k$ determinants generated by $k$ excitation patterns of a single-layer ansatz. Table~\ref{tab:operator_comparison} compares the \methodename pool sizes with those of the QEB pool, which are 4 to 25 times larger. The baseline throughout is batched ADAPT-VQE \cite{sapova2022variational} using the QEB pool, in which the $b$ highest-gradient operators are selected at each iteration. 
The batch sizes $b$ listed in Table~\ref{tab:operator_comparison} are chosen so that further increases
yield no significant improvement in the ADAPT-VQE convergence rate for the
systems studied; the baseline therefore represents the best case with
respect to batch size.

Performance is measured against exact full configuration interaction (FCI) energies using two hardware-resource metrics. The first is the CNOT gate count, which governs circuit depth and hence sensitivity to decoherence. The second
metric is  a hardware-agnostic proxy, the cumulative number of
Pauli-term evaluations, in the figures and tables, cumulative hardware circuit executions
denotes this Pauli-term evaluation count. One energy evaluation  $\langle \psi | H | \psi \rangle$  requires $N_H$ quantum measurements, where $N_H$ is the number of measurable Pauli strings in the Hamiltonian, evaluating one ADAPT gradient also costs $N_H$ measurements per candidate operator, whereas evaluating one overlap gradient in our framework costs a single measurement (Section~\ref{sec:growthencompress}). Energy evaluations
performed by the classical optimizer are included for both
methods. Finally, the ENPT2 correction is computed classically on GPU
with Hyperion and incurs no additional quantum resource
cost.

\subsection{Measurement cost: gradient-free layered growth}
\label{sec:measurement_cost}
At each iteration, the complete \methodename pool (generated as described in Section~\ref{sec:pool}) is appended to the ansatz and the  parameters are re-optimized. Because the pool is physics-informed and complete, the ansatz reaches chemical accuracy without the multi-iteration descent required by batched ADAPT-VQE (Figures~\ref{fig:convergence_beh2}--\ref{fig:convergence_h2o}, right panels, orange squares versus blue curves). For stretched hydrogen chains (H$_6$, H$_8$, H$_{10}$ at d=3.0~\AA{}), chemical accuracy is achieved at the first iteration, for equilibrium geometries, it is reached within the first few iterations. 

The low measurement cost follows from the elimination of Hamiltonian gradient evaluations. In \methodenamens, the only quantum measurements are energy evaluations, each costing $N_H$ Pauli-string measurements. By contrast, each batched ADAPT-VQE iteration evaluates gradients for every operator in a pool 4 to 25 times larger, at a cost of $N_H$ measurements per operator.

As shown in Table~\ref{tab:cnots_cost_comparison}  (Standard column), the measurement cost  at
chemical accuracy is a small fraction of the ADAPT-VQE baseline for all
systems studied of Table~\ref{tab:cnots_cost_comparison}. For the hydrogen chains at 0.85~\AA, the ratio ranges from 8\% (H$_6$) to
18\% (H$_8$), with H$_{10}$ at 10\%, and it is 5\% for H$_2$O.  For
stretched geometries ($d = 3.0$~\AA), where convergence is typically
harder, the advantage is equally pronounced: 10\% for BeH$_2$, 12\% for
H$_6$ and H$_{10}$, and 19\% for H$_8$.

These reductions are a direct consequence of bypassing the gradient selection phase entirely, and are obtained without compression, circuit depth is addressed in the following section.

\subsection{Circuit depth: overlap-guided compression}

While \methodename reduces measurement cost substantially (Section~\ref{sec:measurement_cost}), appending a complete pool at each iteration produces circuits that are deeper than those of batched ADAPT-VQE.
At the first chemically accurate iteration, the CNOT count is 1.3--2.9$\times$
that of the baseline for the hydrogen chains at 0.85~\AA\ and H$_2$O, and
1.0--1.9$\times$ for the stretched geometries (Table~\ref{tab:cnots_cost_comparison}, Standard column).
We now show how overlap-guided compression reduces this depth penalty, and
removes it at stretched geometries.

\begin{figure*}[htbp]
    \centering

    \begin{minipage}[t]{0.49\textwidth}
        \centering
        \includegraphics[width=\linewidth]{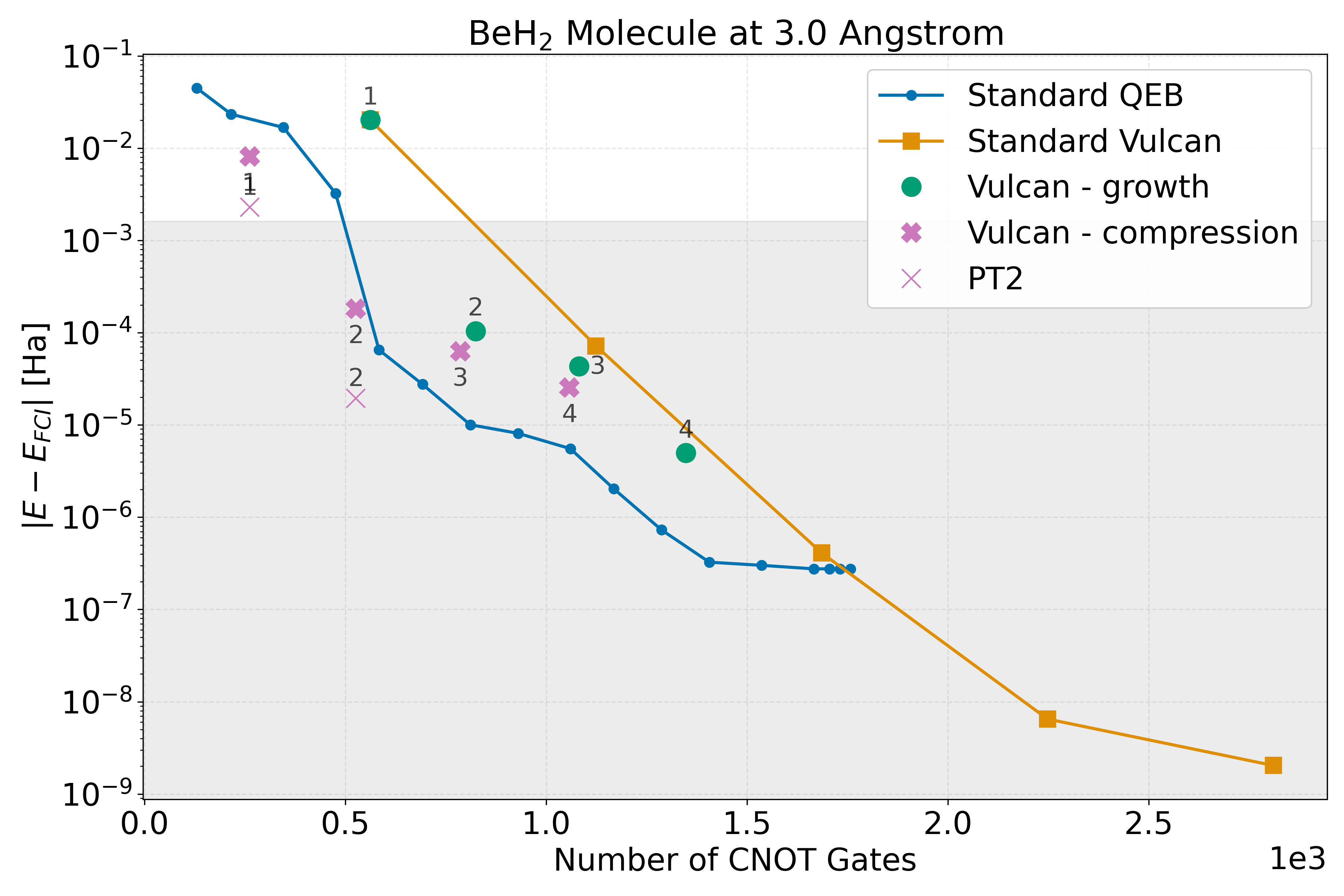}
    \end{minipage}
    \hfill
    \begin{minipage}[t]{0.49\textwidth}
        \centering
        \includegraphics[width=\linewidth]{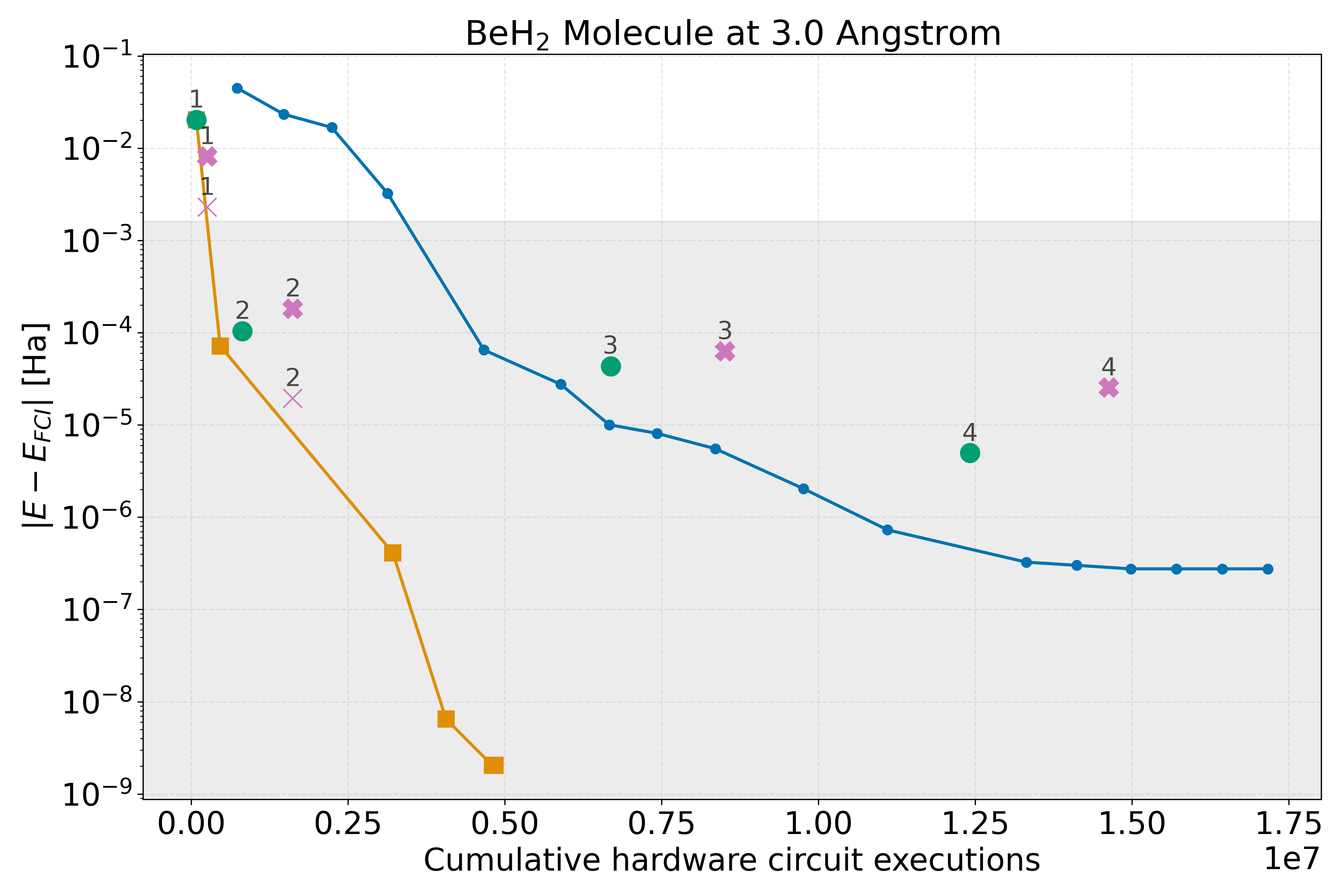}
    \end{minipage}

\caption{Energy error with respect to the FCI energy as a function of the
number of CNOTs (left panel) and of the cumulative hardware circuit executions
(right panel) for the BeH$_2$ molecule at an internuclear distance
$d = 3.0$~\AA. Blue circles (``Standard QEB''): batched ADAPT-VQE with the QEB
pool; orange squares (``Standard Vulcan''): \methodename without compression; green
circles: \methodename growth iterations; large pink `X': compressed iterations; thin
`$\times$': PT2-corrected energies. Numbers label the iterations. The grey band
indicates chemical accuracy.}
    \label{fig:convergence_beh2}
\end{figure*}

\begin{figure*}[htbp]
    \centering
    \begin{minipage}[t]{0.49\textwidth}
        \centering
        \includegraphics[width=\linewidth]{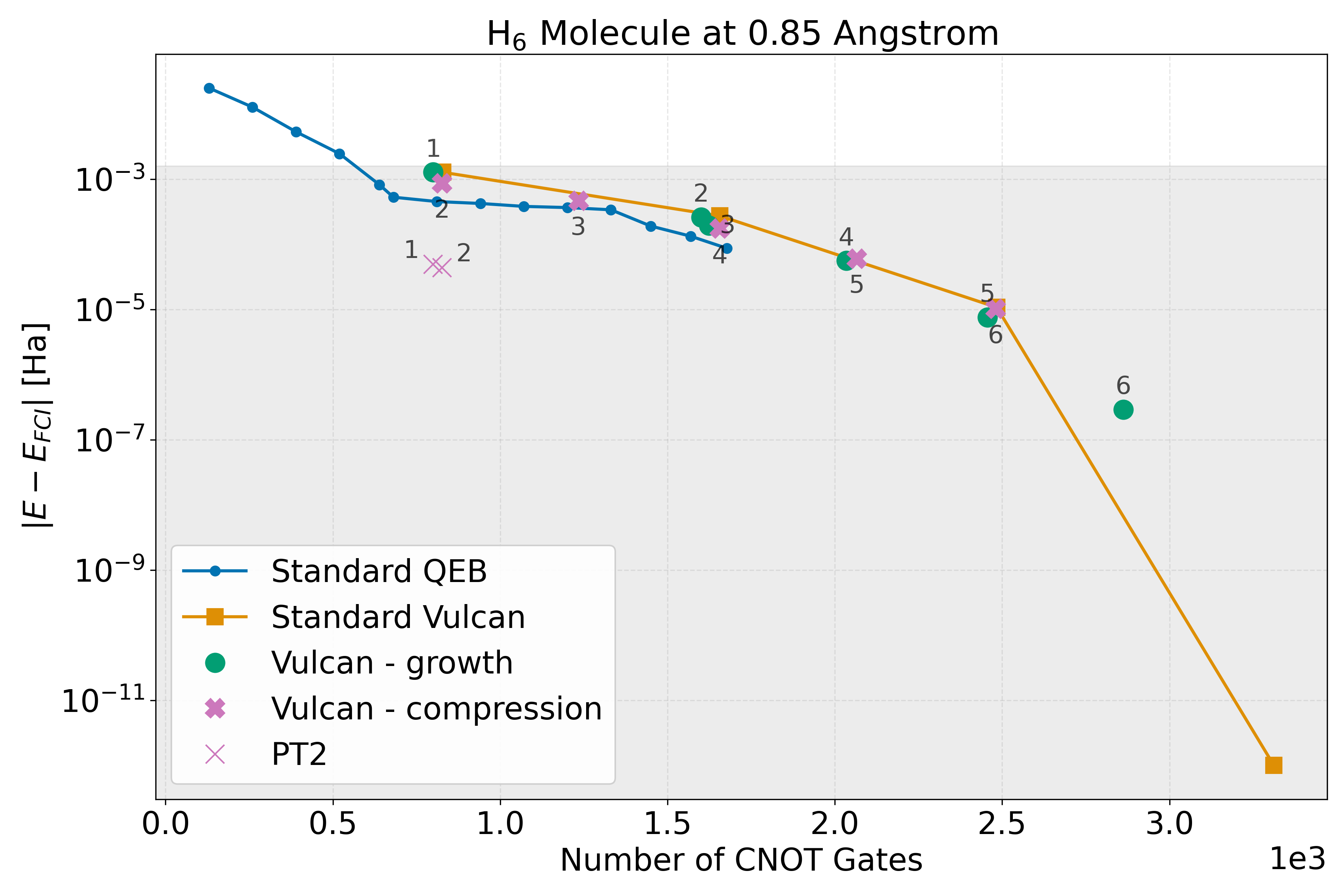}
    \end{minipage}
    \hfill
    \begin{minipage}[t]{0.49\textwidth}
        \centering
        \includegraphics[width=\linewidth]{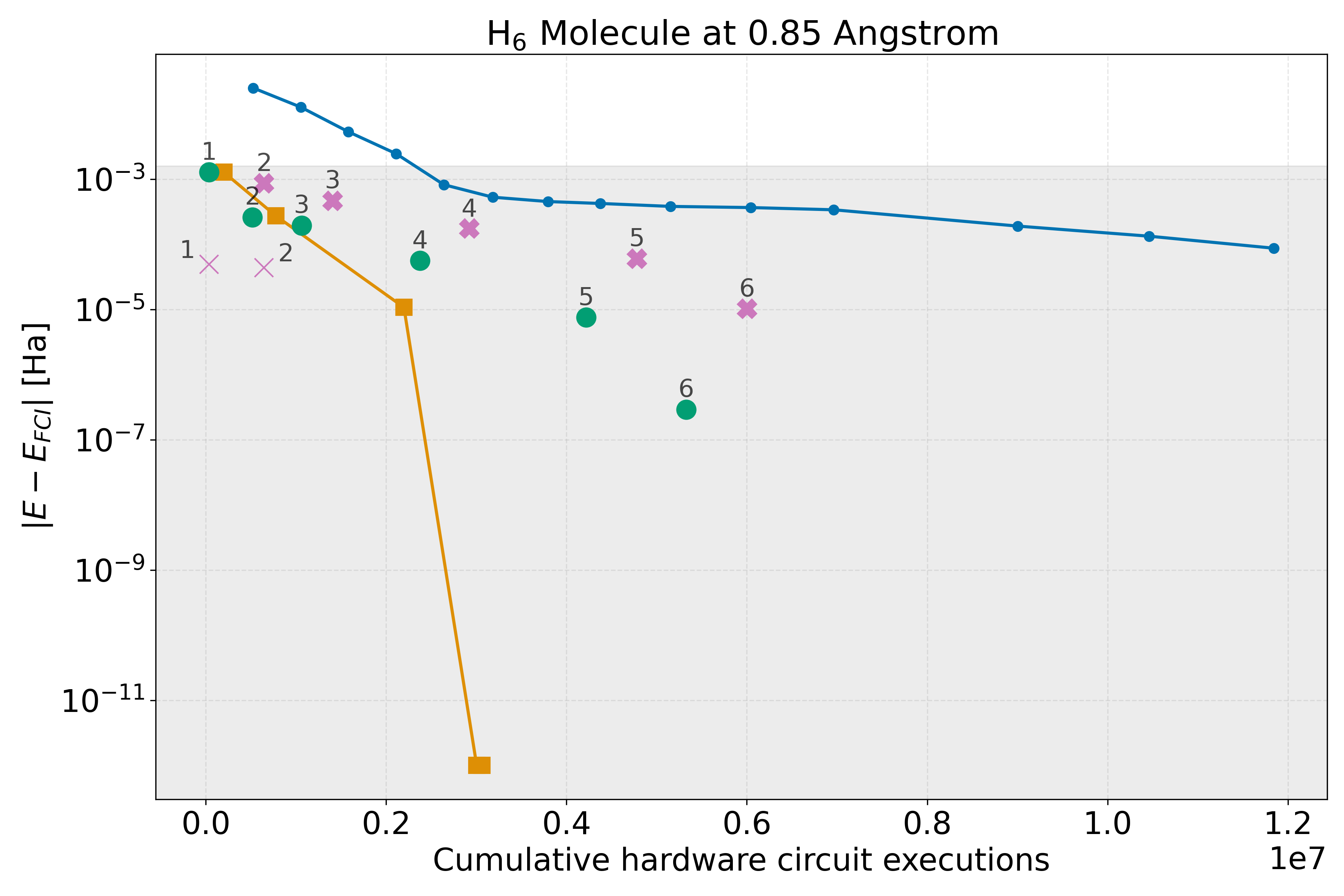}
    \end{minipage}
    \centering
    \begin{minipage}[t]{0.49\textwidth}
        \centering
        \includegraphics[width=\linewidth]{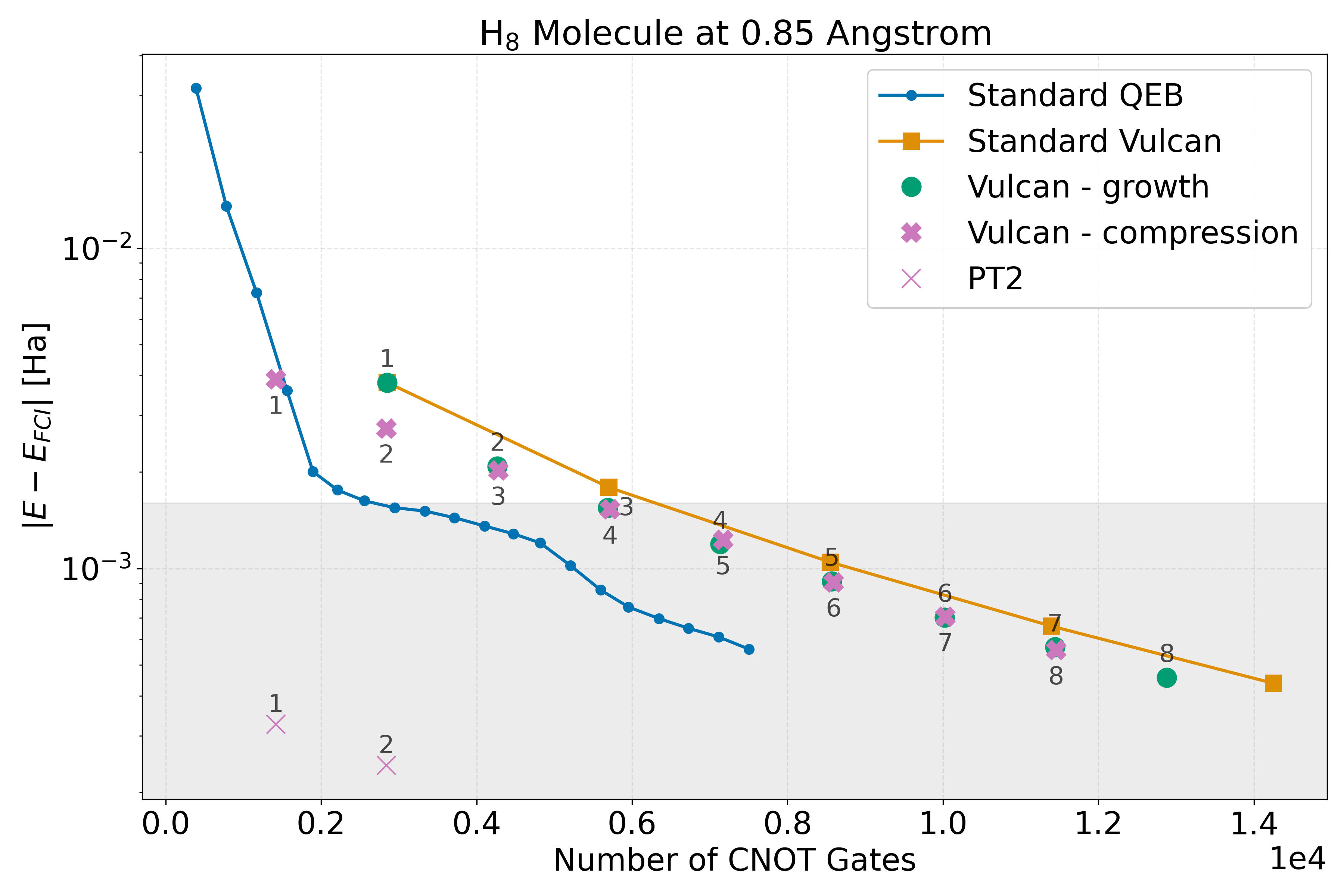}
    \end{minipage}
    \hfill
    \begin{minipage}[t]{0.49\textwidth}
        \centering
        \includegraphics[width=\linewidth]{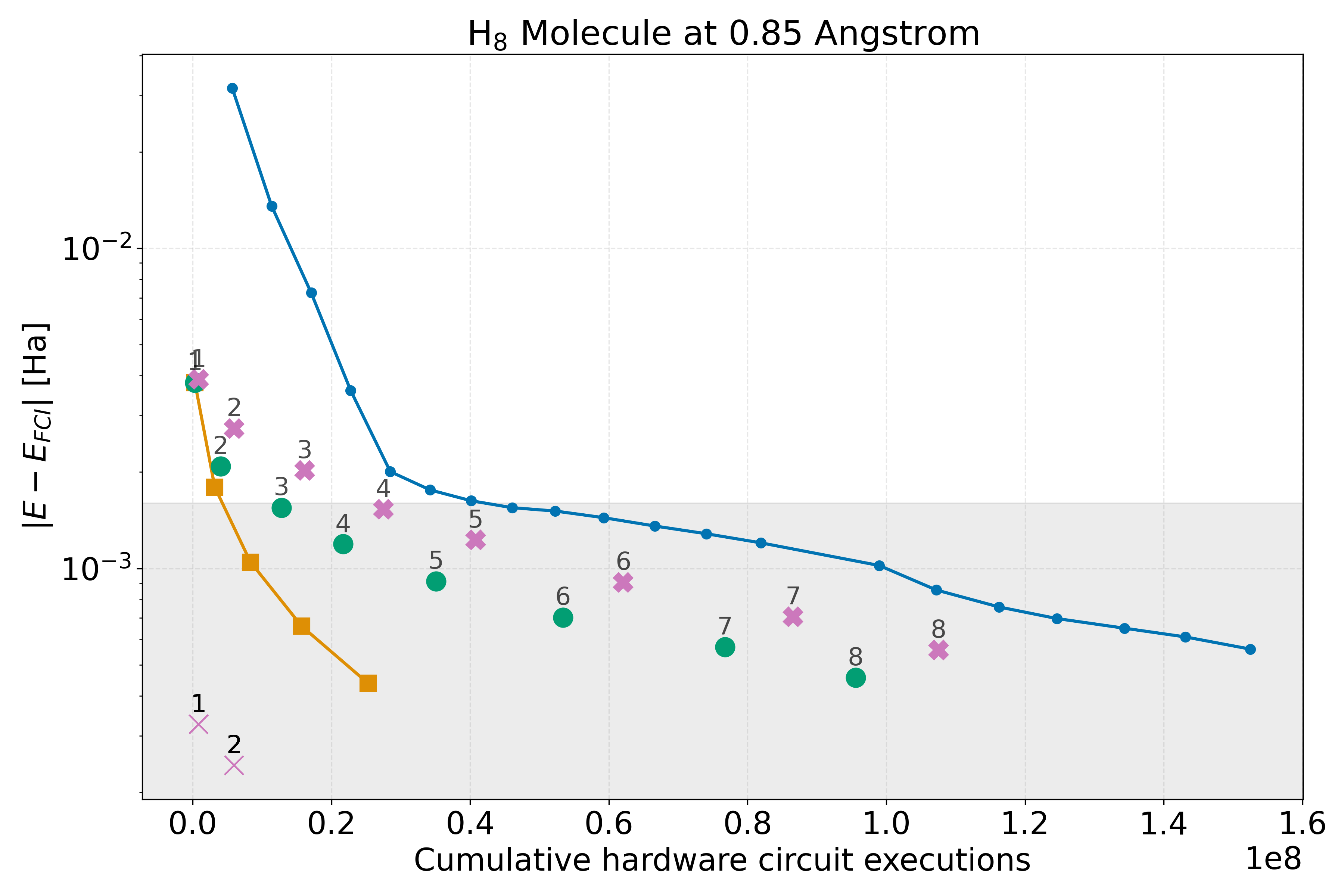}
    \end{minipage}
    \vspace{0.5cm}
    \begin{minipage}[t]{0.49\textwidth}
        \centering
        \includegraphics[width=\linewidth]{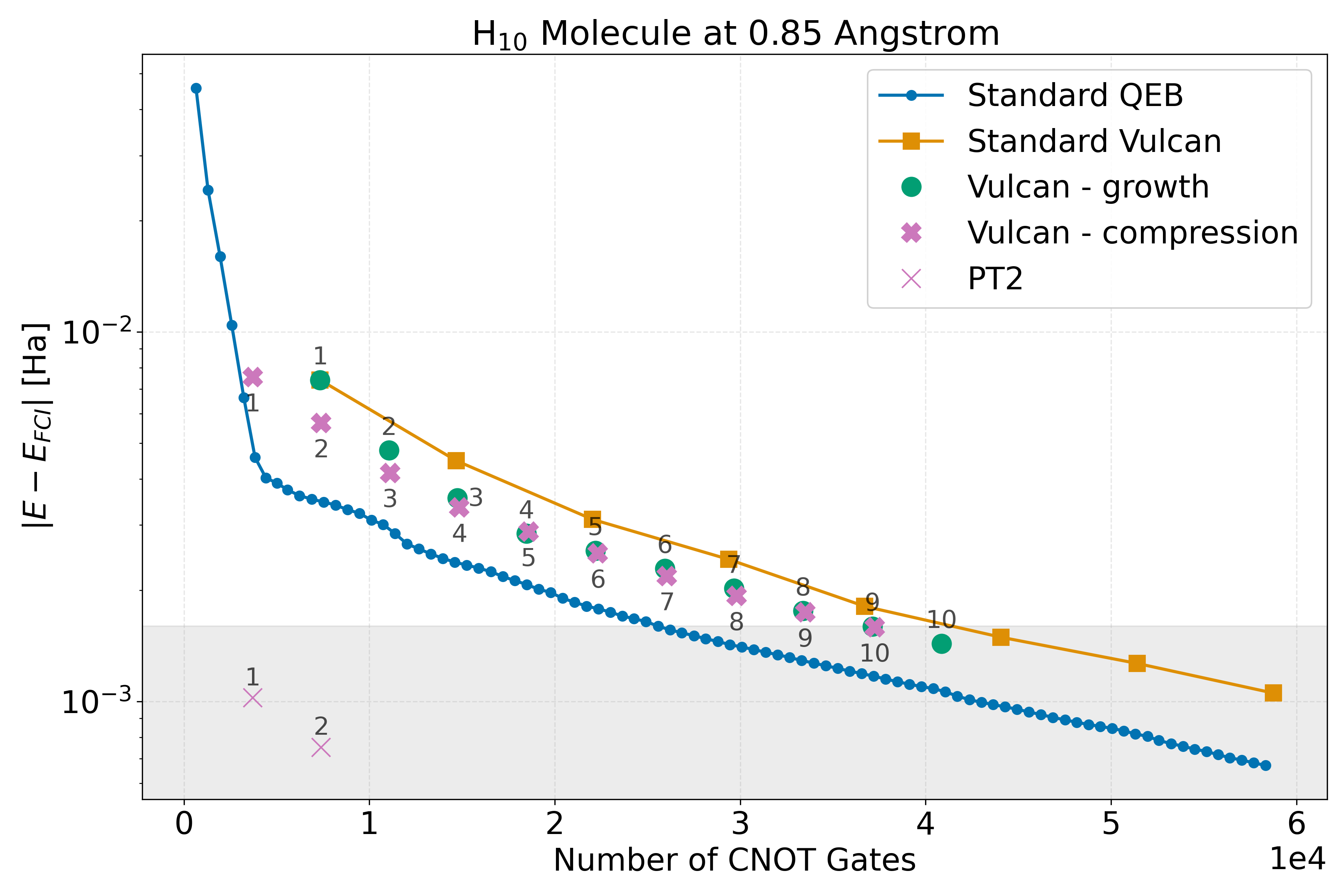}
    \end{minipage}
    \hfill
    \begin{minipage}[t]{0.49\textwidth}
        \centering
        \includegraphics[width=\linewidth]{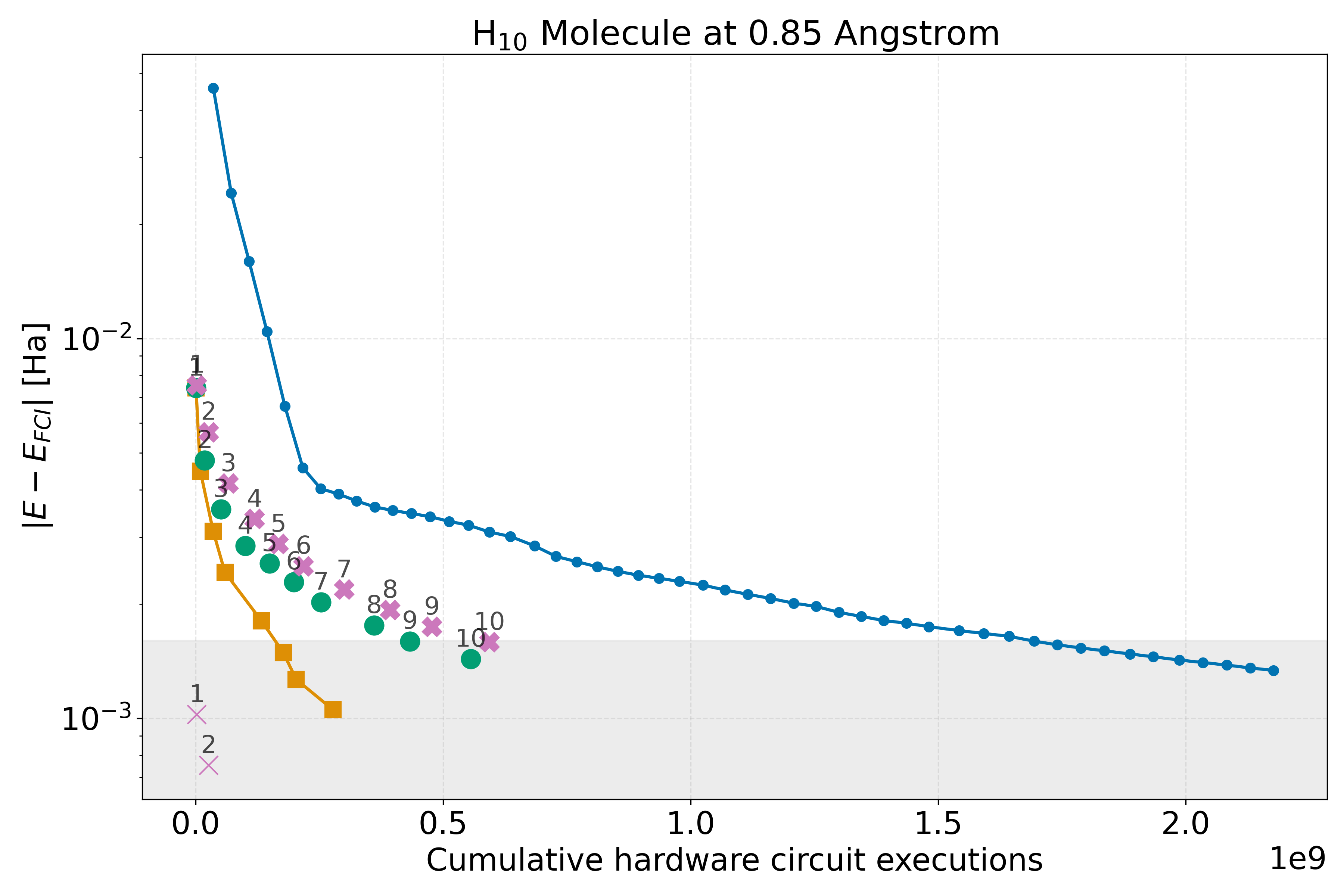}
    \end{minipage}
\caption{Same as Fig.~\ref{fig:convergence_beh2} for the hydrogen chains H$_6$, H$_8$ and H$_{10}$ (top
to bottom) at an internuclear distance $d = 0.85$~\AA.}
    \label{fig:convergence_eq}
\end{figure*}

\begin{figure*}[htbp]
    \centering
    \begin{minipage}[t]{0.49\textwidth}
        \centering
        \includegraphics[width=\linewidth]{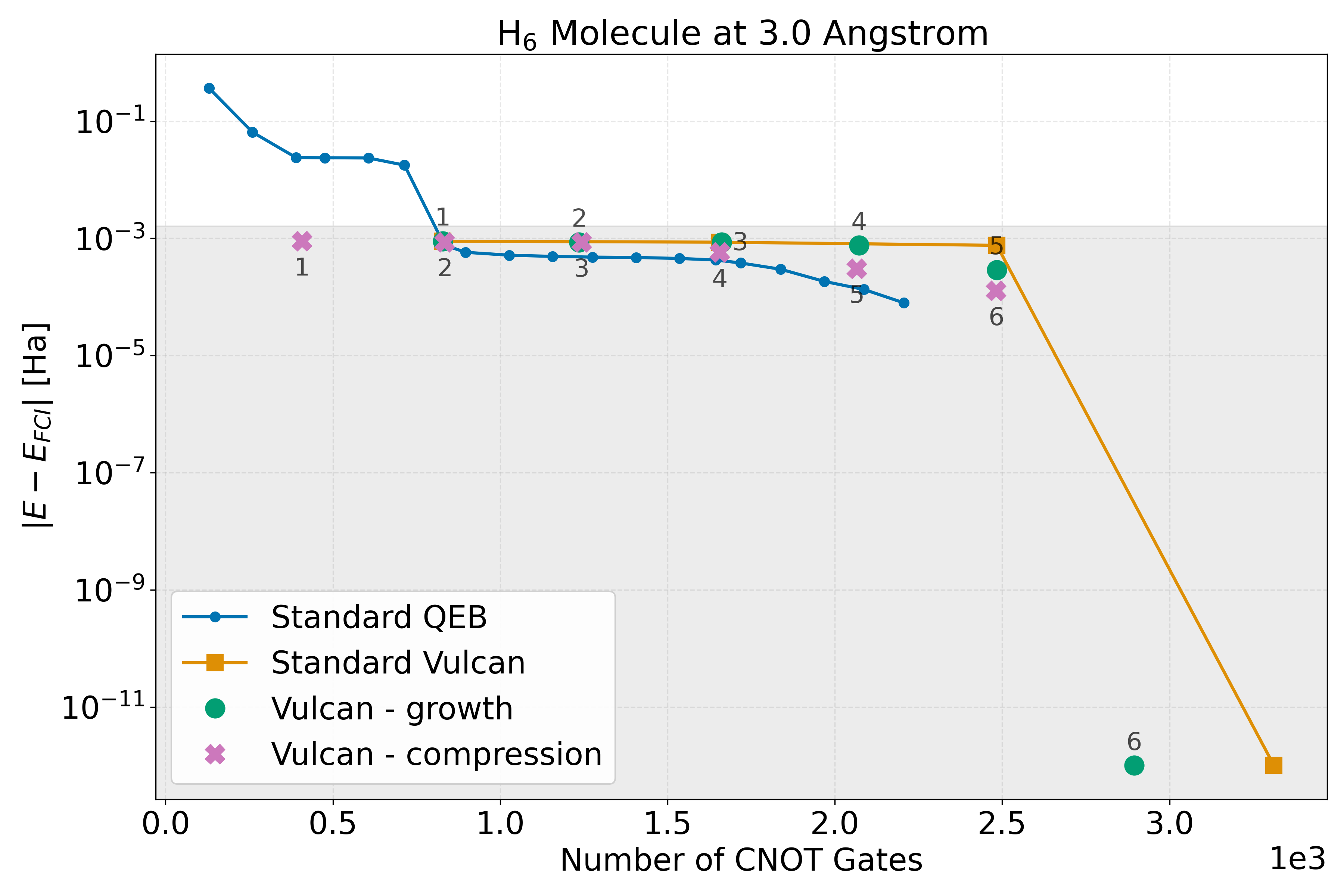}
    \end{minipage}
    \hfill
    \begin{minipage}[t]{0.49\textwidth}
        \centering
        \includegraphics[width=\linewidth]{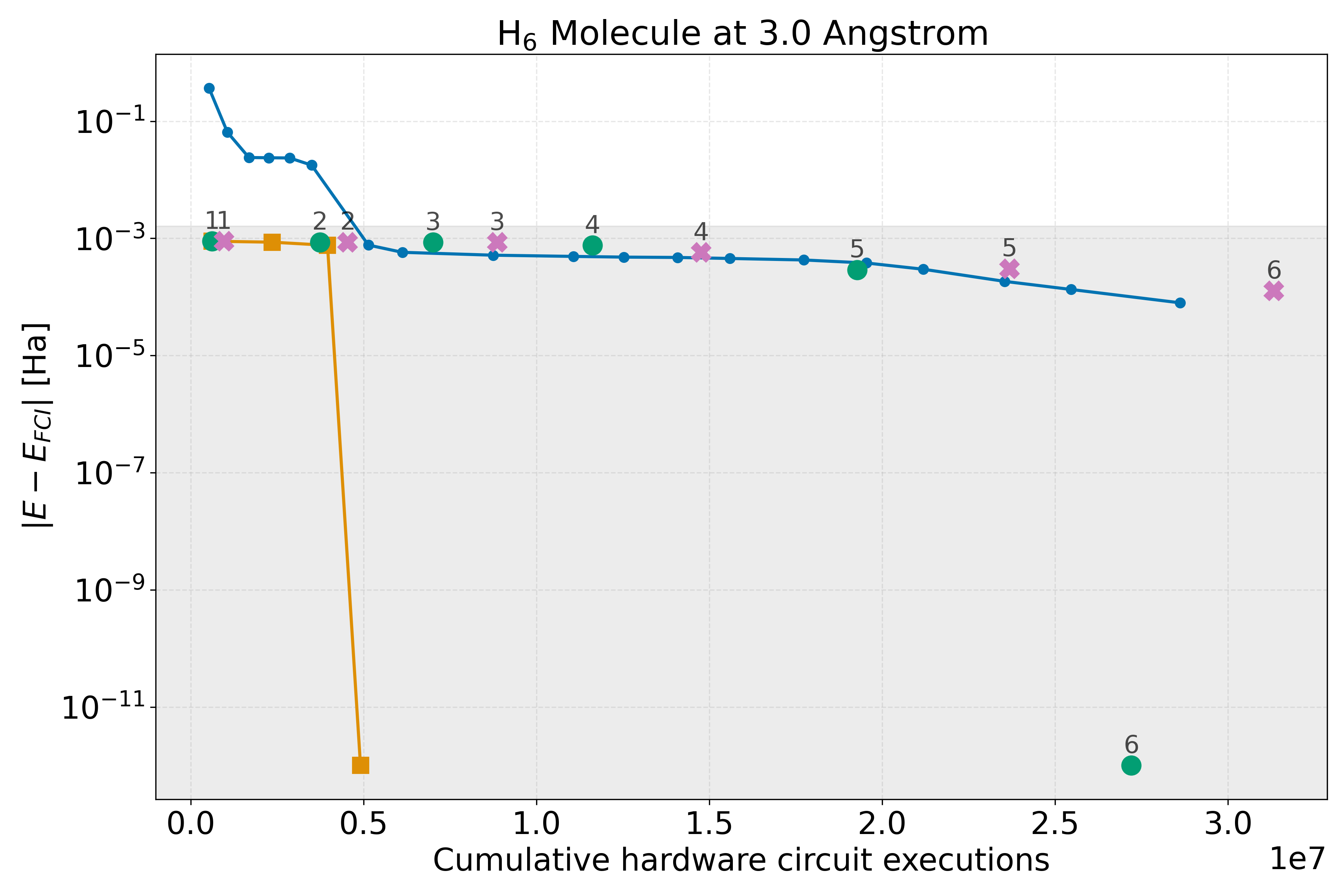}
    \end{minipage}
    \vspace{0.5cm}
    \centering
    \begin{minipage}[t]{0.49\textwidth}
        \centering
        \includegraphics[width=\linewidth]{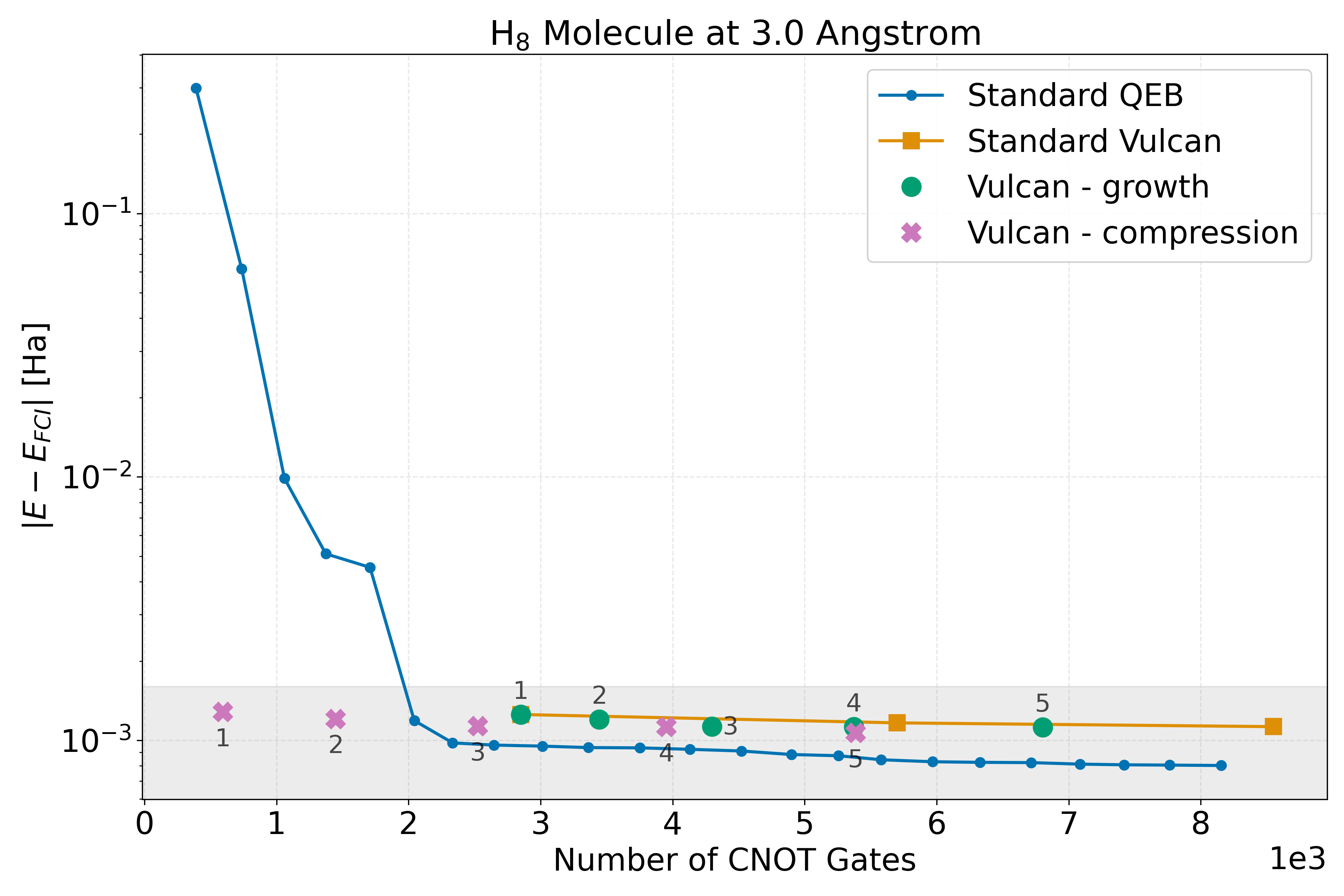}
    \end{minipage}
    \hfill
    \begin{minipage}[t]{0.49\textwidth}
        \centering
        \includegraphics[width=\linewidth]{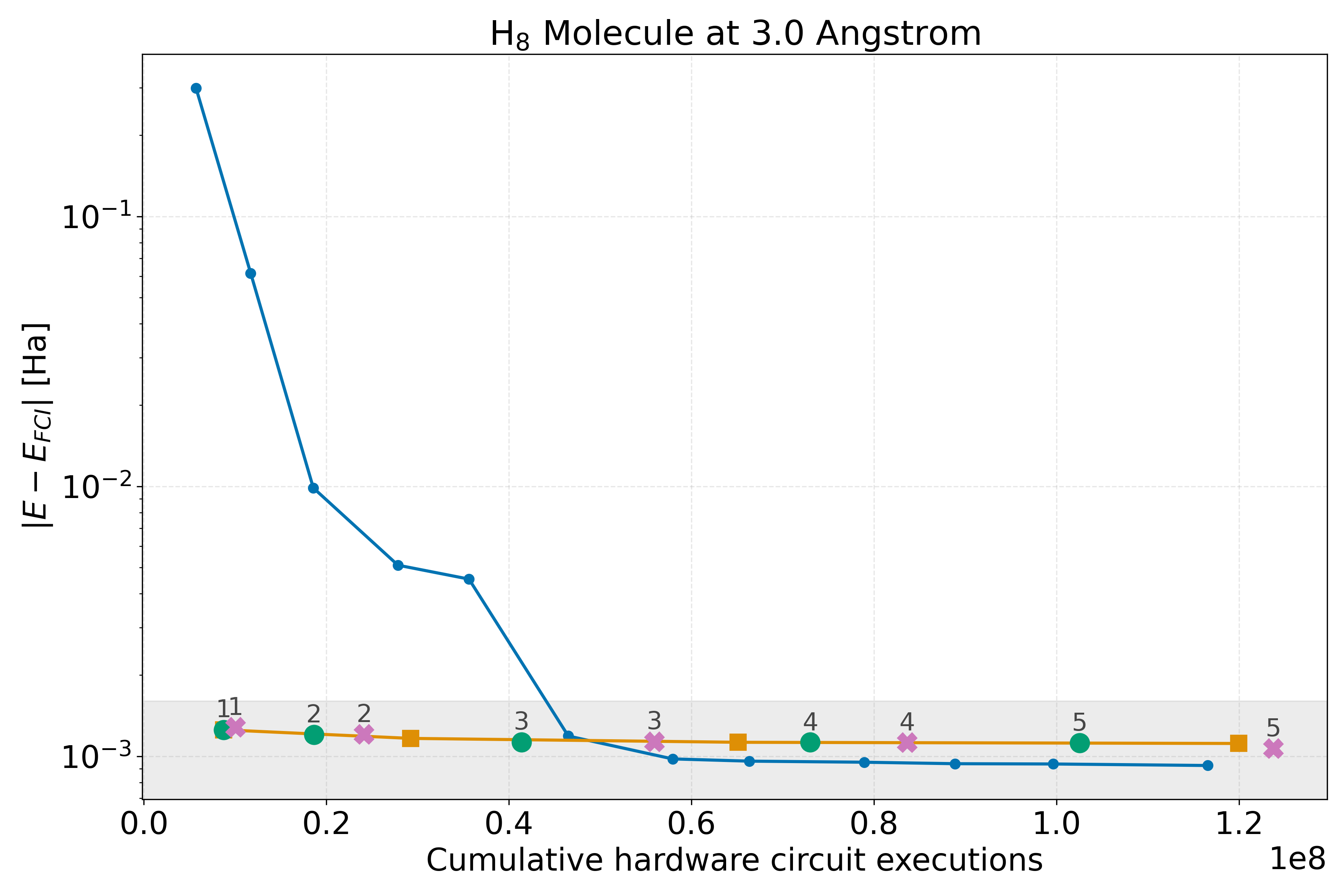}
    \end{minipage}
    \vspace{0.5cm}
    \begin{minipage}[t]{0.49\textwidth}
        \centering
        \includegraphics[width=\linewidth]{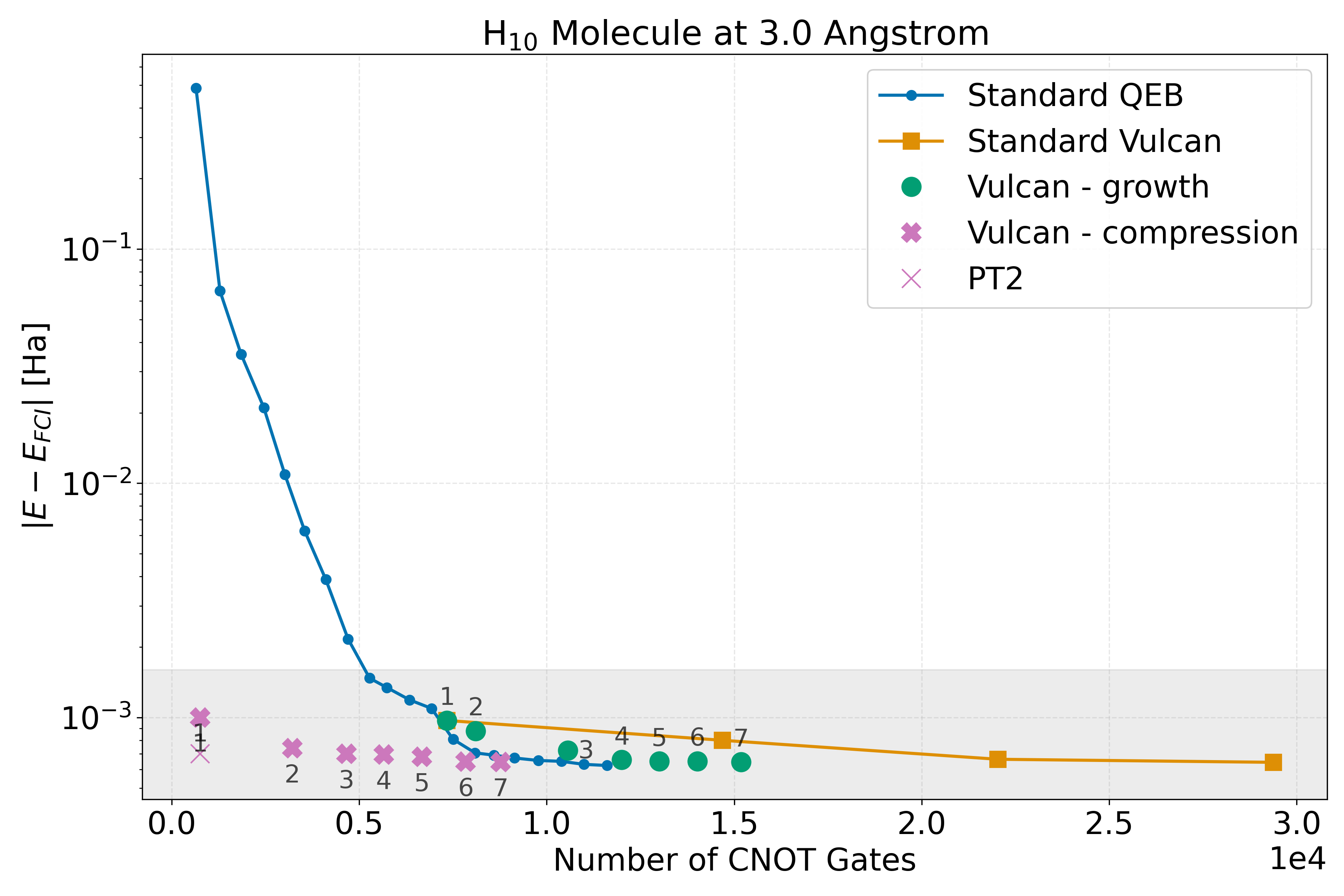}
    \end{minipage}
    \hfill
    \begin{minipage}[t]{0.49\textwidth}
        \centering
        \includegraphics[width=\linewidth]{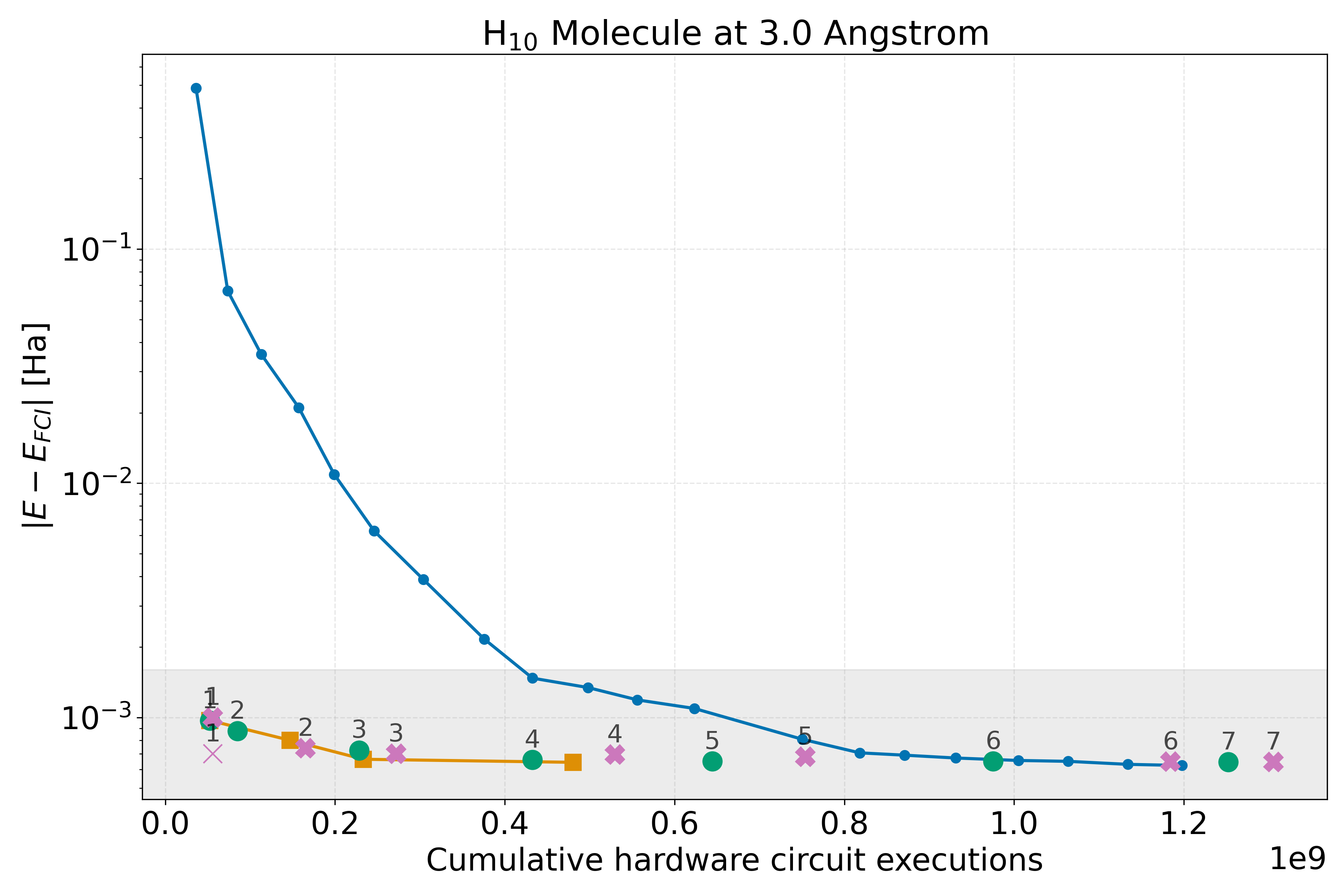}
    \end{minipage}
\caption{Same as Fig.~\ref{fig:convergence_beh2} for the hydrogen chains H$_6$, H$_8$ and H$_{10}$ (top
to bottom) at an internuclear distance $d = 3.0$~\AA.}
    \label{fig:convergence_str}
\end{figure*}

\begin{figure*}[htbp]
    \centering

    \begin{minipage}[t]{0.49\textwidth}
        \centering
        \includegraphics[width=\linewidth]{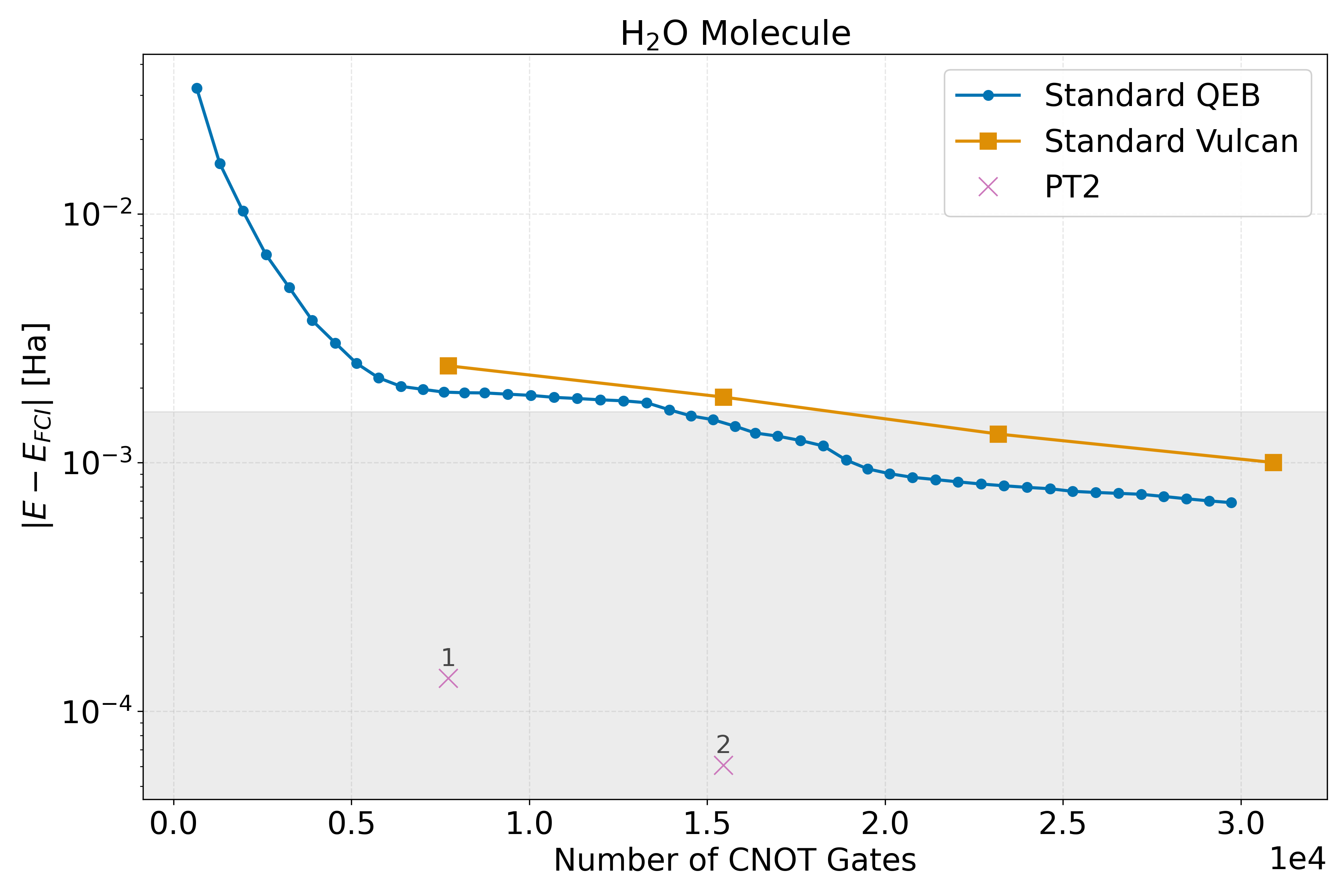}
    \end{minipage}
    \hfill
    \begin{minipage}[t]{0.49\textwidth}
        \centering
        \includegraphics[width=\linewidth]{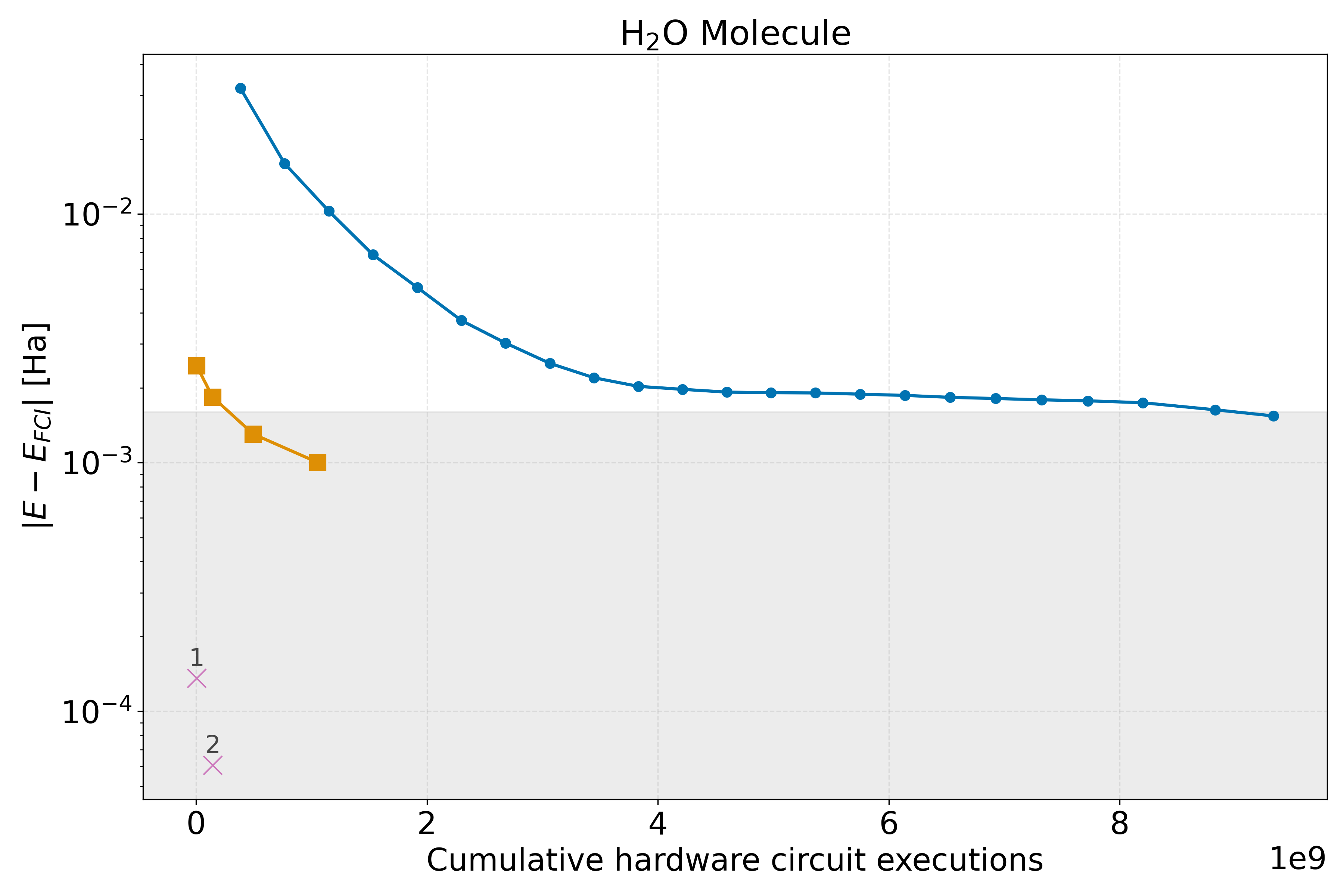}
    \end{minipage}

\caption{Same as Fig.~\ref{fig:convergence_beh2} for the H$_2$O molecule in the 6-31G basis
(26 qubits).}
    \label{fig:convergence_h2o}
\end{figure*}

Compression operates in two phases, as illustrated in Figures~\ref{fig:convergence_beh2}--\ref{fig:convergence_str}: a growth phase (green circles), where the complete pool is appended and optimized, and a compression phase (large filled pink `X' symbols), where operators are re-selected by overlap gradient and the ansatz is pruned to produce a shallower circuit. Numerical markers in the figures link each compression target to its corresponding compressed ansatz. Compression inserts generators one at a time until the infidelity gain
becomes negligible or the compressed layer reaches a set fraction of the
grown layer (typically 50\%). The initial growth iteration (labeled 1) corresponds to the first uncompressed \methodename iteration. Details of the overlap compression algorithm are given in Section~\ref{sec:growthencompress}.

For the equilibrium hydrogen chains, compression does not bring the CNOT
count below the baseline (H$_6$: 1.29, H$_8$: 1.94, H$_{10}$: 1.45; Table~\ref{tab:cnots_cost_comparison},
Compressed column). Compression is more effective for the
stretched geometries: the ratio drops to 0.90 for BeH$_2$, 0.49 for H$_6$,
0.29 for H$_8$ and 0.14 for H$_{10}$. At stretched geometries, the chemically accurate states obtained can be
spin-contaminated, and compression reproduces any contamination of its
target.

It is worth noting that compression is not free in measurement cost. The overlap-guided compression adds circuit executions, raising the cumulative measurement cost from the uncompressed values reported in Section~\ref{sec:measurement_cost}: for example, H$_8$ at 0.85~\AA{} goes from 0.18 to 0.59 of the baseline, and H$_{10}$ from 0.10 to 0.35. Crucially, the compressed measurement cost remains below the batched ADAPT-VQE baseline in every case tested. The user therefore faces a tunable trade-off: stop after growth for minimal measurement cost, or  compress for a shallower circuit at a higher, but still sub-baseline, measurement budget. Overlap compression reduces gate count at fixed qubit number. For systems whose register size itself exceeds the available hardware, a complementary reduction is needed: projecting the Hamiltonian onto a smaller subspace before the grow-then-compress loop begins.

\begin{figure*}[htbp]
    \centering

    \begin{minipage}[t]{0.49\textwidth}
        \centering
        \includegraphics[width=\linewidth]{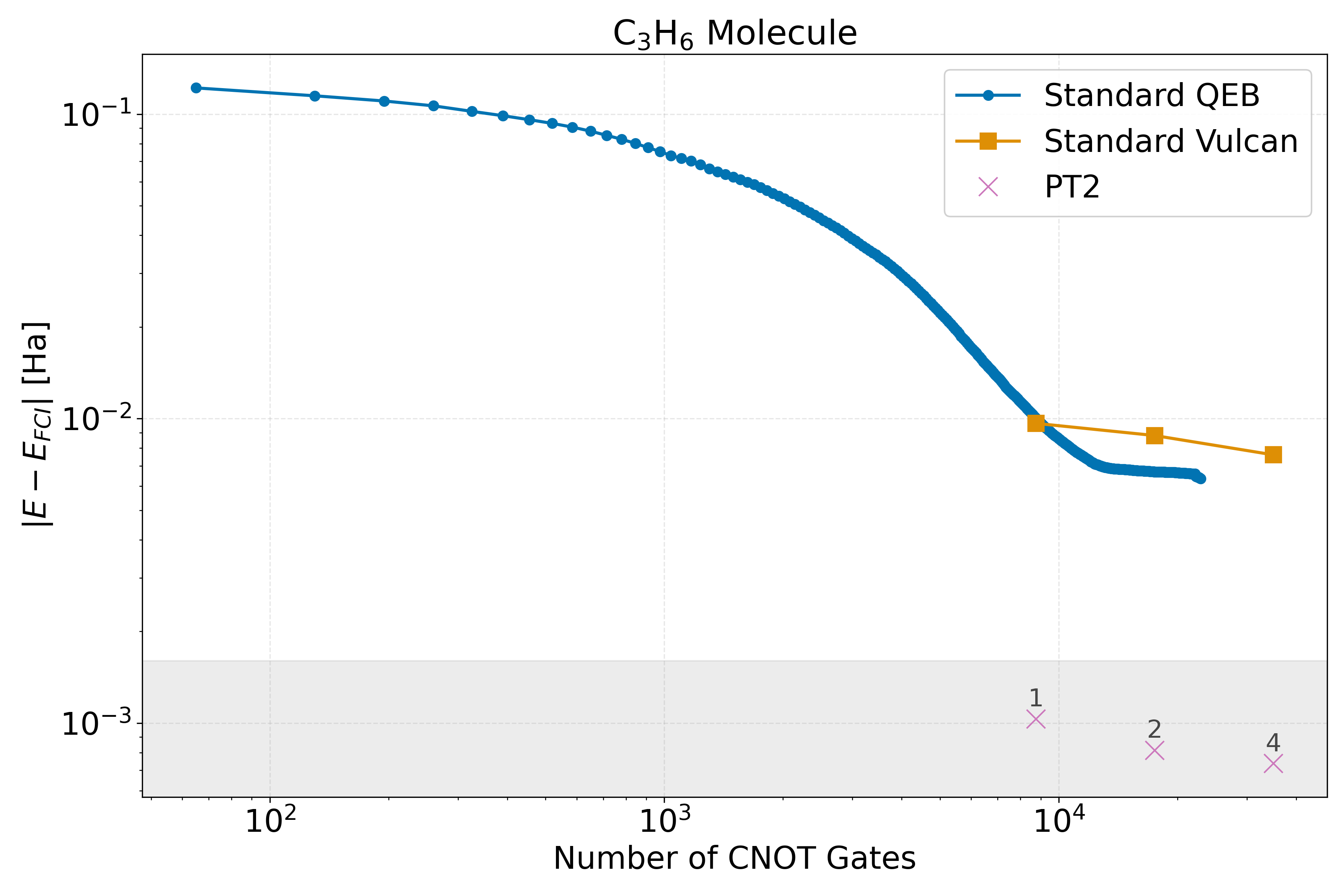}
    \end{minipage}
    \hfill
    \begin{minipage}[t]{0.49\textwidth}
        \centering
        \includegraphics[width=\linewidth]{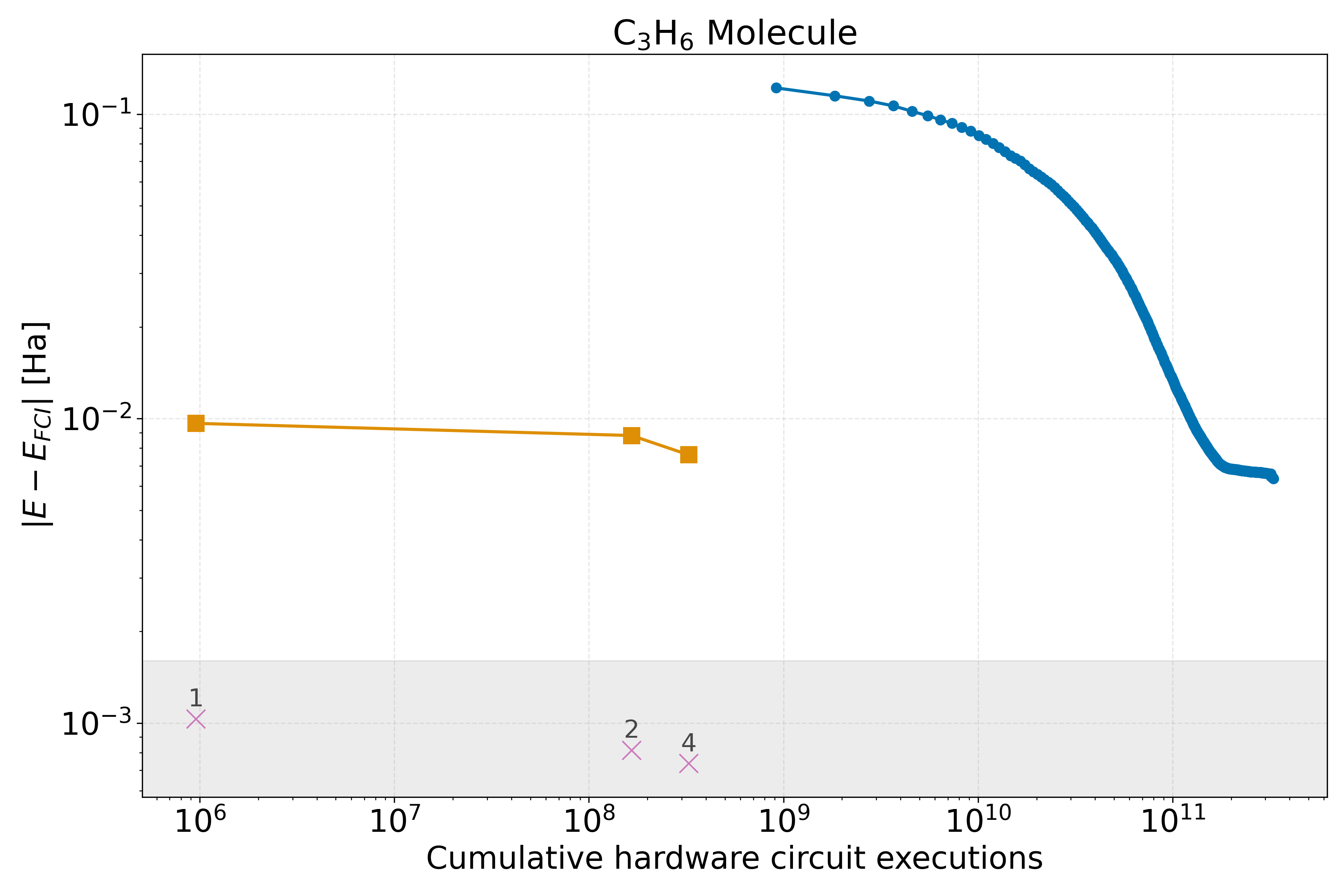}
    \end{minipage}

\caption{Same as Fig.~\ref{fig:convergence_beh2} for the C$_3$H$_6$ molecule in the STO-3G basis. The
calculations use an active space of 18 electrons in 16 spatial orbitals
(32 qubits). The FCI reference energy is
computed within the same active space.}
    \label{fig:convergence_C3H6}
\end{figure*}

\begin{table*}[htbp]
\centering

\begin{ruledtabular}
\begin{tabular}{lcccccc}
\textbf{Molecule} &
\multicolumn{3}{c}{\textbf{CNOTs}} &
\multicolumn{3}{c}{\textbf{Cumulative hardware executions}} \\
\cline{2-4}\cline{5-7}
&
\textbf{Standard} &
\textbf{Compressed} &
\textbf{PT2} &
\textbf{Standard} &
\textbf{Compressed} &
\textbf{PT2} \\
\hline
H$_6$ - $0.85$~\AA & $1.30$ & $1.29$ & $1.29$ & $\boldsymbol{0.08}$ & $\boldsymbol{0.24}$ & $\boldsymbol{0.24}$ \\
H$_8$ - $0.85~\AA$ & $2.90$ & $1.94$& $\boldsymbol{0.48}$ & $\boldsymbol{0.18}$ & $\boldsymbol{0.59}$ & $\boldsymbol{0.02}$ \\
H$_{10}$ - $0.85~\AA$ & $1.72$ & $1.45$ & $\boldsymbol{0.14}$ & $\boldsymbol{0.10}$ & $\boldsymbol{0.35}$ & $\boldsymbol{0.002}$ \\
BeH$_2$ - $3.0~\AA$ & $1.92$ & $\boldsymbol{0.90}$ & $\boldsymbol{0.90}$ & $\boldsymbol{0.10}$ & $\boldsymbol{0.35}$ & $\boldsymbol{0.35}$ \\
H$_6$ - $3.0~\AA$  & $\boldsymbol{0.99}$ & $\boldsymbol{0.49}$ & - & $\boldsymbol{0.12}$ & $\boldsymbol{0.19}$ & - \\
H$_8$ - $3.0~\AA$  & $1.39$ & $\boldsymbol{0.29}$ & - & $\boldsymbol{0.19}$ & $\boldsymbol{0.22}$ & - \\
H$_{10}$ - $3.0~\AA$  & $1.39$ & $\boldsymbol{0.14}$ & - & $\boldsymbol{0.12}$ & $\boldsymbol{0.13}$ & - \\
H$_{2}$O (6-31G) & $1.59$ & - & $\boldsymbol{0.53}$ & $\boldsymbol{0.05}$ & - & $\boldsymbol{0.0005}$ \\
\end{tabular}
\end{ruledtabular}
\caption{Fraction of CNOT counts and cumulative hardware executions at the
first iteration where chemical accuracy is reached for each method with
respect to the same metric with batched ADAPT-VQE with the QEB pool. Standard
means the standard \methodenamens, and Compressed means the compressed
\methodenamens. The PT2 results are calculated using the compressed
\methodenamens, except for H$_2$O, for which the standard \methodename{} was
used. Bold numbers mean that our method needed fewer CNOTs or measurements
than batched ADAPT-VQE with the QEB pool to achieve chemical accuracy. --: not
computed (no compression for H$_2$O) or not needed (PT2 at stretched
geometries, where chemical accuracy is reached without correction).}
\label{tab:cnots_cost_comparison}
\end{table*}

\subsection{Qubit count: subspace projection}
\label{ref::project}
Given the operator pool from Phase 1, we select $k$ excitation patterns from
the most important generators of the first-layer ansatz. The $K = 2^k$
determinants obtained by applying all their combinations to the
Hartree--Fock state span the subspace, and $H_{\rm proj}$ is the
corresponding $K\times K$ submatrix of $H$, a Hamiltonian on $k$ qubits.

The subspace is
grown one qubit at a time. At each $k$, a second-order
Epstein--Nesbet estimate of the energy contribution
from the determinants excluded by the projection is
computed. Two variants of this estimate are available:
one that takes the exact ground state of the current
$H_{\mathrm{proj}}$ as its reference, and a cheaper
one that uses the Hartree--Fock state directly. When
the estimate falls below a tolerance $\tau$, the
growth stops (Section~\ref{sec:hproj}).

\begin{figure*}[t]
  \centering
  \begin{subfigure}[t]{0.49\textwidth}
    \centering
    \includegraphics[width=\linewidth]{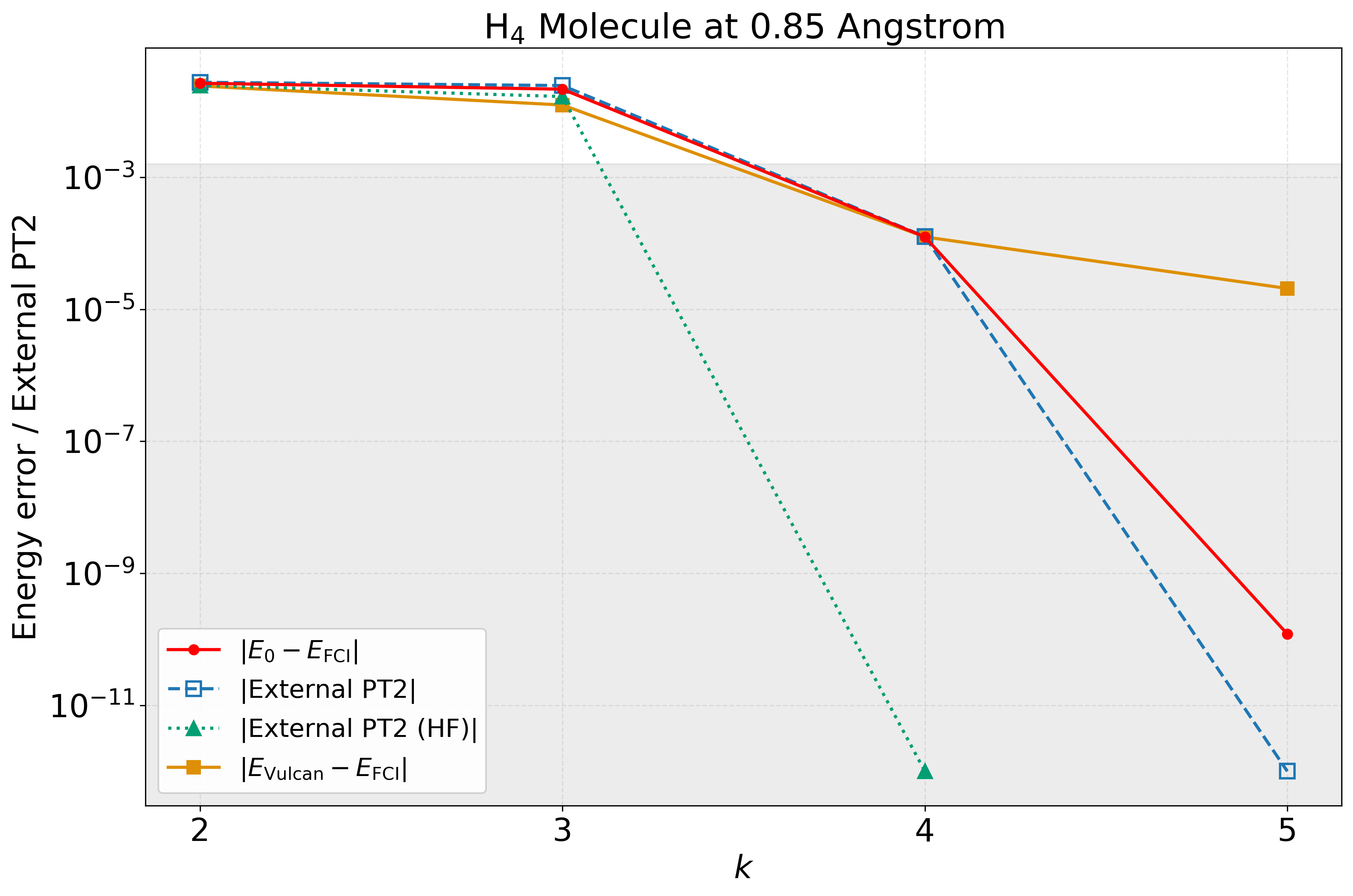}
    \caption{Exact diagonalization vs $k$}
    \label{fig:hproj_H4_k}
  \end{subfigure}\hfill
  \begin{subfigure}[t]{0.49\textwidth}
    \centering
    \includegraphics[width=\linewidth]{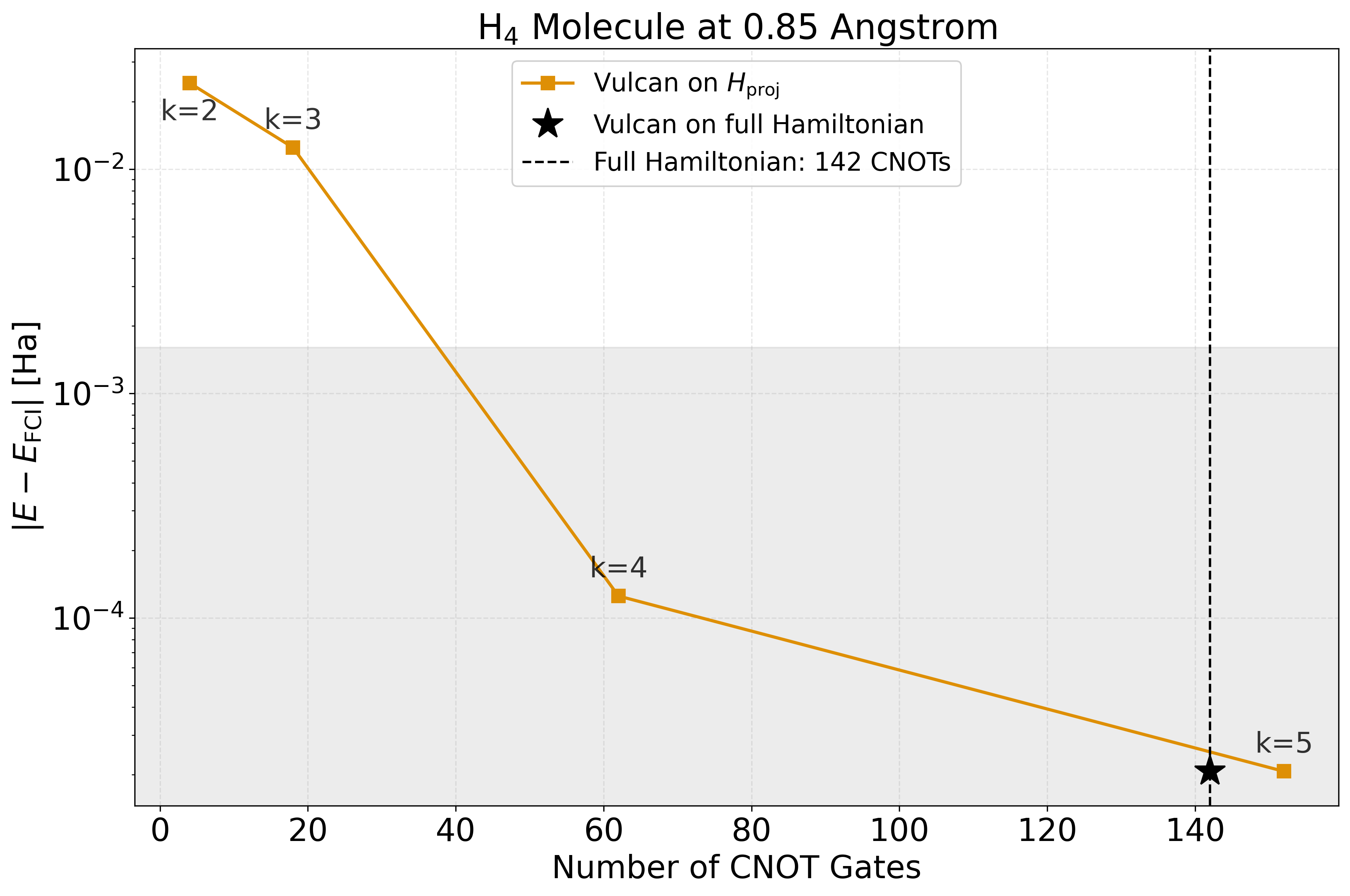}
    \caption{\methodename{} vs CNOTs}
    \label{fig:hproj_H4_cnots}
  \end{subfigure}

  \vspace{0.5em}
  \begin{subfigure}[t]{0.49\textwidth}
    \centering
    \includegraphics[width=\linewidth]{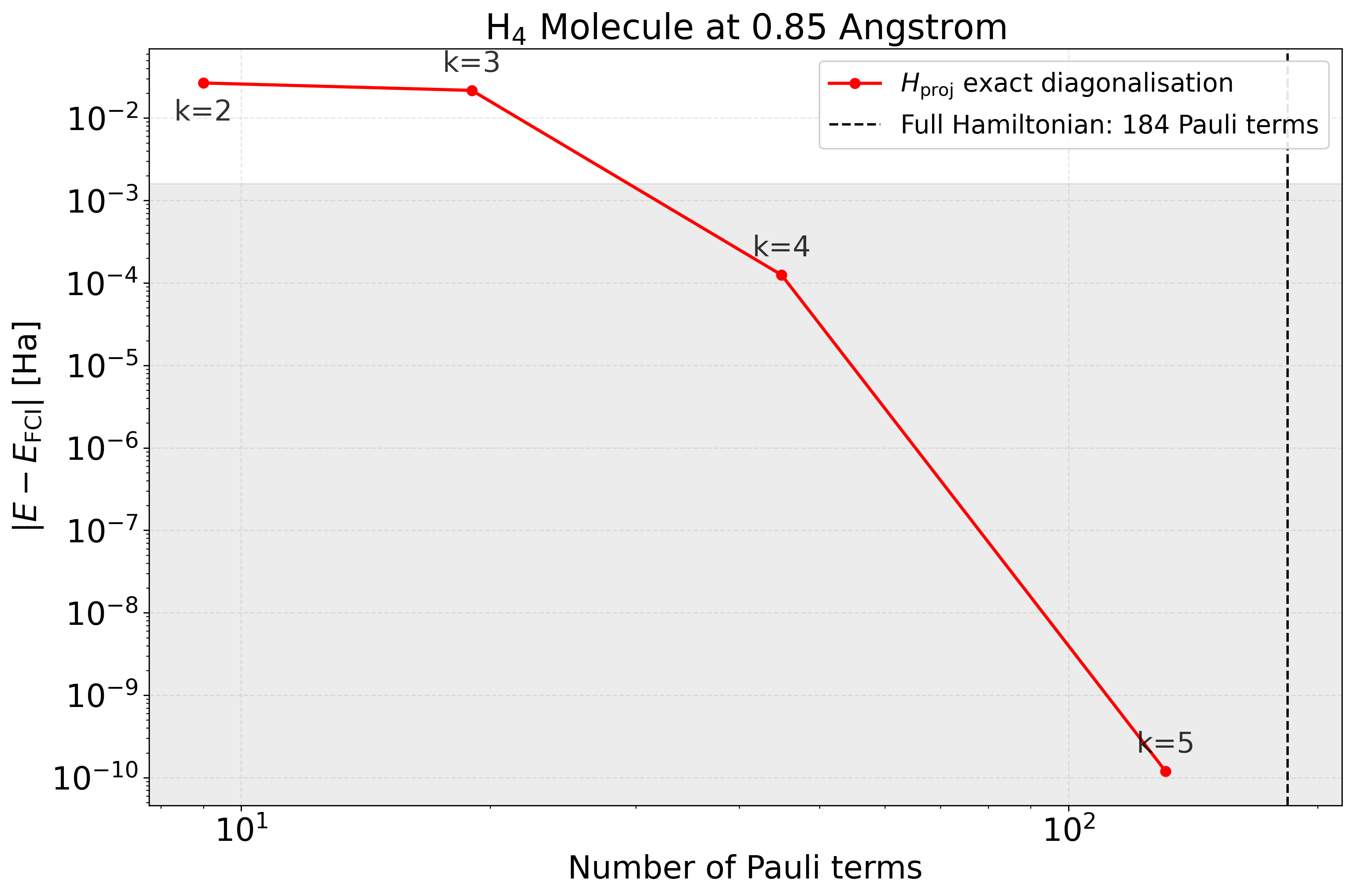}
    \caption{Exact diagonalization vs Pauli terms}
    \label{fig:hproj_H4_paulis}
  \end{subfigure}
\caption{Adaptive-$k$ projection for the H$_4$ molecule at 0.85~\AA\ (STO-3G,
8 qubits). (a) As a function of the number of qubits $k$ kept by the
projection: energy error of the ground state of the projected Hamiltonian
$H_{\rm proj}$ ($|E_0 - E_{\rm FCI}|$), PT2 estimate of the energy lost by the
projection with the exact subspace ground state (``External PT2'') or the
Hartree--Fock state (``External PT2 (HF)'') as reference, and error of one
\methodename{} iteration on $H_{\rm proj}$. (b) Error of one \methodename{}
iteration on $H_{\rm proj}$ as a function of the number of CNOTs; the star
marks one \methodename{} iteration on the full Hamiltonian, and the dashed line
its CNOT count. (c) Exact-diagonalization error as a function of the number of
Pauli terms of $H_{\rm proj}$; the dashed line marks the number of terms of the
full Hamiltonian. The grey band indicates chemical accuracy.}
  \label{fig:hproj_H4}
\end{figure*}

\begin{figure}[htbp]
  \centering
  \begin{subfigure}[t]{0.49\textwidth}
    \centering
    \includegraphics[width=\linewidth]{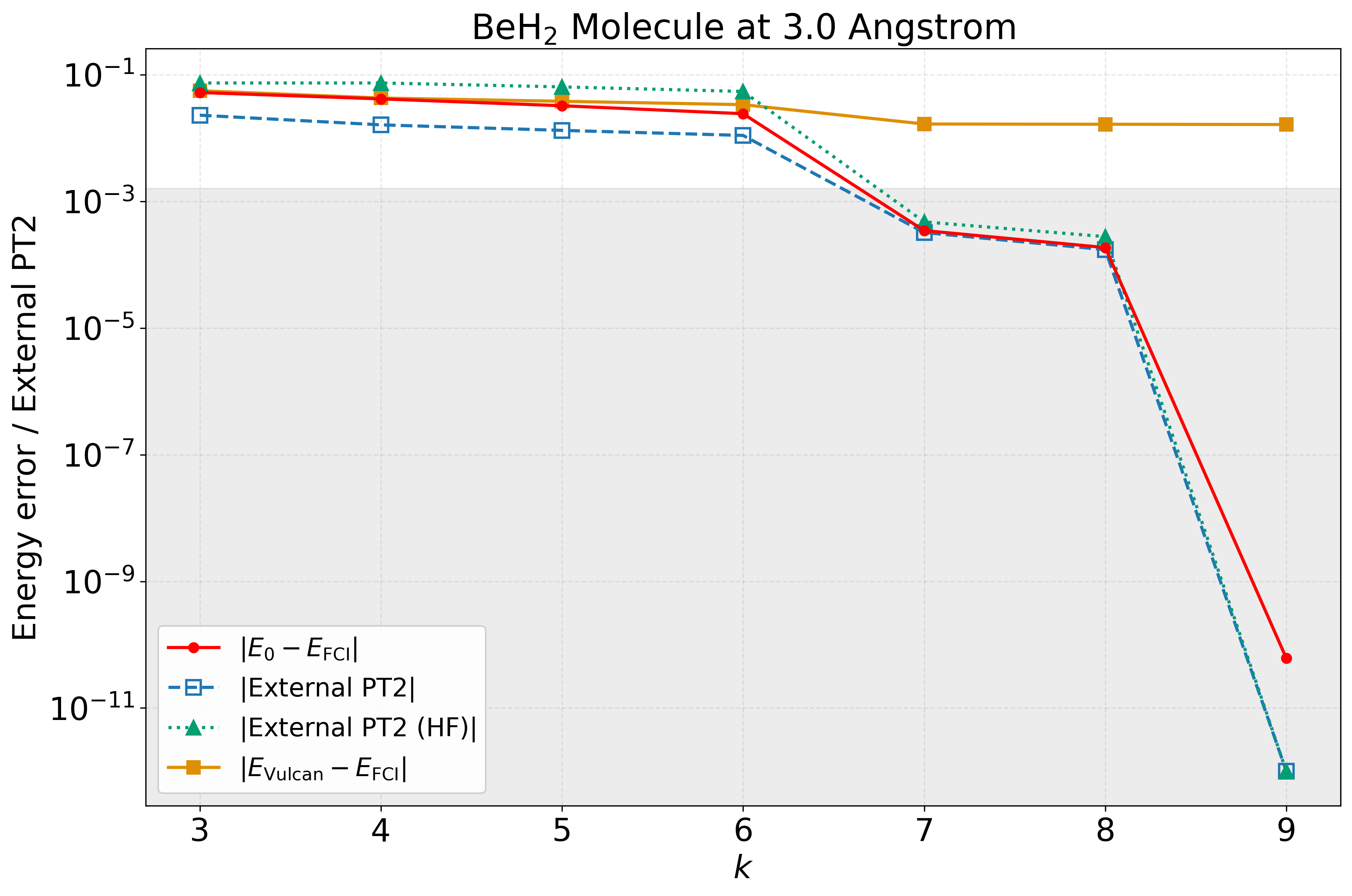}
    \caption{Exact diagonalisation vs $k$}
    \label{fig:hproj_BeH2_k}
  \end{subfigure}\hfill
  \begin{subfigure}[t]{0.49\textwidth}
    \centering
    \includegraphics[width=\linewidth]{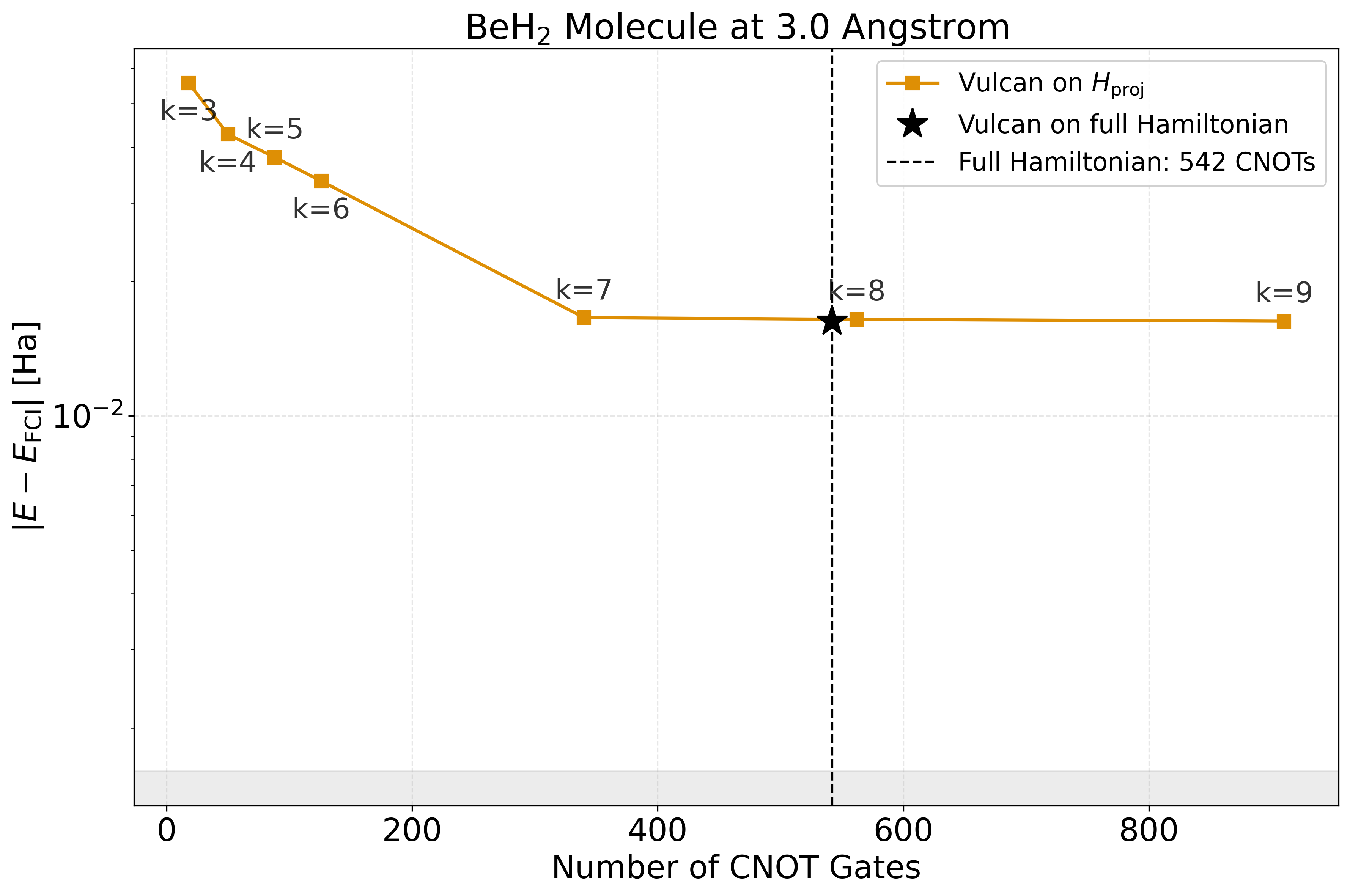}
    \caption{\methodename vs CNOTs}
    \label{fig:hproj_BeH2_cnots}
  \end{subfigure}\hfill
  \begin{subfigure}[t]{0.49\textwidth}
    \centering
    \includegraphics[width=\linewidth]{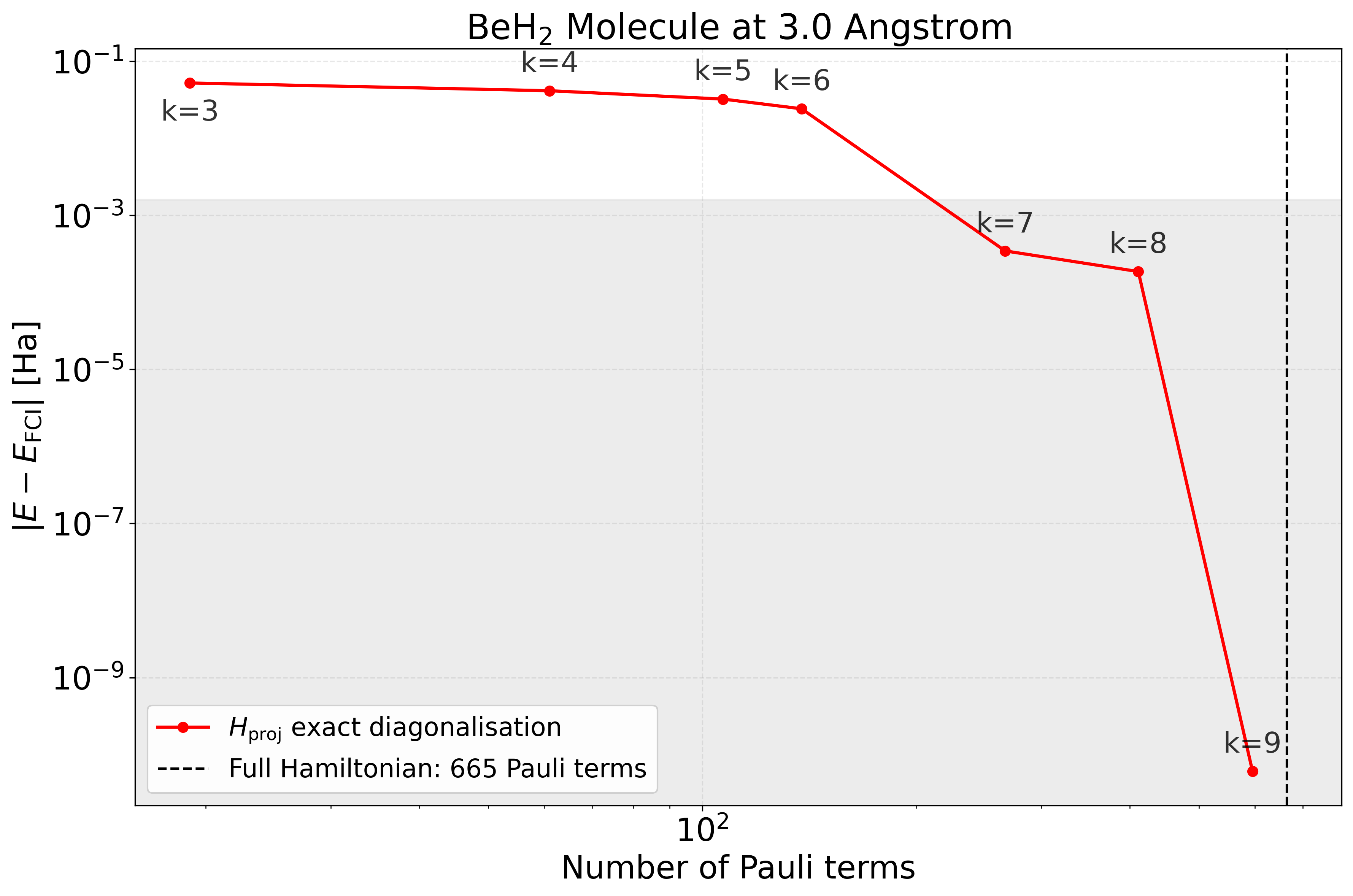}
    \caption{Exact diagonalisation vs Pauli terms}
    \label{fig:hproj_BeH2_paulis}
  \end{subfigure}
  \caption{Same as Fig.~\ref{fig:hproj_H4} for the BeH$_2$ molecule at 3.0~\AA{} (STO-3G, 14 qubits).}
  \label{fig:hproj_BeH2}
\end{figure}

\begin{figure}[htbp]
  \centering
  \begin{subfigure}[t]{0.49\textwidth}
    \centering
    \includegraphics[width=\linewidth]{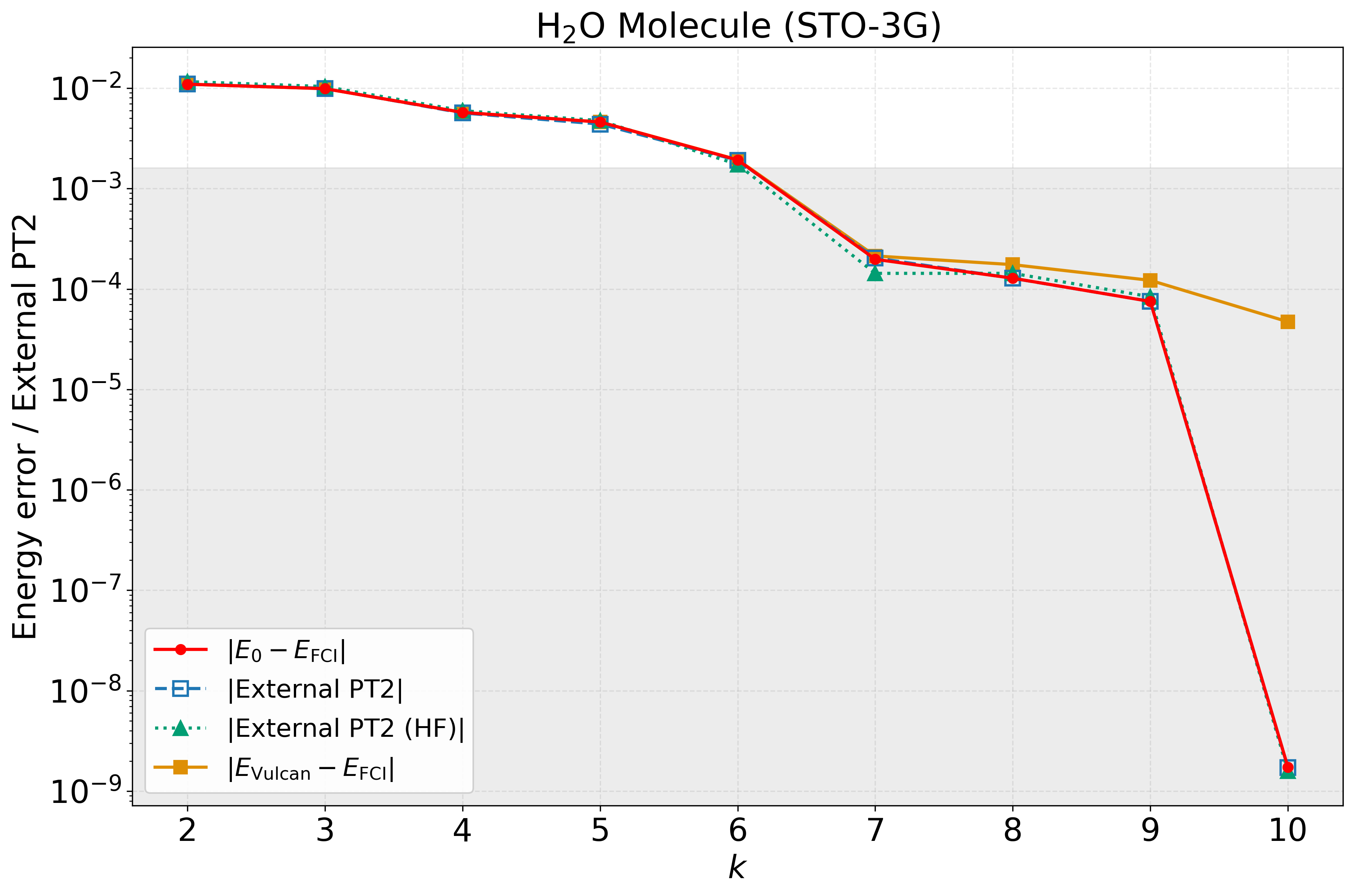}
    \caption{Exact diagonalisation vs $k$}
    \label{fig:hproj_H2O_sto3g_k}
  \end{subfigure}\hfill
  \begin{subfigure}[t]{0.49\textwidth}
    \centering
    \includegraphics[width=\linewidth]{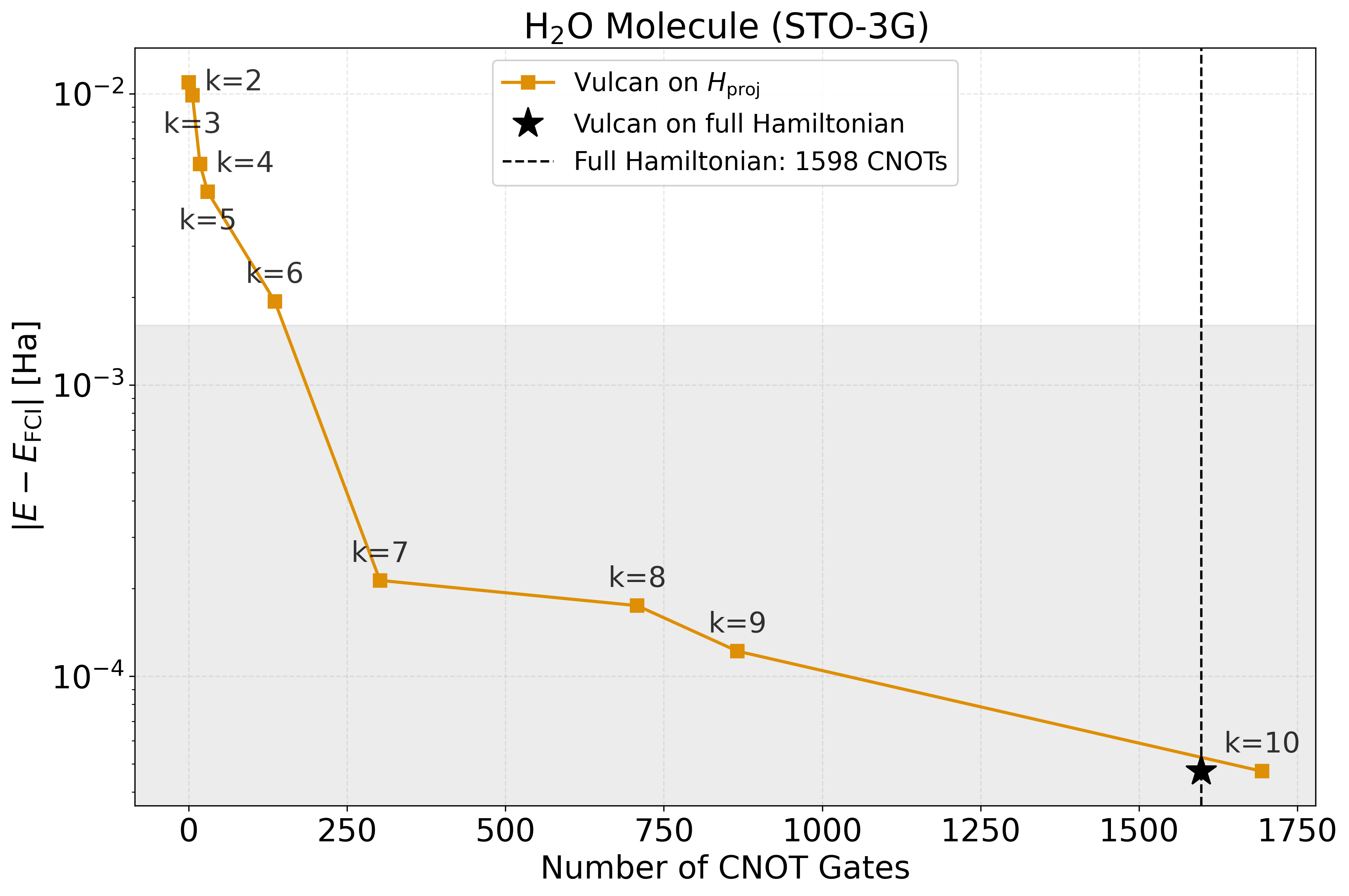}
    \caption{\methodename vs CNOTs}
    \label{fig:hproj_H2O_sto3g_cnots}
  \end{subfigure}\hfill
  \begin{subfigure}[t]{0.49\textwidth}
    \centering
    \includegraphics[width=\linewidth]{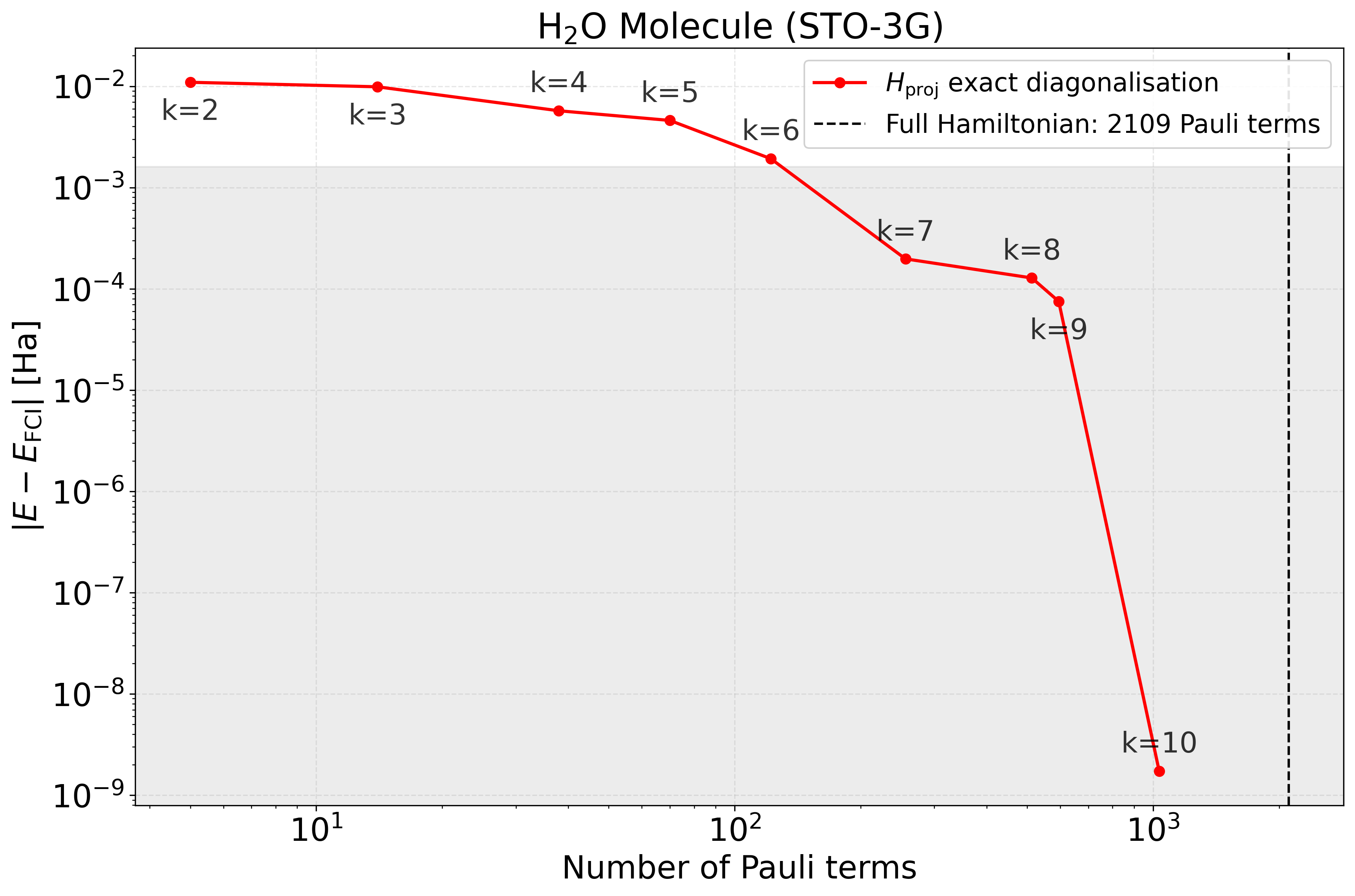}
    \caption{Exact diagonalisation vs Pauli terms}
    \label{fig:hproj_H2O_sto3g_paulis}
  \end{subfigure}
  \caption{Same as Fig.~\ref{fig:hproj_H4} for the H$_2$O molecule   (STO-3G, 14 qubits).}
  \label{fig:hproj_H2O_sto3g}
\end{figure}

\begin{figure}[htbp]
  \centering
  \begin{subfigure}[t]{0.49\textwidth}
    \centering
    \includegraphics[width=\linewidth]{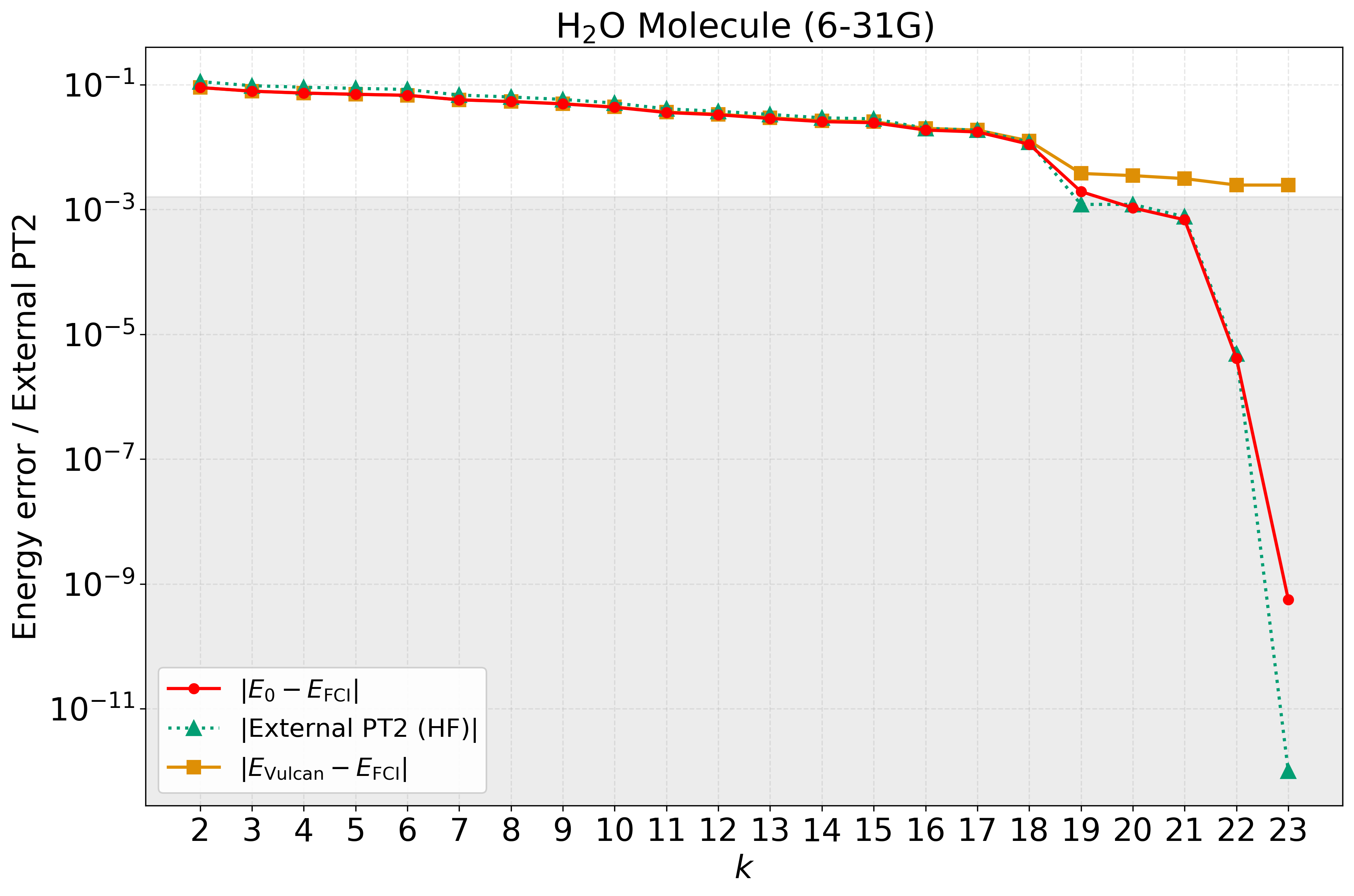}
    \caption{Exact diagonalisation vs $k$}
    \label{fig:hproj_H2O_631g_k}
  \end{subfigure}\hfill
  \begin{subfigure}[t]{0.49\textwidth}
    \centering
    \includegraphics[width=\linewidth]{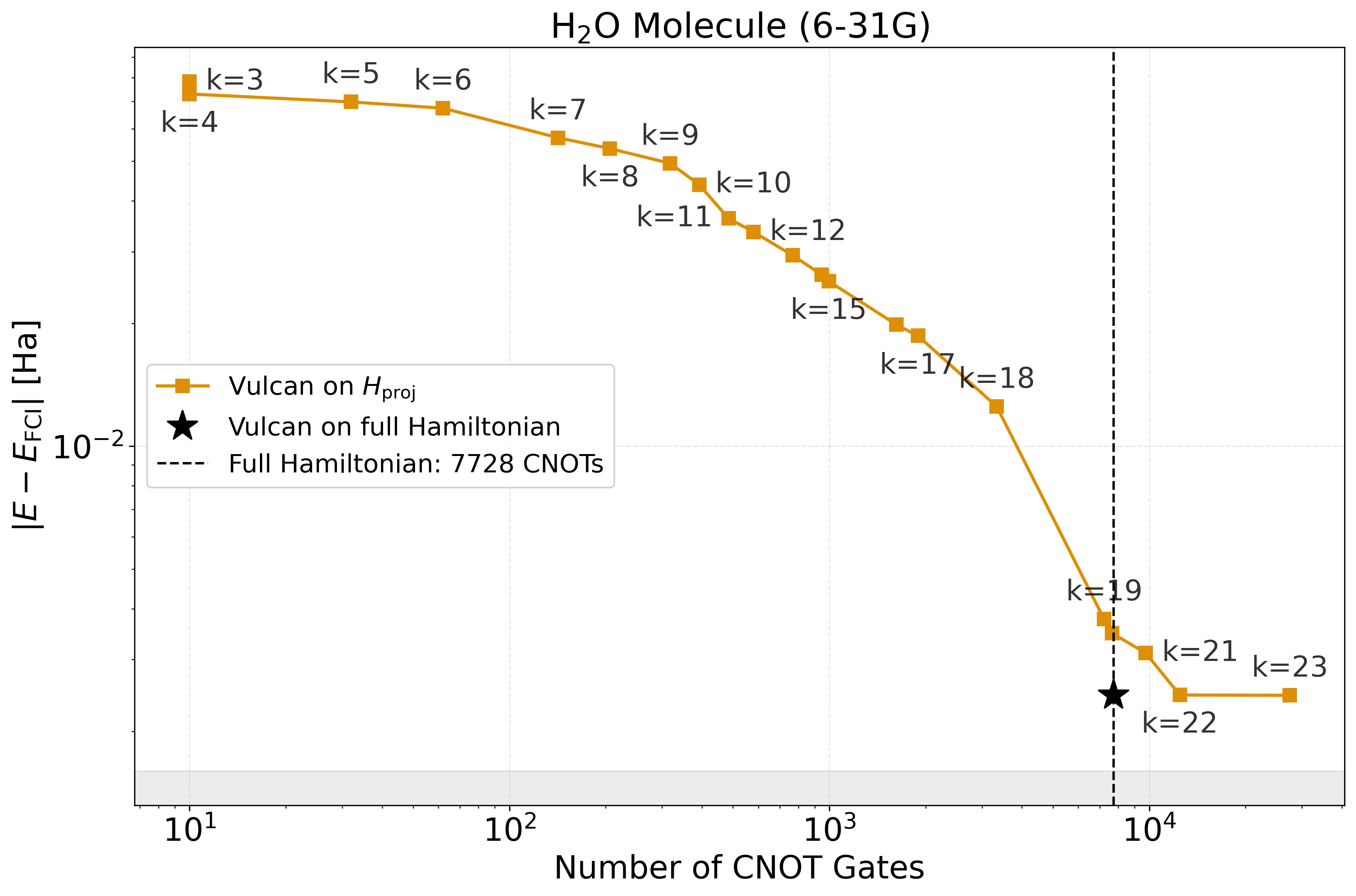}
    \caption{\methodename vs CNOTs}
    \label{fig:hproj_H2O_631g_cnots}
  \end{subfigure}\hfill
  \begin{subfigure}[t]{0.49\textwidth}
    \centering
    \includegraphics[width=\linewidth]{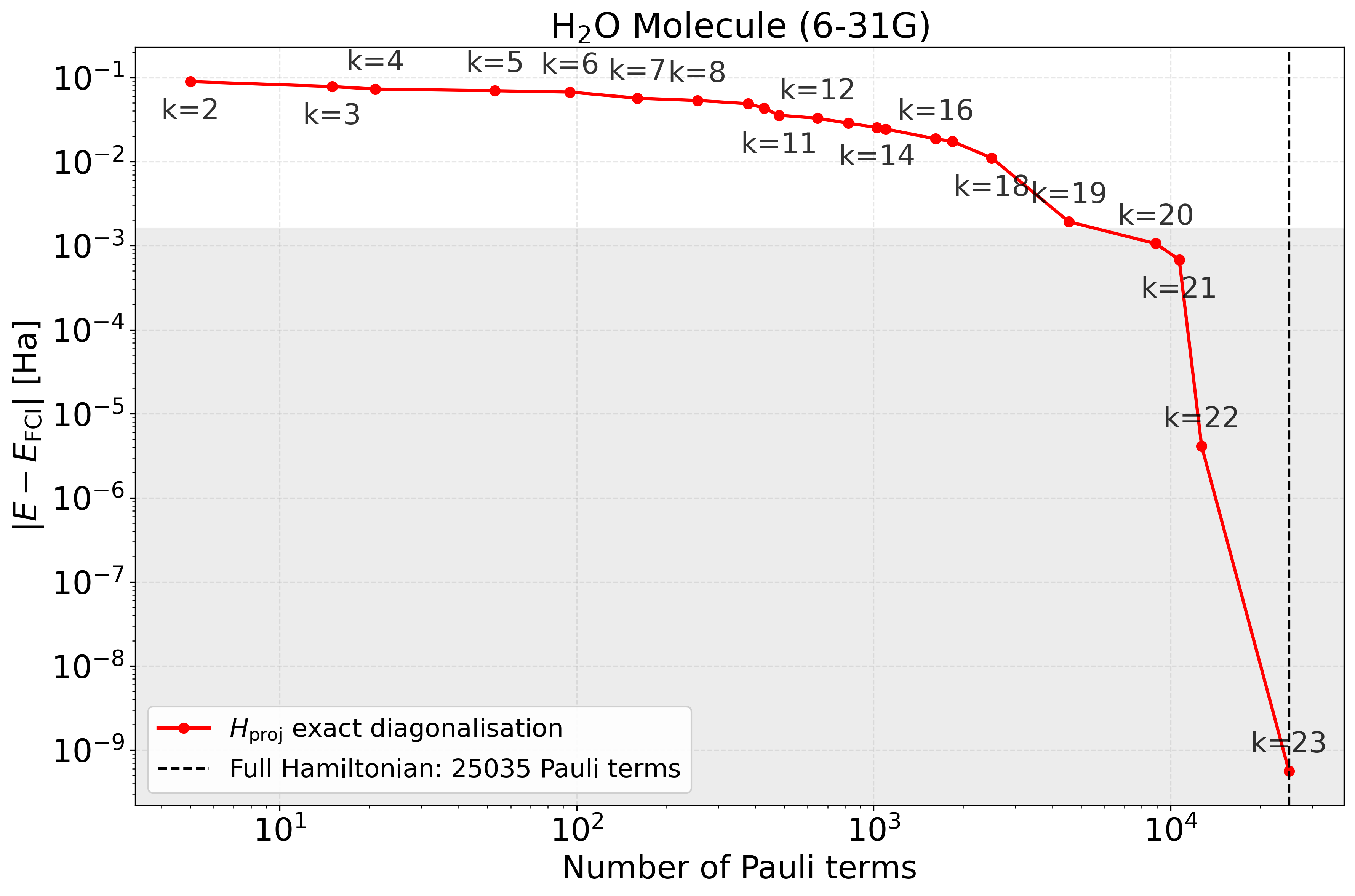}
    \caption{Exact diagonalisation vs Pauli terms}
    \label{fig:hproj_H2O_631g_paulis}
  \end{subfigure}
  \caption{Same as Fig.~\ref{fig:hproj_H4} for the H$_2$O molecule  (6-31G, 26 qubits).}
  \label{fig:hproj_H2O_631g}
\end{figure}

Figures~\ref{fig:hproj_H4}--\ref{fig:hproj_H2O_631g} show
the results for H$_4$ (STO-3G, 8~qubits), BeH$_2$
(STO-3G, 14~qubits),  H$_2$O (STO-3G, 14~qubits), and H$_2$O (6-31G, 26~qubits).
Each figure contains three panels.

Panel~(a) shows the energy error of the exact
diagonalization of $H_{\mathrm{proj}}$ as a function of
subspace size $k$, together with the two PT2 estimates
and the error of a single \methodename iteration on
$H_{\mathrm{proj}}$. In all three STO-3G systems, the
exact-diagonalization error drops rapidly with $k$ and
enters the chemical accuracy band well before the full
register is reached: at $k = 4$ out of 8 for
H$_4$, $k = 7$  out of 14 for BeH$_2$, and
$k = 7 $  out of 14 for H$_2$O. 
The PT2 correction estimates the energy lost by the projection. With the
exact subspace ground state as reference, this estimate agrees closely with
the actual energy loss at every $k$ for H$_4$ and H$_2$O, and for BeH$_2$
once the dominant determinants are included ($k \geq 7$); below that it
underestimates the loss by about a factor of two. The Hartree--Fock-based
estimate is cheaper, since it does not require diagonalizing
$H_{\mathrm{proj}}$, and follows the same trend, although it can lie above
or below the actual loss. The exact-reference estimate provides an automatic stopping criterion. The
Hartree--Fock-based one can stop too early: for H$_4$, it vanishes at $k = 4$
while $1.3\times10^{-4}$~Ha is still missing.

Panel~(b) shows the error of a single \methodename{}
iteration on $H_{\mathrm{proj}}$ as a function of CNOT
count. The star marks the same single-iteration
calculation on the full Hamiltonian.  For H$_4$, the projected calculation
reaches chemical accuracy at a fraction of the
full-register gate cost. For stretched BeH$_2$, a single \methodename iteration does not reach chemical accuracy on the projected register, and increasing k beyond 7 does not help: the exact error at k = 7 is already 3.4×$10^{-4}$ Ha, so the remaining error comes from the single iteration. A second iteration on the 7-qubit register reaches chemical accuracy
($3.8\times10^{-4}$~Ha) with 680 CNOTs, against 1084 CNOTs for two
iterations on the full 14-qubit register.

These circuits can be shortened further without loss of accuracy. Choosing
the basis of the subspace that minimizes the CNOT count of the projected
pool, and compressing the projected layer with the overlap-guided procedure
(see Supplementary Material), reduces them by 40--74\% at the same energy
(Table~\ref{tab:proj}). The resulting circuits require 1.4 to 6 times fewer
CNOTs than ADAPT-VQE with the QEB pool on the full register, and 2.6 to 10
times fewer than \methodename{} on the full register, on a smaller register
(half the qubits for the STO-3G systems, 20 of 26 for H$_2$O in 6-31G).

Panel~(c) shows the exact-diagonalization error as a
function of the number of Pauli terms in
$H_{\mathrm{proj}}$. As $k$ decreases, the projected
Hamiltonian becomes substantially simpler than the
full Hamiltonian (dashed line). For H$_2$O, the full
Hamiltonian contains 2109 Pauli terms, while
$H_{\mathrm{proj}}$ at $k = 7 $  contains only
256. This reduces 
the per-evaluation measurement cost, since each energy
measurement scales with the number of Hamiltonian
terms.

Figure~\ref{fig:hproj_H2O_631g} extends the projection
to H$_2$O in the 6-31G basis, a 26-qubit system.
Panel~(a) shows the exact-diagonalization error
reaching chemical accuracy at $k = 20$ out of 26
qubits, confirming that the projected subspace
captures the essential correlation energy even for
this larger system. Panel~(c) confirms the
Hamiltonian simplification: the full system contains
25035 Pauli terms, reduced to 8911 at $k = 20$.

Panel~(b) shows the circuit cost. A single \methodename iteration on the projected register at k = 20 requires 7658 CNOTs, about the same as the 7728 CNOTs of the same iteration on the full 26-qubit register, and the cost grows quickly with k: at k = 23, where the projection keeps the whole space and recovers the full-register energy, the same iteration requires 27456 CNOTs. Fewer qubits therefore do not give a shallower circuit here. This is not due to the pool, since each projected generator performs the same rotation as on the full register, but to the cost of each generator: on the full register a generator acts on at most four spin orbitals, whereas on the projected register it acts on a growing fraction of the qubits. This cost can be reduced by choosing a different basis for the same subspace
and compressing the layer: at $k = 20$, the circuit then requires 2012 CNOTs,
against 4043 for ADAPT-VQE with the QEB pool at the same accuracy (Table \ref{tab:proj}).

Projection consistently reduces the qubit register and the Hamiltonian
complexity. For the STO-3G systems, it also gives shallower circuits; for
H$_2$O in 6-31G, this requires the optimized basis. Once the basis is
optimized and the circuit compressed, the projected circuit is shorter than
ADAPT-VQE with the QEB pool at the same accuracy in all four systems
(Table~\ref{tab:proj}), and the PT2 estimate provides a stopping criterion for the
growth of the subspace in all of them.


\begin{table*}[]
\centering
\begin{ruledtabular}
\begin{tabular}{lccccc}
\textbf{System} & \textbf{Qubits} & \textbf{\methodename} & \textbf{QEB} & \textbf{Original basis} & \textbf{Optimized basis} \\
       & full $\to k$ & (full register) & (full register) & (no compression) & (+ compression)\vspace{0.5mm} \\
\hline & \\[-1.9ex]
H$_4$            & $8 \to 4$  & 142 ($2.1\times10^{-5}$)  & 130 ($1.3\times10^{-4}$) & 62 ($1.3\times10^{-4}$)  & \textbf{22} ($1.3\times10^{-4}$) \\
BeH$_2$ (3.0~\AA) & $14 \to 7$ & 1084 ($8.4\times10^{-5}$) & 593 ($8.5\times10^{-5}$) & 680 ($3.8\times10^{-4}$) & \textbf{410} ($3.8\times10^{-4}$) \\
H$_2$O (STO-3G)  & $14 \to 7$ & 1598 ($4.7\times10^{-5}$) & 338 ($2.1\times10^{-4}$) & 302 ($2.1\times10^{-4}$) & \textbf{152} ($2.2\times10^{-4}$) \\
H$_2$O (6-31G)   & $26 \to 20$ & 7728 ($2.5\times10^{-3}$) & 4043 ($3.5\times10^{-3}$) & 7658 ($3.5\times10^{-3}$) & \textbf{2012} ($3.5\times10^{-3}$) \\
\end{tabular}%
\end{ruledtabular}
\caption{CNOT counts, with the energy error with respect to FCI (Ha)
in parentheses (STO-3G basis unless stated).
  \emph{Qubits}: size of the full register and of the projected register $k$.
  \emph{\methodename{} (full register)}: uncompressed \methodename{} on the
  full register, with the same number of iterations as the original-basis
  column (for one iteration, the star in panel (b) of
  Figs.~\ref{fig:hproj_H4}--\ref{fig:hproj_H2O_631g}), given with its own
  error; unlike the other columns, it is not taken at the accuracy of the
  original-basis column.
  \emph{QEB (full register)}: ADAPT-VQE with the QEB pool, one operator added
  per iteration, stopped at the first iteration whose error is at most that
  of the original-basis column.
  \emph{Original basis (no compression)}: one \methodename{} iteration on the
  $k$-qubit register with the projected pool (two iterations for BeH$_2$,
  where a single iteration stops at $1.7\times10^{-2}$~Ha); its error sets the
  accuracy at which the QEB and optimized-basis columns are compared.
  \emph{Optimized basis (+ compression)}: the same iterations in the optimized
  basis of the subspace, rebuilt by overlap-guided compression on the $k$
  qubits and stopped at the first circuit reaching the error of the
  original-basis column (Supplementary Material). Bold: fewest CNOTs at equal
  accuracy.}
  \label{tab:proj}
\end{table*}

\subsection{Early termination: Stochastic PT2}

As the number of ansatz layers increases, the marginal energy improvement from each additional layer diminishes. Rather than deepening the circuit further, we apply a second-order Epstein--Nesbet perturbation theory (ENPT2) correction classically, using only the energy and wavefunction already obtained from the preceding quantum optimization. The ENPT2 correction adds small contributions of determinants
outside the variational state, as in selected configuration
interaction (sCI) methods. Because this step adds no quantum circuit executions, it effectively shifts the remaining error onto classical GPU hardware.

PT2-corrected energies are indicated in Figures~\ref{fig:convergence_beh2}--\ref{fig:convergence_h2o} by thin pink `$\times$'
markers, distinct from the larger filled `X' of the compressed iterations, with numerical labels identifying the compressed  \methodename iteration used as the seed. For equilibrium geometries, the correction consistently brings the energy below chemical accuracy using circuits from the first or second compressed  iteration. The corresponding measurement and CNOT costs are reported in Table~\ref{tab:cnots_cost_comparison} (PT2 columns) and reflect only the quantum resources of the seed iteration, not the classical PT2 cost. PT2 enables early termination of the quantum loop, reaching chemical accuracy from a shallower seed circuit with far less accumulated measurement cost. For H$_8$ at 0.85~\AA{}, chemical accuracy via compression alone requires iterating until 0.59 of the baseline measurement budget has been spent, applying PT2 to the first compressed iteration instead reaches the same accuracy at a cumulative cost of only 0.02. For H$_{10}$ at 0.85~\AA{}, the corresponding reduction is from 0.35 to 0.002. For stretched geometries, where chemical accuracy is already reached without correction, PT2 is omitted.

For larger systems, exact evaluation of the ENPT2 sum is prohibitive due to exponential scaling. We employ a stochastic variant based on importance sampling (Section~\ref{sec::PT2}), which evaluates only a small fraction of the perturbative terms.

We test the stochastic variant on two larger systems:
H$_2$O in the 6-31G basis, a 26-qubit problem
(Figure~\ref{fig:convergence_h2o}), and C$_3$H$_6$
(Figure~\ref{fig:convergence_C3H6}).

For H$_2$O, the ENPT2 sum contains up to $2^{26}$ terms. The stochastic method samples a small subset of
these terms, with probabilities proportional to the squared couplings
$|H_{0j}|^2$, and estimates the total correction from that subset alone. As
shown in Table~\ref{tab:stochastic_pt2}, evaluating fewer than 0.01\% of the
terms (about 1700 to 6700 denominators) is sufficient: for the first
\methodename iteration, the estimated correction stabilizes at approximately
$-2.5\times10^{-3}$~Ha with a standard error of a few $10^{-5}$~Ha, and
varies by about $10^{-4}$~Ha across subset sizes, well below chemical
accuracy. Chemical accuracy is reached already at the first \methodename
iteration combined with this correction, at a measurement cost of 0.0005 of the batched ADAPT-VQE baseline. The second-iteration corrections show comparable
convergence. This is the same 26-qubit system for which subspace
projection (Figure~\ref{fig:hproj_H2O_631g}) reduces the register
to $k = 20$ qubits while preserving chemical accuracy
in exact diagonalization. 
On that register, the circuit in the optimized basis needs fewer CNOTs than
one  \methodename iteration on the full register (2012 against 7728, Table~\ref{tab:proj}), but
stops at $3.5\times10^{-3}$~Ha.

For C$_3$H$_6$ (STO-3G, active space of 18 electrons in 16 orbitals,
32 qubits), evaluating only 1484 denominators, the stochastic correction
reduces the energy error to $1.0\times10^{-3}$~Ha at the first \methodename
iteration, after $9.6\times10^{5}$ circuit executions (Figure~\ref{fig:convergence_C3H6}). Batched
ADAPT-VQE remains $6.4\times10^{-3}$~Ha above the FCI energy after 359
iterations and $3.3\times10^{11}$ circuit executions. The computational cost
of this step is reported in Section~\ref{sec::PT2}.

The combination of a single \methodename layer with stochastic ENPT2 therefore
provides a practical route to chemical accuracy, within the emulation regime,
for systems where the full grow-then-compress loop would be expensive: the
quantum circuit remains shallow, and importance sampling reduces the number
of classical denominator evaluations, although memory still scales with the
full state vector (Section~\ref{sec::PT2}).

One can see that the formula for the PT2 (Eq.~\ref{eq:enpt2}) diverges if the denominator becomes too small. Although we did not observe this divergence in any of the systems tested, we also implemented the Diagonal Unitary Correction (DUC) formula (Eq.~25 of~\cite{Ryabinkin2021}) to prevent any potential instabilities. The DUC yields identical results to Eq.~\ref{eq:enpt2}, even when combined with the stochastic method.

\begin{table}[h]

\centering

\begin{tabular*}{0.7\columnwidth}{@{\extracolsep{\fill}}ccc}
\hline\hline
\multicolumn{3}{c}{\textbf{1st Iteration}} \\ \hline \\[-2.4ex]
$f$ ($\times 10^{-4}$) & $\Delta_E$ ($\times 10^{-3}$) & $\sigma$ ($\times 10^{-5}$) \\ \hline
0.25 & -2.400 & 6.5 \\ \hline
0.5  & -2.468 & 4.6 \\ \hline
0.75 & -2.496 & 3.4 \\ \hline
1.0  & -2.506 & 3.0 \\ \hline\hline
\end{tabular*}

\vspace{0.5cm}

\begin{tabular*}{0.7\columnwidth}{@{\extracolsep{\fill}}ccc}
\hline\hline
\multicolumn{3}{c}{\textbf{2nd Iteration}} \\ \hline \\[-2.4ex]
$f$ ($\times 10^{-4}$) & $\Delta_E$ ($\times 10^{-3}$) & $\sigma$ ($\times 10^{-5}$) \\ \hline
0.25 & -1.663 & 6.5 \\ \hline
0.5  & -1.716 & 4.8 \\ \hline
0.75 & -1.796 & 3.8 \\ \hline
1.0  & -1.729 & 4.9 \\ \hline\hline
\end{tabular*}

\caption{Stochastic ENPT2 correction for H$_2$O (6-31G, 26 qubits), seeded
by the first and second \methodename iterations. $f$ is the fraction of the
$2^{26}$ terms of Eq.~\eqref{eq:enpt2} whose denominators are evaluated, $\Delta E$ is the
estimated correction in Ha, and $\sigma$ is the standard error of the
estimator (Algorithm~\ref{alg:stochastic_enpt2}).}
\label{tab:stochastic_pt2}
\end{table}

\section{Discussion}

In this work, \methodename offers an alternative approach to the gradient-based operator selection of ADAPT-VQE with a physics-derived, complete pool whose generators are produced in seconds on a classical computer. The cost of this method is a set of trade-offs: between measurement budget, circuit depth, qubit count, and classical post-processing, that the user navigates according to whichever hardware resource is the scarcest. 

When the dominant constraint is measurement budget, the uncompressed layered ansatz is the appropriate operating point: it reaches chemical accuracy at 5--20\% of the baseline measurement cost, but with deeper circuits. When the circuit depth is the binding constraint, overlap compression recovers depths comparable or shorter than batched ADAPT-VQE for four of the seven systems, at a measurement cost that remains below the baseline in every case tested. When the required number of qubits exceeds the available register, subspace projection reduces the register before any quantum circuit is executed, with the PT2-guided stopping criterion on the dimension of the projected subspace. 

Finally, when GPU resources are available, stochastic ENPT2 permits early termination of the quantum loop entirely: for H$_2$O in the 6-31G basis, a single layer suffices once the perturbative correction is applied classically.

Projection, compression, and perturbative correction act on different
resources and are complementary. For H$_2$O in 6-31G, projection reduces the
26-qubit register to 20 qubits (Figure~10), and stochastic ENPT2 separately
reaches chemical accuracy after one layer on the full register (Figure~\ref{fig:convergence_h2o}). On
the projected registers, choosing the basis of the subspace and compressing the
projected layer gives circuits shorter than both \methodename{} and ADAPT-VQE with the
QEB pool on the full register, for all four systems tested (H$_4$, BeH$_2$,
H$_2$O in STO-3G and 6-31G, Table~\ref{tab:proj}).

We benchmarked \methodename{} on H$_6$, H$_8$, H$_{10}$, BeH$_2$, H$_2$O and
C$_3$H$_6$, comparing against batched ADAPT-VQE with QEB pools, an operator
pool 4 to 25 times larger than ours with gradient-based selection optimized to
its best-case batch size. Despite this, \methodename{}  reached chemical accuracy at
5--20\% of the baseline measurement cost for all systems of Table~\ref{tab:cnots_cost_comparison}. After
overlap compression, the CNOT count fell below the baseline for four of the
seven systems, all at stretched geometries, and for stretched hydrogen chains
the improvement grew with chain length, reaching 0.14$\times$ for H$_{10}$.
Stochastic ENPT2 extended this to 26 qubits (H$_2$O, 6-31G), where chemical
accuracy was achieved after a single layer at 0.05\% of the baseline
measurement cost, and to 32 qubits (C$_3$H$_6$), where batched ADAPT-VQE does
not reach chemical accuracy.

\textit{\textbf{Outlook}.} 
Although VULCAN is ultimately aimed at hardware implementations, all presented results are based on noiseless state-vector emulations, with costs reported as Pauli-term evaluations rather than shot counts. The behavior of \methodename under shot noise and hardware error, particularly the optimization of layers containing up to $\sim10^3$ parameters, remains to be established. Because reproducing all proposed VULCAN features demands significant quantum resources and careful, hardware-dependent measurement management, this will be the focus of a separate hardware implementation study. Even without these incoming enhancements and as a preliminary proof of concept, we evaluated the total energy of dihydrogen (H$_2$) on the Quantinuum System Model H2 trapped-ion (see Supplementary Material) and found a good match ($-1.12666$~Ha) with the emulated theoretical values ($-1.12836$~Ha), $1.7\times10^{-3}$~Ha above the FCI energy, close to chemical accuracy
($1.6\times10^{-3}$~Ha).

Several technical aspects will also require dedicated studies beyond the scope of the present work. First, the ordering of Pauli-string exponentials within a layer modifies the ansatz and may impact convergence; exploring this dependence will be a valuable direction. Second, the single-Pauli \methodename pool does not explicitly conserve particle number, $S_z$, or total spin. While this has no visible effect at equilibrium geometries, in the strongly correlated regime of stretched chains, states achieving chemical accuracy in energy can exhibit spin contamination and partially deviate from the $(N_\alpha, N_\beta)$ sector of the reference state. Such spin contamination is a well-known phenomenon not specific to VULCAN, appearing similarly in other symmetry-unadapted methods like unrestricted Hartree--Fock or non-spin-adapted DMRG. A rigorous analysis of these effects in challenging, highly correlated benchmark systems will be highly useful.  Third, all energies above are converged within a finite basis set. Reaching chemical accuracy with respect to the complete basis set additionally requires a
basis-set correction. In the Supplementary Material (Section~2), we combine
VULCAN with the density-based basis-set correction (DBBSC)~\cite{traore2024shortcut}, which
estimates the missing energy from the density of the quantum state at no
additional quantum cost. Results demonstrate the full applicability of the DBBSC approach to VULCAN as it was demonstrated for ADAPT-VQE\cite{traore2024shortcut}. 

Finally, although an extensive classical ENPT2 correction is a highly desirable strategy for early termination of the quantum sequence, developing a fully quantum ENPT2 evaluation method will become increasingly relevant as quantum processors advance toward fault-tolerant quantum computing (FTQC).

\textit{\textbf{Concluding Remarks}.} Overall, future work will focus on three primary directions: (1) deploying \methodename on physical NISQ hardware to directly test noise resilience; (2) integrating projection with the full grow-then-compress loop for systems beyond minimal basis sets; and (3) restoring the symmetries of the state. This last objective does not require altering the pool or the circuit architecture; instead, penalty terms can be incorporated into the cost function to be minimized, analogous to constrained VQE.

The following table, Table~\ref{tab:summary}
 summarizes:i) the usual VQE bottlenecks addressed in this work when dealing with quantum chemistry; ii) the \methodename component that solves each of them, and iii) the remaining open questions. Overall, VULCAN is a step forward in bridging the gap between current quantum algorithms limitations and the demanding requirements of scalable, high-accuracy quantum chemistry computations towards meaningful applications\cite{convergence2026}.

\begin{table*}[]
\begin{ruledtabular}
\begin{tabular}{l|l|l}
\multicolumn{1}{c|}{VQE Bottleneck}                                                                                       & \multicolumn{1}{c|}{Technical solution (VULCAN)}                                                                                                                      & \multicolumn{1}{c}{Results in this work}                                                                                                                                                          \\  \hline
\begin{tabular}[c]{@{}l@{}}Energy-gradient measurements \\ to select operators (ADAPT-VQE)\end{tabular}               & \begin{tabular}[c]{@{}l@{}}Whole pool appended as one layer, \\ no operator selection\end{tabular}                                                 & \begin{tabular}[c]{@{}l@{}}Chemical accuracy at 5--20\% of the \\ measurement cost of batched ADAPT-VQE \\ (hydrogen chains, BeH$_2$, H$_2$O, \\ up to 26 qubit simulations)\end{tabular}       \\ \hline
Choice of starters for MCPs                                                                                           & \begin{tabular}[c]{@{}l@{}}Physically motivated starters generated \\ from the Hamiltonian by the VDBF flow \\ equation, completed into an MCP\end{tabular} & \begin{tabular}[c]{@{}l@{}}Pool produced with no preliminary VQE \\ runs, 4-25 times smaller than the QEB pool\end{tabular}                                                                          \\ \hline
Circuit depth of layered ansätze                                                                                      & State-overlap compression                                                                                                                          & \begin{tabular}[c]{@{}l@{}}CNOT count below batched ADAPT-VQE \\ for four of the seven tested systems\end{tabular}                                                                               \\ \hline
Number of qubits of the register                                                                              & \begin{tabular}[c]{@{}l@{}}Projection of the Hamiltonian onto a\\ dominant-determinant subspace\end{tabular}                                                                                                            &\begin{tabular}[c]{@{}l@{}}Chemical accuracy in exact diagonalization\\ with 23--50\% fewer qubits and 60--88\% fewer\\ Pauli terms (H$_4$, BeH$_2$, H$_2$O; up to 26 qubits)\end{tabular}

\\ \hline
Circuit cost on the projected register                                                                                & \begin{tabular}[c]{@{}l@{}}Optimized subspace basis and \\ cost-aware compression\end{tabular}                                                     & \begin{tabular}[c]{@{}l@{}}40--74\% fewer CNOTs than in the original basis;\\ 1.4--6 times fewer than ADAPT-VQE (QEB)\\ on the full register, at the same accuracy.\end{tabular}                                                                                        \\ \hline
\begin{tabular}[c]{@{}l@{}}Number of iterations and measurements\\ to reach chemical accuracy\end{tabular} & ENPT2 correction                                                                                                                                   & \begin{tabular}[c]{@{}l@{}}Chemical accuracy after a single \methodename layer:\\ H$_2$O (6-31G, 26 qubits) at 0.05\% of the cost of\\ batched ADAPT-VQE, C$_3$H$_6$ (32 qubits), where\\ batched ADAPT-VQE does not reach it\end{tabular} \\ \hline
\begin{tabular}[c]{@{}l@{}} High number of terms in the ENPT2\\ correction for larger systems\end{tabular} & Stochastic ENPT2                                                                                                                                   & \begin{tabular}[c]{@{}l@{}}Fewer than 0.01\% of the terms evaluated for H$_2$O\\ (6-31G); 1484 denominators for C$_3$H$_6$\end{tabular}        
\end{tabular}
\end{ruledtabular}
\caption{Summary of the VQE bottlenecks in quantum chemistry addressed by \methodename and corresponding results obtained in this work.}
\label{tab:summary}
\end{table*}

\section{Methods}
\label{sec:Methods}
\subsection{The layered ansatz}
The \methodename ansatz is defined in Eq.~\ref{eq-ans}.
Each product
$\prod_{j=1}^{M}\exp(i\theta_{j,k}P_j)$ constitutes
one layer. Layers are appended one at a time. When a new layer $k$ is added, all parameters of the ansatz are re-optimized,
starting from their previous values for the earlier layers and from zero for
the new one. The number of free parameters therefore grows with the number of
layers, by $M$ per layer.

In the compressed variant, each grown layer is replaced by a shorter layer
built by overlap-guided insertion (Section~\ref{sec:growthencompress} ). The compressed ansatz
retains the product-of-exponentials form, with a   sequence
$(j_{k,1},\dots,j_{k,L_k})$ of $L_k$ pool indices, in which an index may
appear more than once:
\begin{equation}
  |\psi\rangle = \prod_{k=1}^{n}\left(\prod_{l=1}^{L_k}
  \exp\!\left(i\theta_{l,k}P_{j_{k,l}}\right)\right)|\psi_0\rangle ,
  \label{eq:compressed_ansatz}
\end{equation}
where the most recently inserted generator acts last, and $L_k \le \max(1,\lfloor fM\rfloor)$, where $f$ is the compression
fraction (typically $f = 0.5$).

\subsection{Operator pool generation}
\label{sec:pool}
The necessity of constructing operator pools based on
the MCP framework, defined as the minimal generating
set for a target Lie algebra, is driven by the severe
measurement overhead inherent in gradient-based
adaptive techniques. In standard protocols, iterative
operator selection requires evaluating an energy
gradient for every pool element, a process that scales
naively $\mathcal{O}(N^4)$ with the number of qubits
and rapidly becomes computationally intractable.
Because MCPs scale linearly with the number of qubits
$N$, they have the potential to mitigate this
bottleneck.

\subsubsection{Existing pools and the starter challenge}

The expressivity and convergence rate of adaptive
variational algorithms depend on the choice of the
operator pool. Traditionally, these pools are
constructed using heuristic or physically motivated
criteria to align with the target problem's structure.
Prominent examples in the literature include:

\begin{itemize}
\item \textbf{Fermionic pools}~\cite{Grimsley2019}:
Comprise physically motivated operators, typically
the set of all unique, spin-complemented one- and
two-body excitations.

\item \textbf{Qubit pool}~\cite{tang2021qubit}: Derived by
decomposing the Fermionic-ADAPT operators via the
Jordan-Wigner mapping, treating the individual Pauli
strings as the fundamental operators of the pool.

\item \textbf{Qubit-Excitation-Based (QEB)
pools}~\cite{Yordanov2021}: Employ exponential
single- and double-qubit excitation evolutions,
expressed directly through qubit creation and
annihilation operators mapping.

\item \textbf{Coupled Exchange Operator (CEO)
pool}~\cite{ramoa2025reducing}: Constructed from linear
combinations of qubit excitations that act on the
exact same set of spin-orbitals, designed to strictly
preserve specific symmetries.
\end{itemize}

A limitation of these pools is that their size scales
polynomially or worse with respect to the number of
qubits. To overcome this, recent works
\cite{tang2021qubit,Viswanathan2026,Shkolnikov2023,Haidar2025,sapova2022variational}
have introduced MCPs. These pools scale linearly with
the number of qubits yet retain sufficient expressivity
to reach any state of the Hilbert space. However,
selecting the optimal elements to seed an MCP remains
a challenge. Existing strategies such as the Unitary
Selective Coupled-Cluster (USCC) approach~\cite{fedorov2022unitary}
isolate physically relevant starters via amplitude
thresholds and symmetry filtering from a preliminary
VQE run. While effective, this requires computing and
storing optimized amplitudes before screening can
begin~\cite{Haidar2025,Viswanathan2026}, replacing one
bottleneck with another.

\subsubsection{Dynamic generation via flow equations}

To overcome the limitations of static pre-screening,
we dynamically synthesize the most relevant operators
using flow generators. By
leveraging algebraic commutators derived from the
target Hamiltonian itself, we generate a compact,
highly expressive operator pool without the need for
prior classical optimizations.

The foundation of our pool generation relies on the
Variational Double Bracket Flow (VDBF) framework.
Originally introduced by Wegner~\cite{wegner1994flow} to
diagonalize Hamiltonians, double bracket flows have
recently been adapted by Shrikhande et
al.~\cite{shrikhande2025rapid} as a means to efficiently
estimate ground state energies via classical Pauli
propagation. The flow is defined by the following
differential equation
\begin{equation}
\frac{d}{ds}H(s) = [G(s), H(s)]
= [[H_D(s), H(s)], H(s)]
\end{equation}
where $G(s)$ is an anti-Hermitian flow generator,
describing the evolution of the Hamiltonian along a
flow parameter $s$. To drive the system towards a
specific target state rather than achieving full
diagonalization, one defines $H_D = \rho$, where
$\rho$ is the projector onto a reference computational
basis state. Taking the zero state
$\rho = |0^N\rangle\langle 0^N|$ and expanding the
projector in the Pauli basis gives
\begin{equation}
|0\rangle\langle 0| = \frac{I + Z}{2},
\end{equation}
so that for an $N$-qubit system
\begin{equation}
|0^n\rangle\langle 0^n|  \propto I + \sum_i Z_i
+ \sum_{i<j} Z_i Z_j + \cdots.
\end{equation}
To circumvent the exponential cost of evaluating the
full projector, we truncate it to retain only the
one-body $Z_i$ terms, yielding the state-specific flow
generator
\begin{equation}
\label{eq:comm}
G(s) = \sum_i \left[ Z_i, H(s) \right]
\end{equation}

In practice we use this generator only at $s = 0$ and with the projector
truncated to its one-body terms; no flow is integrated. At this order the
Pauli support of $G(0)$ consists of the strings $P Z_i$ with $P \in H$ and
$\{P, Z_i\} = 0$, i.e.\ the Hamiltonian terms that couple the reference to
other determinants. This screening is related to the direct interaction set
used to select generators, as used in \cite{shrikhande2025rapid}, but differs
in that no energy-gradient ranking is performed and the complete set is
appended as a layer. Higher orders ($s > 0$, or two-body projector terms)
would expose further excitations at a higher classical cost and with a
larger pool. Since the purpose of this step is to supply physically relevant
starters for the MCP, the lowest order is sufficient.

Classical VDBF algorithms such as the one proposed
by~\cite{shrikhande2025rapid} remain in the Heisenberg
picture, using $G(s)$ to sequentially rotate the
Hamiltonian via Pauli propagation. Although relying
strictly on this one-body approximated flow for exact
classical Hamiltonian diagonalization can lead to slow
convergence, our method avoids this pitfall. We do not
use $G(s)$ as a rigid evolution operator, but rather as
a mathematical tool to identify a targeted Pauli
operator pool. The subsequent variational relaxation
of the parameters $\theta_{j,k}$ within our layered
quantum ansatz recovers the higher-order correlations
that the truncated classical flow would otherwise
capture too slowly. We therefore remain in the
Schr\"odinger picture, repurposing $G(s)$ not as an
evolution operator but as a generator to construct a
highly expressive variational manifold.

Evaluating the commutator in the equation above yields
a linear combination of Pauli operators. These
operators are isolated as individual Pauli strings and
added to our candidate pool. In classical VDBF
implementations, the coefficients of the Pauli
rotations are fixed by the numerical integration of
the flow equation. In contrast, by injecting these
operators into our layered VQE framework, their
respective amplitudes $\theta_{j,k}$ become free
variational parameters.

The resulting candidate operators are then symmetry-filtered with respect to
the Hartree--Fock reference: a candidate is retained only if, applied to
$|\psi_0\rangle$, it yields a determinant with the same numbers of $\alpha$
and $\beta$ electrons. This restricted filter is used for H$_2$O and
C$_3$H$_6$, for the hydrogen chains and BeH$_2$, a looser filter is used that
only preserves the parity of $N_\alpha$ and $N_\beta$. Some of them serve as starters for a minimal complete pool built with the construction algorithm of~\cite{Viswanathan2026}, which guarantees that the final pool generates the full target Lie algebra, the remaining candidates are added to it, as detailed below. The restricted filter guarantees the correct sector for the first excitation out of the reference only. An individual Pauli string does not commute with the particle-number or spin operators, and acting on other determinants it can change $N_{\alpha}$ and $N_{\beta}$. The pool therefore  does not preserve the total  particle number, $S_z$ or total spin. The state may thus acquire weight outside the ($N_{\alpha}$, $N_{\beta}$) sector of the reference, as well as spin contamination.

\textit{Pool structure and size.} The pool is the union of two sets. The
first is a minimal complete pool (MCP) of $2N-4$ generators, constructed with the selected starters
following Ref.~\cite{Viswanathan2026}; it scales linearly with $N$ and guarantees that the pool
generates the target Lie algebra. The second consists of the  remaining VDBF excitations
added to the MCP, which make the ansatz converge
faster; their number is bounded by the number of Hamiltonian Pauli strings
and grows as $O(N^4)$. Table~\ref{tab:operator_comparison} reports both. The linear scaling therefore
refers to the MCP; the layer size is dominated by the added excitations and
remains 4 to 25 times smaller than the QEB pool.

\subsection{Grow-then-compress loop}
\label{sec:growthencompress}
The ansatz is constructed iteratively in a layered
fashion, drawing operators from the complete pool
generated in Section~\ref{sec:pool}. Each iteration
proceeds through three steps: target acquisition,
overlap-guided compression, and energy re-optimization. 

\paragraph{Target acquisition.}
At iteration $n$, the algorithm appends the full pool
as a new layer to the current ansatz and re-optimizes all parameters of the ansatz. The resulting state defines the uncompressed
target $|\Phi_{\mathrm{target}}\rangle$ for the
subsequent compression.

\paragraph{Overlap-guided compression.} The compressed layer is built on top of
the frozen base $|\psi_{\rm base}\rangle$, which is the last accepted
compressed ansatz, or $|\psi_0\rangle$ at the first iteration. All states
are real, and the generators
$A_k = iP_k$ are real antisymmetric matrices ($P_k$ contains an odd number
of $Y$ operators), so that $\exp(i\theta P_k) = \exp(\theta A_k)$ maps real
states to real states. We therefore minimize the infidelity
\begin{equation}
C(\boldsymbol\theta) = 1 - \langle \Phi_{\rm target} |
\psi(\boldsymbol\theta)\rangle^2 ,
\end{equation}
where $|\psi(\boldsymbol\theta)\rangle$ is the base followed by the
generators inserted so far. New generators are always inserted at the
outermost position (acting last) with $\theta = 0$. For such an insertion,
\begin{equation}
\left.\frac{\partial C}{\partial\theta_k}\right|_{\theta_k=0}
= -\langle\Phi_{\rm target}|\psi\rangle\, g_k ,
\qquad
g_k = 2\langle \Phi_{\rm target}|A_k|\psi\rangle ,
\end{equation}
and since $\langle\Phi_{\rm target}|\psi\rangle$ is common to all candidates,
ranking by the overlap gradient $|g_k|$ is equivalent to ranking by the
infidelity gradient.

Each insertion step proceeds as follows. (i) The overlap gradient $g_k$ is
evaluated for every generator of the full pool; the pool is restored at every
step, so a generator may be inserted more than once. (ii) Candidates are
tried in order of decreasing $|g_k|$: each is placed at the outermost
position with $\theta = 0$, and all parameters of the compressed layer are
re-optimized with the L-BFGS-B algorithm to minimize $C$. (iii) A candidate is rejected,
and the next one tried, if its optimized angle satisfies
$|\theta| < \theta_{\rm rej}$. The pass ends when the infidelity gain of an
insertion falls below $\Delta C_{\min}$; when $\max_k |g_k| < g_{\rm null}$,
i.e.\ no generator improves the fidelity to first order; after $n_{\rm rej}$
consecutive rejections; or when the compressed layer reaches its maximum length
$\max(1, \lfloor f M \rfloor)$, where $f$ is the compression fraction
(typically $f = 0.5$).

\paragraph{Cost of the overlap gradient.} In standard ADAPT-VQE, the energy
gradient of a single candidate costs $N_H$ measurements. The overlap gradient bypasses
the Hamiltonian entirely. Because $A_k$ and the states are real,
$\langle\Phi_{\rm target}|A_k|\psi\rangle$ is real, and since
$e^{\frac{\pi}{2}A_k} = A_k$ for a Pauli generator,
\begin{equation}
|g_k|^2 = 4\left|\langle\Phi_{\rm target}|
e^{\frac{\pi}{2}A_k}|\psi\rangle\right|^2 .
\end{equation}
The magnitude of each overlap gradient is therefore obtained from a single
overlap evaluation, reducing the  cost from $N_H$  measurements to 1. The
sign of $g_k$, which is not needed for ranking, is not recovered. In
emulation, all $g_k$ are computed exactly.

\paragraph{Energy re-optimization.}
The compressed ansatz undergoes a final energy optimization over all its
parameters, the base is held fixed only during the compression step. The
outer loop then returns to target acquisition, using
the compressed ansatz as the new baseline, and
terminates when the energy change between iterations
falls below a convergence threshold $\varepsilon$ or a
maximum iteration count is reached.

\subsection{Subspace projection}
\label{sec:hproj}
For systems whose qubit count exceeds the available
register, we project the Hamiltonian onto a smaller
subspace before the grow-then-compress loop begins.
 
\subsubsection{Subspace construction}

Starting from the Hartree--Fock determinant
$|\psi_0\rangle$, we define the set of all
determinants reachable by repeated application of the
excitation patterns present in the Hamiltonian's Pauli
strings. 

We select $k$ linearly independent excitation patterns
$B_0, \dots, B_{k-1}$ and define the projected
subspace as the $K = 2^k$ determinants obtained by
applying all possible combinations of these patterns
to $|\psi_0\rangle$:
\begin{equation}
\mathcal{S}_k = \mathrm{span}\{|D_1\rangle,
|D_2\rangle, \dots, |D_K\rangle\},
\end{equation}
where each $|D_i\rangle$ is a computational-basis
determinant obtained by applying a combination of $k$
linearly independent excitation patterns
$B_0, \dots, B_{k-1}$ to $|\psi_0\rangle$.
For example, with $k = 2$ excitation patterns $B_0$
and $B_1$, the subspace contains four determinants: $|\psi_0\rangle$ itself,
$B_0|\psi_0\rangle$, $B_1|\psi_0\rangle$, and
$B_0B_1|\psi_0\rangle$. Although each pattern is chosen in the sector of the reference, combinations
of patterns can produce determinants with different numbers of $\alpha$ and
$\beta$ electrons. For H$_4$ at $k=4$, for instance, $B_0$
($1\alpha 1\beta \to 2\alpha 2\beta$) and $B_3$
($0\alpha 0\beta \to 2\alpha 2\beta$) each conserve $N_\alpha = N_\beta = 2$,
but applying both to $|\psi_0\rangle$ flips orbitals $2\alpha 2\beta$
twice and empties $0\alpha 0\beta 1\alpha 1\beta$, giving a determinant with no electrons; 4 of the 16 determinants of $S_4$
lie outside the $(N_\alpha, N_\beta)$ sector.
$S_k$ therefore contains such determinants, and $H_{\mathrm{proj}}$ is
defined on all of them.

The projected Hamiltonian, $H_{\mathrm{proj}}$, is obtained by computing
the matrix elements of $H$ in the basis
$\{|D_1\rangle, \dots, |D_K\rangle\}$ and
re-expressing the resulting $K \times K$ operator as
a Pauli sum on $k$ qubits. The projected
$H_{\mathrm{proj}}$ typically contains far fewer
Pauli terms than the original $H$. 
The reduced
problem can then be passed to the grow-then-compress
loop (Section~\ref{sec:growthencompress}) or solved
by exact diagonalization for validation.

\subsubsection{Selection of excitation patterns}

The quality of the projected subspace depends on which
$k$ excitation patterns are chosen. We rank candidates
using a two-stage procedure: a greedy sweep that
determines optimal angles, followed by a beam search that ranks the patterns by importance.

\paragraph{Step 1: greedy sweep.}
A single application of $H$ to $|\psi_0\rangle$ yields the coupling
$|\langle\psi_0|H A_j|\psi_0\rangle|$ for every pool generator
$A_j = iP_j$. The generators are
sorted by decreasing coupling magnitude. This
ordering defines the sequence in which the generators
appear in the product of exponentials.

Starting from $|\psi\rangle = |\psi_0\rangle$,
we sweep through the sorted generators one by one,
performing a greedy energy minimization inspired by
greedy VQE. For each generator $A_j$:
\begin{enumerate}
\item Compute $|\phi\rangle = A_j|\psi\rangle$ and orthogonalize against the
current state: $|\phi_\perp\rangle = (|\phi\rangle -
\langle\psi|\phi\rangle|\psi\rangle)/\|\cdot\|$. If the perpendicular norm is
below $10^{-8}$, the generator is assigned $\theta_j = 0$. 
\item Minimize the energy exactly in the
two-dimensional space
$\{|\psi\rangle, |\phi_\perp\rangle\}$:
$2\theta_j = \mathrm{atan2}(C, -B)$, where
$B = (H_{00} - H_{11})/2$ and $C = H_{01}$.
\item Update $|\psi\rangle \leftarrow
\cos\theta_j\,|\psi\rangle
- \sin\theta_j\,|\phi_\perp\rangle$ so that
subsequent generators are evaluated against the
improved state.
\end{enumerate}
Each generator requires one application of $H$, so
the full sweep costs $M + 1$ $H$-applications. The
coupling determines the order; the greedy sweep
determines the angles. No angle is ever revisited.

\paragraph{Step 2: beam search and importance
ranking.}
The state has the form of a product
$\prod_j(\cos\theta_j\, I - \sin\theta_j\, A_j)|\psi_0\rangle$.

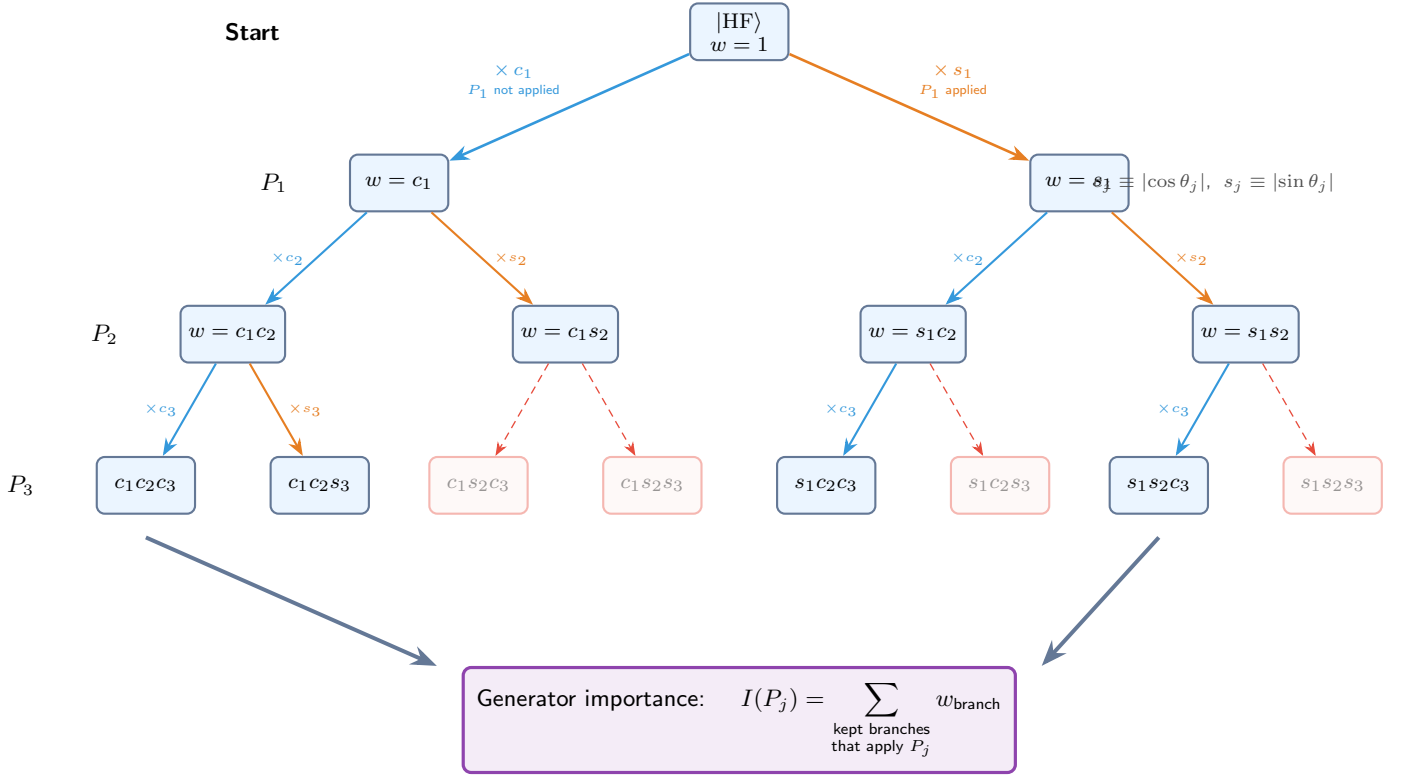
\begin{figure*}[htbp]
\centering
 
\definecolor{prunecol}{RGB}{231, 76, 60}
\definecolor{coscolor}{RGB}{52, 152, 219}
\definecolor{sincolor}{RGB}{230, 126, 34}
\definecolor{keepfill}{RGB}{235, 245, 255}
\definecolor{keepdraw}{RGB}{100, 120, 150}
\definecolor{scorecol}{RGB}{142, 68, 173}
 
\begin{tikzpicture}[
    font=\sffamily\footnotesize,
    >=Stealth,
    kept/.style={
        rectangle,
        rounded corners=3pt,
        draw=keepdraw,
        line width=0.8pt,
        fill=keepfill,
        align=center,
        minimum height=0.75cm,
        minimum width=1.3cm,
        inner sep=3pt
    },
    pruned/.style={
        rectangle,
        rounded corners=3pt,
        draw=prunecol,
        line width=0.8pt,
        fill=prunecol!8,
        align=center,
        minimum height=0.75cm,
        minimum width=1.3cm,
        inner sep=3pt,
        opacity=0.4
    },
    scorebox/.style={
        rectangle,
        rounded corners=3pt,
        draw=scorecol,
        line width=1.2pt,
        fill=scorecol!10,
        align=center,
        minimum height=0.7cm,
        inner sep=5pt,
        font=\sffamily\small
    },
    coslbl/.style={font=\sffamily\scriptsize, coscolor,
                   align=center},
    sinlbl/.style={font=\sffamily\scriptsize, sincolor,
                   align=center},
    steplbl/.style={font=\sffamily\bfseries\small,
                    anchor=east}
]
 
\def\dy{-2.0cm}
\def\dxA{4.5cm}
\def\dxB{2.2cm}
\def\dxC{1.15cm}
 
\node[kept] (root) at (0,0)
    {$|\mathrm{HF}\rangle$\\[-1pt]$w = 1$};
 
\node[steplbl] at ($(root.west) + (-5.3, 0)$) {Start};
 
\node[kept] (l1a) at (-\dxA, \dy)
    {$w = c_1$};
\node[kept] (l1b) at (\dxA, \dy)
    {$w = s_1$};
 
\draw[->, coscolor, line width=1pt] (root) -- (l1a)
    node[midway, above left, coslbl]
    {$\times\,c_1$\\[-1pt]\tiny $P_1$ not applied};
\draw[->, sincolor, line width=1pt] (root) -- (l1b)
    node[midway, above right, sinlbl]
    {$\times\,s_1$\\[-1pt]\tiny $P_1$ applied};
 
\node[steplbl] at ($(l1a.west) + (-0.7, 0)$) {$P_1$};
 
\node[anchor=west, font=\sffamily\scriptsize, text=black!70]
    at (\dxA + 1.8, \dy)
    {$c_j \equiv |\!\cos\theta_j|$,\;
     $s_j \equiv |\!\sin\theta_j|$};
 
\node[kept] (l2a) at (-\dxA-\dxB, 2*\dy)
    {$w = c_1 c_2$};
\node[kept] (l2b) at (-\dxA+\dxB, 2*\dy)
    {$w = c_1 s_2$};
\node[kept] (l2c) at (\dxA-\dxB, 2*\dy)
    {$w = s_1 c_2$};
\node[kept] (l2d) at (\dxA+\dxB, 2*\dy)
    {$w = s_1 s_2$};
 
\draw[->, coscolor, line width=0.8pt] (l1a) -- (l2a)
    node[midway, left, coslbl, xshift=-1pt] {\tiny$\times c_2$};
\draw[->, sincolor, line width=0.8pt] (l1a) -- (l2b)
    node[midway, right, sinlbl, xshift=1pt] {\tiny$\times s_2$};
\draw[->, coscolor, line width=0.8pt] (l1b) -- (l2c)
    node[midway, left, coslbl, xshift=-1pt] {\tiny$\times c_2$};
\draw[->, sincolor, line width=0.8pt] (l1b) -- (l2d)
    node[midway, right, sinlbl, xshift=1pt] {\tiny$\times s_2$};
 
\node[steplbl] at ($(l2a.west) + (-0.7, 0)$) {$P_2$};
 
 
\node[kept] (l3a) at (-\dxA-\dxB-\dxC, 3*\dy)
    {$c_1 c_2 c_3$};
\node[kept] (l3b) at (-\dxA-\dxB+\dxC, 3*\dy)
    {$c_1 c_2 s_3$};
 
\node[pruned] (l3c) at (-\dxA+\dxB-\dxC, 3*\dy)
    {$c_1 s_2 c_3$};
\node[pruned] (l3d) at (-\dxA+\dxB+\dxC, 3*\dy)
    {$c_1 s_2 s_3$};
 
\node[kept] (l3e) at (\dxA-\dxB-\dxC, 3*\dy)
    {$s_1 c_2 c_3$};
\node[pruned] (l3f) at (\dxA-\dxB+\dxC, 3*\dy)
    {$s_1 c_2 s_3$};
 
\node[kept] (l3g) at (\dxA+\dxB-\dxC, 3*\dy)
    {$s_1 s_2 c_3$};
\node[pruned] (l3h) at (\dxA+\dxB+\dxC, 3*\dy)
    {$s_1 s_2 s_3$};
 
\draw[->, coscolor, line width=0.8pt] (l2a) -- (l3a)
    node[midway, left, coslbl, xshift=-1pt] {\tiny$\times c_3$};
\draw[->, sincolor, line width=0.8pt] (l2a) -- (l3b)
    node[midway, right, sinlbl, xshift=1pt] {\tiny$\times s_3$};
\draw[->, prunecol, line width=0.5pt, densely dashed] (l2b) -- (l3c);
\draw[->, prunecol, line width=0.5pt, densely dashed] (l2b) -- (l3d);
\draw[->, coscolor, line width=0.8pt] (l2c) -- (l3e)
    node[midway, left, coslbl, xshift=-1pt] {\tiny$\times c_3$};
\draw[->, prunecol, line width=0.5pt, densely dashed] (l2c) -- (l3f);
\draw[->, coscolor, line width=0.8pt] (l2d) -- (l3g)
    node[midway, left, coslbl, xshift=-1pt] {\tiny$\times c_3$};
\draw[->, prunecol, line width=0.5pt, densely dashed] (l2d) -- (l3h);
 
\node[steplbl] at ($(l3a.west) + (-0.7, 0)$) {$P_3$};
 
\node[scorebox, below=2.0cm of root |- l3a.south] (scores)
    {Generator importance:
     \quad $I(P_j) = \displaystyle\sum_{\substack{
       \text{kept branches}\\
       \text{that apply } P_j}} w_{\text{branch}}$};
 
\draw[->, keepdraw, line width=1.5pt]
    ($(l3a.south) + (0, -0.3)$) -- ($(scores.north) + (-4, 0)$);
\draw[->, keepdraw, line width=1.5pt]
    ($(l3g.south) + (0, -0.3)$) -- ($(scores.north) + (4, 0)$);
 
 
\end{tikzpicture}
 
\caption{Beam search for scoring generator importance,
illustrated with $M = 3$ generators and beam width
$2^k = 4$ ($k = 2$). At each generator $P_j$, every
surviving branch splits into two: one where $P_j$ is
applied (orange, weight $\times\,|\!\sin\theta_j|$)
and one where it is not (blue, weight
$\times\,|\!\cos\theta_j|$). Only the $2^k$ heaviest
branches survive (solid); the rest are pruned (dashed).
Each generator's importance is the total weight of the
surviving branches that include it. The generators are
ranked by this score, and the first $k$ in-sector,
linearly independent excitation patterns define the
basis of the projected subspace.}
\label{fig:beam_search}
\end{figure*}

To score each generator's contribution, we expand
this product approximately using a beam search of
width $2^k$ (Figure~\ref{fig:beam_search}). At each
generator, every surviving branch splits into two:
one where the generator is applied (weight
$\times\,|\sin\theta_j|$) and one where it is not
(weight $\times\,|\cos\theta_j|$). Only the $2^k$
heaviest branches are kept. Generators with
$\theta_j = 0$ are skipped. Each generator's importance $I(P_j)$ is the total
weight of the surviving branches in which $P_j$ was
applied. The generators are sorted by decreasing
importance, and the first $k$ whose
excitation patterns are linearly independent are
selected. Only the excitation pattern of each
selected generator is used, the subspace
$\mathcal{S}_k$ is the full span of all $2^k$
combinations of these $k$ patterns applied to
$|\psi_0\rangle$.

The sweep runs once regardless of $k$. When the
subspace is grown adaptively, only the beam search
is repeated at each new $k$, since the beam width
$2^k$ must match the current subspace size.
In Figure~\ref{fig:beam_search}, we give an example of this branching  process  
for a three-generator
example with $k = 2$.

\subsubsection{Adaptive-$k$ growth with PT2 stopping}

Rather than fixing $k$ in advance, we grow the
subspace one excitation pattern at a time. At each
$k$, we evaluate how much energy is missing from the
subspace using a second-order Epstein--Nesbet
estimate:
\begin{equation}
\Delta E^{(2)}_k = \sum_{D \notin \mathcal{S}_k}
\frac{|\langle D|H|\psi_{\mathrm{ref}}\rangle|^2}
{E_{\mathrm{ref}} - \langle D|H|D\rangle}.
\end{equation}
The numerators are obtained from a single sparse
application of $H$ to $|\psi_{\mathrm{ref}}\rangle$
within the reachable determinant set. The diagonal
energies $\langle D|H|D\rangle$ are read from the
diagonal terms of $H$.

Two choices of reference are available:

\begin{itemize}
\item \textbf{Exact reference:}
$|\psi_{\mathrm{ref}}\rangle$ is the ground state of
$H_{\mathrm{proj}}$ at the current $k$, obtained by
dense diagonalization of the $2^k \times 2^k$
projected Hamiltonian. This captures the full energy
missing from the subspace but requires diagonalization
and one $H$-application at every $k$.

\item \textbf{Hartree--Fock reference:}
$|\psi_{\mathrm{ref}}\rangle = |\psi_0\rangle$.
The couplings and diagonal energies are computed once
and reused at every $k$, requiring no diagonalization.
This variant only sees determinants directly coupled
to HF and gives a smaller estimate, but is
substantially cheaper.
\end{itemize}

The growth stops when $|\Delta E^{(2)}_k|$ falls below
a tolerance $\tau$ or a maximum $k_{\max}$ is reached.

The exact-reference estimate requires diagonalizing $H_{\rm proj}$ at every
$k$; it is used here to validate the stopping criterion. When it is used,
the projected problem has in effect been solved classically, and the
subsequent \methodename run on $H_{\rm proj}$ serves to assess circuit cost. When $H_{\mathrm{proj}}$ is to be solved on a quantum device, the
Hartree--Fock-reference estimate is the applicable criterion: it is cheaper,
but it can lie above or below the missing energy
(Figures~\ref{fig:hproj_H4}--\ref{fig:hproj_H2O_631g}), so a tighter
tolerance $\tau$ is advisable.

\subsubsection{Solution on the projected register}

On the projected register we do not build a new operator pool: we reuse the
pool of Section~\ref{sec:pool}. Each generator of that pool is an excitation
that turns one determinant into another. We keep only the generators whose
excitation can be written as a combination of the $k$ selected patterns
$B_0,\dots,B_{k-1}$. Such a generator never takes the state outside the
subspace $S_k$, so it can be rewritten as a Pauli string on the $k$ qubits,
in the same way as $H_{\mathrm{proj}}$ itself. The other generators would
leave $S_k$ and are discarded. For H$_4$ at $k=4$, 14 of the 27 generators
are kept.

Within the subspace, each kept generator performs exactly the same rotation
as on the full register. The ansatz on the $k$ qubits is therefore the
original ansatz restricted to $S_k$. Like the original pool
(Section~\ref{sec:pool}), it does not conserve particle number or spin: the
subspace contains determinants outside the $(N_\alpha, N_\beta)$ sector of
the reference, and the projected generators can couple them. This projected
pool is then used as usual: we either diagonalize $H_{\mathrm{proj}}$
exactly, which gives the best energy available in the subspace, or run
\methodename iterations (Section~\ref{sec:growthencompress}) on the $k$ qubits. Both
are reported in Figures~\ref{fig:hproj_H4}--\ref{fig:hproj_H2O_631g}.

Fewer qubits do not automatically mean cheaper gates. For H$_4$ at $k = 4$,
a double excitation acting on 4 of the 8 spin orbitals becomes a string
acting on 3 of the 4 projected qubits; for H$_2$O (6-31G) at $k = 20$, a
projected generator costs about 19 CNOTs on average, against about 6 on the
full register.

This cost depends on the choice of patterns, not on the subspace itself and therefore can be reduced: any
other basis of the same span describes the same determinants and leaves every
energy unchanged, but changes how many qubits each generator acts on. We
therefore choose the basis that minimizes the CNOT count of the projected
pool, and compress the resulting layer with the overlap-guided procedure of
Section~\ref{sec:growthencompress}, ranking candidates by gradient per added
CNOT. Details of this method are given in the Supplementary Material.

\subsection{Classical Post-Processing: ENPT2 and stochastic evaluation }
\label{sec::PT2}

We apply a second-order Epstein--Nesbet perturbation
theory (ENPT2) correction classically to the VQE
energy, following the formulation of Ryabinkin et
al.~\cite{Ryabinkin2021}. This type of correction has
been widely used in the closely related family of
selected Configuration Interaction (sCI)
methods~\cite{chien2018excited, garniron2019quantum},
where it refines the energy of a truncated
wavefunction and vanishes identically when the full
configuration space is included.

\subsubsection{Formulation}

Let $\mathcal{H}$ denote the Hilbert space of $N$ qubits and $H$ the Hamiltonian
acting on it. Let $U = \prod_j e^{i\theta_j P_j}$ be the optimized ansatz
unitary of Eq.~(1), so that $|\Psi\rangle = U|\psi_0\rangle$ is the VQE state,
and define the dressed Hamiltonian $\tilde H = U^\dagger H U$. The reference
energy is the VQE energy,
\begin{equation}
E_0 = \langle\psi_0|\tilde H|\psi_0\rangle = \langle\Psi|H|\Psi\rangle .
\end{equation}

Let $\mathcal{P}_X = \{I, X\}^{\otimes N}$ be the set of all $2^N$ Pauli
strings built from the identity and $\sigma_x$. Each $X_j \in \mathcal{P}_X$
maps $|\psi_0\rangle$ onto a determinant $|D_j\rangle = X_j|\psi_0\rangle$,
whose diagonal energy and coupling are
\begin{align}
E_j &= \langle\psi_0|X_j \tilde H X_j|\psi_0\rangle, \\
H_{0j} &= \langle\psi_0|\tilde H X_j|\psi_0\rangle .
\end{align}

The ENPT2 correction is then
\begin{equation}
\label{eq:enpt2}
\Delta E^{(2)}_{\mathrm{ENPT}}
= -\sum_{X_j \in \mathcal{P}_X \setminus
\{I^{\otimes N}\}}
\frac{|H_{0j}|^2}{E_j - E_0}.
\end{equation}

For all systems considered in this work, we verified
that the Diagonal Unitary Correction (DUC) formula
(Eq.~25 of~\cite{Ryabinkin2021}) gives equivalent
results.

\subsubsection{Efficient evaluation via the auxiliary
vector}

The key computational insight is that all $2^N$
coupling coefficients $H_{0j}$ can be extracted
simultaneously. Define the auxiliary state vector

\begin{equation}
|w\rangle = U^\dagger H|\Psi\rangle = \tilde H|\psi_0\rangle.
\end{equation}

Since $|D_j\rangle$ is a computational basis state, the coupling reduces to
\begin{equation}
H_{0j} = \langle D_j|w\rangle .
\end{equation}

A single evaluation of $|w\rangle$ therefore yields
every numerator in Eq.~\ref{eq:enpt2}. The
denominators $E_j - E_0$ need only be computed for
indices $j$ where $|H_{0j}|^2$ exceeds a threshold
$\epsilon$, each requiring one application of
$\tilde H$. The cost is dominated by the number
of such active terms $N_{\mathrm{active}}$, which
motivates the stochastic approach below when
$N_{\mathrm{active}}$ is large.

Algorithm~\ref{alg:enpt2} summarizes the deterministic
evaluation.

\begin{algorithm}[H]
\caption{Deterministic ENPT2}
\label{alg:enpt2}
\begin{algorithmic}[1]
\Require Hamiltonian $H$, ansatz $U$, Hartree–Fock state $|\psi_0\rangle$, threshold $\eta$
\Ensure ENPT2 correction $\Delta E$
\State $|\Psi\rangle \leftarrow U|\psi_0\rangle$, $E_0 \leftarrow \langle\Psi|H|\Psi\rangle$,
       \; $E_0 \gets \langle\psi_0|H|\psi_0\rangle$
\State$|w\rangle \leftarrow U^\dagger H|\Psi\rangle$
       \Comment{all couplings at once}
\State $\Delta E \gets 0$
\For{$j = 1$ to $2^N - 1$}
  \If{ $|\langle D_j|w\rangle|^2 > \eta$}
    \State $E_j \leftarrow \langle\psi_0|X_j \tilde H X_j|\psi_0\rangle$
    \State $\Delta E \leftarrow \Delta E - |\langle D_j|w\rangle|^2/(E_j - E_0)$
  \EndIf
\EndFor
\State \Return $\Delta E$
\end{algorithmic}
\end{algorithm}

\subsubsection{Stochastic evaluation}

For large systems, the number of active terms grows
prohibitively even after filtering out zero couplings.
We estimate the sum in Eq.~\ref{eq:enpt2}
stochastically using importance sampling with the
distribution
\begin{equation}
q_j = \frac{|H_{0j}|^2}{\sum_i |H_{0i}|^2}
\end{equation}
where only nonzero couplings are included. This
distribution is near-optimal, since the variance of
the estimator is minimized when $q_j$ is proportional
to the absolute value of the summand (see
Supplementary Material for the
derivation). Defining
\begin{equation}
f_j = -\frac{|H_{0j}|^2}{E_j - E_0},
\end{equation}
the ENPT2 correction is $\Delta E^{(2)}_{\mathrm{ENPT}}
= \sum_j f_j$, estimated by drawing $N_s$ samples from
$q$ and computing
\begin{equation}
\hat{S}_{N_s} = \frac{1}{N_s}\sum_{m=1}^{N_s} \frac{f_{I_m}}{q_{I_m}}.
\end{equation}
The variance of this estimator provides the error
estimate $\sigma$ reported in
Table~\ref{tab:stochastic_pt2}.

All couplings $H_{0j}$ are obtained from the auxiliary
vector $|w\rangle$ as in the deterministic case. Only
the sampled denominators $E_j - E_0$ need to be
evaluated individually, which is what makes the
stochastic approach efficient: the expensive per-term
computation is restricted to a small random subset.

Algorithm~\ref{alg:stochastic_enpt2} summarizes the
procedure.

\begin{figure}[H]
\begin{algorithm}[H]
\caption{Stochastic ENPT2}
\label{alg:stochastic_enpt2}
\begin{algorithmic}[1]
\Require Hamiltonian $H$, ansatz  $U$, batch size
$\Delta N_s$, target tolerance $\epsilon$, $\psi_0$ (Hartree–Fock state)
\Ensure Correction $\Delta E$, standard error $\sigma$
\State $|\Psi\rangle \leftarrow U|\psi_0\rangle$, $E_0 \leftarrow \langle\Psi|H|\Psi\rangle$,
       \; $E_0 \gets \langle\psi_0|H|\psi_0\rangle$
\State  $|w\rangle \leftarrow U^\dagger H|\Psi\rangle$
       \Comment{all couplings at once}
\State $W \leftarrow \sum_{j>0} |\langle D_j|w\rangle|^2$, $q_j \leftarrow |\langle D_j|w\rangle|^2/W$
       \; $q_j \gets |\langle j|w\rangle|^2 / W$
\State $\mathcal{S} \gets \emptyset$,
       \; $\Delta E \gets 0$
\Repeat
  \State $\Delta E_{\mathrm{prev}} \gets \Delta E$
  \State Draw $\Delta N_s$ samples from $q$ and append
         to $\mathcal{S}$
  \State $N_s \gets |\mathcal{S}|$
  \For{each unique state $j$ in $\mathcal{S}$ not yet
       evaluated}
    \State $E_j \leftarrow \langle\psi_0|X_j \tilde H X_j|\psi_0\rangle$
  \EndFor
  \State $m_j \gets$ occurrences of $j$ in
         $\mathcal{S}$
  \State $v_j \gets -W / (E_j - E_0)$
         \Comment{value per draw of state $j$}
  \State $\Delta E \gets
        \frac{1}{N_s}\sum_j m_j v_j$
  \State $\sigma \gets
      \sqrt{\frac{1}{N_s(N_s-1)}\sum_j m_j (v_j-\Delta E)^2}$
\Until{$\sigma \leq \epsilon$ \textbf{and}
       $|\Delta E - \Delta E_{\mathrm{prev}}|
       \leq \epsilon$}
\State \Return $\Delta E$, $\sigma$
\end{algorithmic}
\end{algorithm}
\vspace{0.2cm}
\end{figure}

The loop stops when the standard error $\sigma$ falls below the tolerance
$\epsilon$, or when the draw budget is exhausted, the change of the estimate
between rounds is monitored as a check on $\sigma$.

Several practical refinements are used in our
implementation.  A maximum draw budget $N_s^{\max}$
guarantees termination. When $E_j$ is close to $E_0$,
the ENPT2 term diverges; in such cases we substitute
the finite Diagonal Unitary Correction (DUC) formula
(Eq.~25 of~\cite{Ryabinkin2021}).  

\textit{Cost and scaling.} Both variants require the auxiliary vector
$|w\rangle$ in full, since the couplings are read from all $2^N$ of its
components. Stochastic sampling
therefore reduces the number of denominators evaluated, not the memory, which
remains that of a dense state vector. Each denominator costs roughly one
energy evaluation of the dressed Hamiltonian. For C$_3$H$_6$ (32 qubits) on a single
NVIDIA H100 GPU, $E_0$ took 26~s, $|w\rangle$ 10~s, and each denominator
28.5~s; the 1484 unique denominators took 11.8~h. The run stopped at its draw
budget (5000 unique states) before reaching the target standard error of
$10^{-6}$~Ha; the achieved $\sigma = 4.5\times10^{-5}$~Ha is nonetheless
$\sim$35 times below chemical accuracy. The smallest denominator was 1.30~Ha
and none was negative, so the DUC substitution was not triggered. ENPT2 thus
applies in the regime where the ansatz can be emulated, systems beyond single-GPU memory require a distributed state vector, a
distributed implementation, not used in this work, will be presented in a
forthcoming paper.

\section*{Data availability}
The datasets generated and analyzed during the current study are available from the corresponding author on reasonable request.

\section*{Code availability}
The classical emulation and algorithmic logic code utilized for this study are available from the corresponding author on reasonable request.

\section*{Acknowledgments}
This work has received funding from the European Research Council (ERC) under the European Union’s Horizon 2020 research and innovation program (grant agreement No 810367), project EMC2 (JPP) and from HQI (JPP). This work was also supported by the French State through the France 2030 program (HYPERION project, i-Demo n°2 managed by Bpifrance). Computations have been performed at IDRIS (Jean Zay) on GENCI Grants: no A0150712052 (J.-P. P.) and no A0200815824 (G.M.). The authors thank Baptiste Claudon for his help on the hardware implementation. Computations on the Quantinuum System Model H2 machines have been made possible through the Quantinuum Startup Partner Program.


\section*{Competing interests}
Jean-Philip Piquemal is a shareholder and co-founder of Qubit Pharmaceuticals. All remaining authors declare no competing interests.


\section*{References}
 
\bibliographystyle{unsrt}
\bibliography{biblio}

\end{document}